\documentclass[aps,prd,preprintnumbers,showpacs,nofootinbib,amssymb]{revtex4}
\usepackage{graphicx}
\usepackage{amsmath}
\usepackage{amssymb}
\usepackage{amsfonts}
\usepackage{bm}
\usepackage{color}

\def\be{\begin{equation}}
\def\ee{\end{equation}}
\def\bea{\begin{eqnarray}}
\def\eea{\end{eqnarray}}

\begin{document}

\title{Kinetic theory of stellar systems and two-dimensional vortices}
\author{Pierre-Henri Chavanis}
%\email{chavanis@irsamc.ups-tlse.fr}
\affiliation{Laboratoire de
Physique Th\'eorique, Universit\'e de Toulouse, CNRS, UPS, France}

\begin{abstract}

We discuss the kinetic theory of stellar systems and two-dimensional vortices
and stress their analogies. We recall the derivation of the Landau and
Lenard-Balescu equations from the Klimontovich formalism. These equations
 take into account two-body correlations and are valid at the order $1/N$, where
$N$ is the number of particles in the system. They have the structure of a
Fokker-Planck equation involving a diffusion term and a drift term.  The
systematic drift of a vortex is the counterpart of the dynamical friction
experienced by a star. At equilibrium, the diffusion and the drift
terms balance each other establishing the Boltzmann distribution of
statistical mechanics. We discuss the problem of kinetic blocking in certain
cases and how it can be solved at the order $1/N^2$ by the consideration of
three-body correlations.
We also consider the behavior of the system close to the critical point
following a recent suggestion by Hamilton and Heinemann
(2023). We present a
simple calculation, valid for spatially homogeneous
systems with long-range interactions described by the Cauchy distribution,
showing how the consideration of the Landau modes regularizes the divergence of
the
friction by polarization at the critical point.

\end{abstract}


\maketitle

\section{Introduction}
\label{sec_intro}

There exist beautiful analogies between stellar systems and two-dimensional (2D)
vortices
\cite{these,csr,houchesPH,tcfd}. This is due to the fact that these systems
have ``attractive'' long-range interactions \cite{campabook}.
As a result, they
self-organize
spontaneously into coherent structures: globular clusters and galaxies
\cite{bt,paddy,ijmpb} in astrophysics 
and large-scale vortices (such as Jupiter's Great Red spot) in geophysical and
astrophysical flows \cite{bouchetvenaille,houchesPH}. Similar results are
obtained for other systems with long-range interactions such as the
Hamiltonian mean field (HMF) model
\cite{ar,cvb} or
spins with long-range interactions moving on a sphere (in the axisymmetric
limit) in relation to the
process of vector resonant
relaxation (VRR) in galactic nuclei \cite{fbc}. The relaxation
towards these organized states is
nontrivial and requires to develop a detailed kinetic theory. The literature on
the subject is vast. We draw the reader to the introduction of our
previous
papers \cite{aa,kinfdvortex,kinfd} for  a thorough overview of the history of
the
subject and an
extensive list of
references. We also refer to the review of Campa {\it et al.}
\cite{cdr} on
the physics of systems with long-range interactions and to the tutorial of
Hamilton and Fouvry \cite{hf} on the kinetic theory of stellar systems.

Self-gravitating systems experience two successive types of relaxation. In the
collisionless regime, they are described by the Vlasov-Poisson
equations \cite{jeans,vlasov}. These equations display a process of violent
collisionless relaxation to a metaequilibrium (or quasistationary) state on a
very short timescale of
the order of the dynamical time $t_D$. Lynden-Bell \cite{lb} has attempted
to predict this metaequilibrium state from a statistical theory. This
theory relies on an assumption of ergodicity (efficient mixing) which is not
always fulfilled in
practice. This is the difficult problem of incomplete violent relaxation
\cite{lb,incomplete}.\footnote{There is no Lynden-Bell equilibrium state (no
entropy maximum at fixed mass and energy) for 3D self-gravitating systems so the
violent relaxation is
necessarily incomplete in that case. When a  maximum entropy state in
the sense of Lynden-Bell
exists mathematically (e.g. for 1D or 2D self-gravitating systems or for other
systems
with long-range interactions such as the HMF model) this statistical equilibrium
state  is not necessarily reached by the system because of
incomplete relaxation (inefficient mixing). The statistical mechanics of
Lynden-Bell and the kinetic
theory of collisionless relaxation for systems with long-range interactions,
including the problem of incomplete relaxation, are
reviewed in \cite{lbnew,ewart}.} Then, on a
much longer timescale, increasing algebraically with the number $N$ of
particles, there is
a slow (secular) collisional relaxation due to finite $N$ effects
(granularities) leading, for $t\rightarrow +\infty$, to the Boltzmann
equilibrium
distribution predicted by conventional statistical
mechanics. This is the case for 1D and 2D self-gravitating systems and for other
systems with long-range interactions like the HMF model, VRR in galactic nuclei,
plasmas... For 3D stellar systems, there is no statistical
equilibrium state in a strict sense in an infinite domain and the system either
evaporates \cite{chandra} or undergoes core collapse (gravothermal catastrophe)
\cite{antonov,lbw}.\footnote{A statistical equilibrium state may exist if the
system is artificially enclosed within a box (in order to prevent its
evaporation) and if its energy is sufficiently high (in order to prevent the
gravothermal catastrophe). This equilibrium state has been considered
by Antonov \cite{antonov} and Lynden-Bell and Wood
\cite{lbw}. If the system is unconfined, the evaporation is a slow
process and there can be a form of statistical equilibrium state on an
intermediate timescale. It is described by the Michie-King distribution
\cite{michie,king,katzking,clm1} which takes into account the escape of high
energy stars.}

The kinetic theory of stellar systems was pioneered by Chandrasekhar
\cite{chandra,chandra1,chandra2,chandra3,chandrany,nice} by
analogy with the
theory of Brownian motion \cite{chandrabrown}. The theory of
stochastic gravitational fluctuations was developed in
\cite{cvn0,cvn1,cvn2,cvn3,cvn4,cvn5,kandruprep}.  Chandrasekhar assumed that the
system is  spatially homogeneous. This amounts to making a local
approximation. The
evolution of a test star in a thermal bath 
is
described by a Fokker-Planck equation of the Kramers type involving a diffusion
term and a friction term. The friction and the diffusion coefficients satisfy
the Einstein relation. This  Fokker-Planck equation relaxes
towards the Maxwell-Boltzmann distribution on a timecale $(N/\ln N)\, t_D$. At 
statistical
equilibrium, the diffusion and the friction balance each other.  On the other
hand, the kinetic evolution of a stellar system as a whole is governed by a
self-consistent  Fokker-Planck equation
\cite{rosenbluth,kingkin,henonbinary} which is equivalent to the Landau
\cite{landau} equation introduced in plasma physics. When collective effects are
taken into account one obtains the Lenard-Balescu \cite{lenard,balescu}
equation. The Landau and Lenard-Balescu equations were introduced in plasma
physics but they are valid for all systems with long-range
interactions in any dimension of space $d$  provided that they are
spatially homogeneous and stable \cite{landaud,epjp1,epjp2,epjp3}. 
These equations describe the
collisional evolution of the system at the order $1/N$ due to the
development of two-body
correlations. They conserve mass and energy and 
satisfy an $H$-theorem for the Boltzmann entropy. When $d>1$ they relax towards
the Maxwell-Boltzmann distribution. The relaxation time scales as $N
t_D$.\footnote{For 3D plasmas and 3D stellar systems the relaxation time scales
as
$(N/\ln N) t_D$ because of logarithmic corrections at small scales arising from
the effect of strong collisions. In plasma
physics, $N$
represents the number of charges in the Debye sphere (usually denoted
$\Lambda$). In stellar dynamics, $N$
represents the number of stars in the Jeans sphere which corresponds to the
typical size of the cluster ($R\sim \lambda_J$).}

For spatially homogeneous
systems with long-range interactions in $d=1$ (like the  HMF model),
the Landau and Lenard-Balescu collision terms vanish
identically \cite{epjp1,epjp2,epjp3} so there is no kinetic evolution at the 
order $1/N$. This is a situation of kinetic blocking due to the absence of
resonances. As a result, the system does not reach the Maxwell-Boltzmann
distribution
on a timescale $N t_D$.  In order to
describe the relaxation of the system towards the Maxwell-Boltzmann distribution
it is
necessary to take into account three-body correlations and develop the kinetic
theory at the order $1/N^2$. An explicit kinetic equation
that is valid at the order $1/N^2$ has been obtained recently by Fouvry {\it et
al.} \cite{fbcn2,fcpn2}  for arbitrary homogeneous 1D systems with long-range
interactions in the approximation where collective effects can be neglected.
Remarkably, this equation satisfies an $H$-theorem for the Boltzmann entropy and
relaxes towards the Maxwell-Boltzmann distribution. This implies that the
relaxation time scales as $N^2
t_D$ for homogeneous 1D systems with long-range interactions.

In reality, self-gravitating systems are spatially inhomogeneous.
For 3D self-gravitating systems, the local approximation is relatively accurate
but it leads to a logarithmic divergence at large scales.\footnote{This
logarithmic divergence occurs in the absence of collective effects. The problem
is even more serious when collective effects are taken into account because an
infinite homogeneous system is Jeans unstable (see Sec. \ref{sec_nontb}).} It
also leads to
completely wrong results in the case of stellar discs and 1D self-gravitating
systems that are strongly inhomogeneous. Other systems with long-range
interactions can also be inhomogeneous. This is the case of the HMF model below
a critical
energy $E_c$. Recently, the Lenard-Balescu equation has been generalized to
the case of spatially inhomogeneous systems by
Heyvaerts \cite{heyvaerts} and Chavanis \cite{angleaction2} using angle-action
variables. When collective effects are neglected it reduces to the
inhomogeneous Landau equation \cite{aa}. For 3D gravitational systems, the
proper treatment of spatial inhomogeneity removes the logarithmic divergence at
large scales. The
inhomogeneous  Landau and Lenard-Balescu equations conserve mass and energy and
satisfy an $H$-theorem for the Boltzmann entropy. They usually relax towards the
Boltzmann distribution on a timescale $N\, t_D$ except in the case of unconfined
3D stellar systems.\footnote{There is also a situation of kinetic blocking for
1D inhomogeneous systems with a monotonic frequency profile when ony $1:1$
resonances are permitted as discussed in Sec. \ref{sec_mono}.} In that case,
there is no
statistical equilibrium state
(no maximum entropy state) in a strict sense and the relaxation is hampered by
the phenomena of evaporation and gravothermal catastrophe (core collapse)
\cite{chandra,antonov,lbw}.

Two-dimensional point vortices also experience two successive types of
relaxation. In the collisionless regime, they are described by the Euler-Poisson
equations \cite{newton}. These equations experience a process of violent
collisionless relaxation to a metaequilibrium (or quasistationary) state on a
very short timescale of
the order of the dynamical time $t_D$. This metaequilibrium state can be
predicted by the Miller-Robert-Sommeria (MRS) statistical theory
\cite{miller,rs}, which is the hydrodynamic analogue of the Lynden-Bell
theory \cite{csr}.\footnote{The MRS theory applies either to the 2D point vortex
gas in the collisionless regime  or to continuous 2D incompressible flows in the
inviscid regime.} Again, the prediction of the MRS theory may be affected by
the problem of incomplete relaxation (the kinetic
theory of violent relaxation for 2D vortices and the problem of incomplete
relaxation are discussed in  \cite{rsmepp,csr,rr,csmepp,quasi,vphydro}). Then,
on a
much longer timescale, there is
a slow (secular) collisional relaxation of the point vortex gas due to finite
$N$ effects
(granularities) leading, for $t\rightarrow +\infty$, to the Boltzmann
equilibrium
distribution predicted by conventional statistical
mechanics. The statistical equilibrium state of 2D point vortices described
by the Boltzmann-Poisson equation has been considered by Joyce and
Montgomery \cite{jm,mj}, Kida \cite{kida}, and Pointin and Lundgren \cite{pl,lp}
following the pioneering work of Onsager \cite{onsager}.\footnote{In his seminal
paper on the statistical
mechanics of 2D point vortices, Onsager
\cite{onsager} related the formation of large-scale vortices to the existence of
negative temperature states. Later on, in unpublished notes \cite{esree}, he
developed a mean field theory of 2D point vortices and derived the
Boltzmann-Poisson equation several years before the authors of Refs.
\cite{jm,mj,kida,pl,lp}.}

The kinetic theory of 2D point vortices was investigated by Chavanis
\cite{preR,pre} by analogy with the theory of Chandrasekhar
\cite{chandra,chandra1,chandra2,nice} for stellar systems and the theory of
Brownian motion \cite{chandrabrown}. The theory of stochastic
fluctuations in the point vortex gas was developed in \cite{fluc1,fluc2,fluc3}.
The evolution of a test vortex in a thermal bath is described by a Fokker-Planck
equation of the Smoluchowski type involving a diffusion term and a drift term.
The drift is the counterpart of Chandrasekhar's dynamical friction
\cite{preR,pre}.
The drift and the diffusion coefficients satisfy an Einstein-like relation
involving a possibly negative temperature. This  Fokker-Planck equation relaxes
towards the Boltzmann distribution on a timecale $(N/\ln N)\, t_D$. At 
statistical
equilibrium, the diffusion and the drift balance each other. On the other hand,
the kinetic evolution of a system of point vortices as a whole is governed by a
Lenard-Balescu-like \cite{dubin,dubin2,klim,onsagerkin,kinfdvortex} equation,
reducing to a
Landau-like \cite{pre,cl,bbgky,kindetail,fcp} equation when collective effects
are neglected. These equations describe the
collisional evolution of the system at the order $1/N$ due to the
development of two-body
correlations. They conserve circulation and energy and satisfy an $H$-theorem
for the Boltzmann entropy. However, they do not necessarily relax towards the
Boltzmann distribution. For axisymmetric flows with a monotonic
profile of angular velocity, the Lenard-Balescu collision term
vanishes identically \cite{pre,cl}. Therefore, the system
evolves as long as the profile of angular velocity is nonmonotonic. When the 
profile of angular velocity becomes monotonic, the system stops
evolving at the order $1/N$. This is a situation of kinetic
blocking \cite{cl} due to the absence of
resonances.\footnote{For unidirectional flows made of a single species
system of point vortices, the Lenard-Balescu collision term vanishes
identically so there is no kinetic evolution at all at the order $1/N$
\cite{pre,kinfdvortex}.} As a
result, the system
does not reach the Boltzmann distribution on a timescale $N t_D$. In order to
describe the relaxation of the system towards the Boltzmann distribution it is
necessary to take into account three-body correlations and develop the kinetic
theory of 2D point vortices at the order $1/N^2$.

In this paper, we review the derivation of the Landau and
Lenard-Balescu equations from the Klimontovich formalism. We treat in detail the
case of 
spatially homogeneous stellar systems (or other homogeneous systems with
long-range interactions) and axisymmetric distributions of 2D point vortices.
The case of
inhomogeneous stellar systems is treated in \cite{angleaction2,kinfd} 
and the
case of unidirectional flows is treated in \cite{kinfdvortex}. 
We review the main properties of
these kinetic equations and discuss the problem of
kinetic blocking. We also discuss the close analogy between the kinetic
theories of stellar systems (or other systems with long-range interactions) and
2D point vortices. In particular, the systematic drift of a point vortex in a
``sea'' of field vortices is the counterpart of the dynamical friction
experienced by a star in a stellar system and the angular velocity of an
axisymmetric vortex flow is the counterpart of the frequency profile of an
inhomogeneous stellar system. The kinetic equations of 2D vortices and
inhomogeneous stellar systems have therefore a very similar structure
\cite{kindetail}. Finally, we consider the behavior of the system close to the
critical point following a recent suggestion by Hamilton and Heinemann
\cite{hh}. We present a simple calculation, valid for spatially homogeneous
systems with long-range interactions described by the Cauchy distribution,
showing how the consideration of the Landau modes regularizes the divergence of
the
friction by polarization at the critical point. We mention, however, 
that fluctuations may be very important close to the critical point and that
deterministic kinetic equations for the mean distribution function (such as the
Landau and Lenard-Balescu equations) should be replaced by stochastic kinetic
equations.

\section{Kinetic theory of homogeneous systems with long-range interactions}
\label{sec_inhos1}

\subsection{Basic equations}
\label{sec_inhos2}

We consider a system of material particles of individual mass
$m$ interacting
via a long-range binary potential $u(|{\bf r}-{\bf r}'|)$ decreasing more slowly
than $r^{-\gamma}$ with $\gamma\le d$ in a
space of
dimension
$d$. The
particles may also be subjected to an external
force  (exterior perturbation) arising from a potential $\Phi_e({\bf r},t)$. The
equations of motion of the
particles are
\begin{eqnarray}
\frac{d{\bf r}_{i}}{dt}={\bf v}_i,\qquad \frac{d{\bf
v}_{i}}{dt}=-\nabla\Phi_d({\bf r}_i)-\nabla\Phi_e({\bf r}_i,t),
\label{n1zero}
\end{eqnarray}
where $\Phi_d({\bf r})=m\sum_j u(|{\bf r}-{\bf r}_j|)$ is the exact potential
produced by the particles.  These equations can be written in
Hamiltonian form as $m d{\bf r}_i/dt=\partial{(H_d+H_e)}/\partial {\bf v}_i$ and
$m d{\bf v}_i/dt=-\partial{(H_d+H_e)}/\partial {\bf r}_i$, where
$H_d=(1/2)\sum_i mv_i^2+\sum_{i<j} m^2 u(|{\bf r}_i-{\bf r}_j|)$ is the
Hamiltonian of the particles and $H_e=\sum_i m\Phi_e({\bf r}_i,t)$ is
the Hamiltonian associated with the external force. The discrete (or singular
exact) distribution function
$f_d({\bf r},{\bf
v},t)=m\sum_i \delta({\bf r}-{\bf r}_i(t))\delta({\bf v}-{\bf v}_i(t))$ of the
particles
satisfies the Klimontovich equation \cite{klimontovich}
\begin{eqnarray}
\frac{\partial f_d}{\partial t}+{\bf v}\cdot \frac{\partial f_d}{\partial {\bf
r}}-\nabla(\Phi_d+\Phi_e)\cdot \frac{\partial f_d}{\partial {\bf v}}=0,
\label{n1}
\end{eqnarray}
where
\begin{eqnarray}
\Phi_d({\bf r},t)=\int u(|{\bf r}-{\bf r}'|)\rho_d({\bf r}',t)\, d{\bf
r}'
\label{n2}
\end{eqnarray}
is the potential produced by the discrete density of particles
$\rho_d({\bf
r},t)=\int f_d({\bf r},{\bf
v},t)\, d{\bf v}=m\sum_i \delta({\bf r}-{\bf
r}_i(t))$. Throughout the
paper, $\rho$ and $f$ refer to the {\it mass}
density of particles normalized such that $\int\rho\, d{\bf r}=\int f\, d{\bf
r}d{\bf v}=M$ where $M=Nm$ is the total mass. 

We introduce the mean distribution function $f({\bf r},{\bf v},t)=\langle
f_{d}({\bf r},{\bf v},t)\rangle$ corresponding to an ensemble
average of
$f_{d}({\bf r},{\bf v},t)$. We then write
$f_d({\bf r},{\bf v},t)=f({\bf r},{\bf v},t)+\delta f({\bf r},{\bf v},t)$, where
$\delta f({\bf r},{\bf v},t)$ denotes the fluctuations about the
mean distribution function.
Similarly, we write $\Phi_d({\bf r},t)=\Phi({\bf
r},t)+\delta\Phi({\bf r},t)$, where $\delta\Phi({\bf
r},t)$ denotes the fluctuations about the mean potential $\Phi({\bf
r},t)=\langle \Phi_d({\bf
r},t)\rangle $. Substituting these
decompositions into Eq. (\ref{n1}), we obtain
\begin{equation}
\frac{\partial f}{\partial t}+\frac{\partial\delta f}{\partial t}+{\bf
v}\cdot\frac{\partial f}{\partial {\bf r}}+{\bf v}\cdot\frac{\partial \delta
f}{\partial {\bf r}}-\nabla\Phi\cdot \frac{\partial f}{\partial {\bf
v}}-\nabla\Phi\cdot \frac{\partial \delta f}{\partial {\bf v}}-\nabla
(\delta\Phi+\Phi_e)\cdot\frac{\partial f}{\partial {\bf
v}}-\nabla(\delta\Phi+\Phi_e)\cdot
\frac{\partial \delta f}{\partial {\bf v}}=0.
\label{n3}
\end{equation}
Taking the ensemble  average of this equation, we get
\begin{equation}
\frac{\partial f}{\partial t}+{\bf v}\cdot\frac{\partial f}{\partial {\bf
r}}-\nabla\Phi\cdot \frac{\partial f}{\partial {\bf v}}=\frac{\partial}{\partial
{\bf v}}\cdot \left\langle \delta f\nabla(\delta\Phi+\Phi_e)\right\rangle.
\label{n4}
\end{equation}
This equation governs the evolution of the mean distribution function. Its
right hand side
can be interpreted as a ``collision'' or ``correlational'' 
term arising from the
granularity of the system (finite $N$
effects).
Subtracting
this expression from Eq. (\ref{n3}), we obtain the equation for the
 fluctuations of the distribution function
\begin{equation}
\frac{\partial \delta f}{\partial t}+{\bf v}\cdot\frac{\partial \delta
f}{\partial {\bf r}}-\nabla\Phi\cdot \frac{\partial \delta f}{\partial {\bf
v}}-\nabla(\delta\Phi+\Phi_e)\cdot \frac{\partial f}{\partial {\bf
v}}=\frac{\partial}{\partial {\bf v}}\cdot  (\delta
f\nabla(\delta\Phi+\Phi_e))-\frac{\partial}{\partial {\bf v}}\cdot \left\langle
\delta
f\nabla(\delta\Phi+\Phi_e)\right\rangle.
\label{n5}
\end{equation}
The foregoing equations are exact since no
approximation has been made for the
moment.

We now assume that  the
fluctuations $\delta\Phi({\bf r},t)$ of the long-range potential  created by
the particles are weak.
Since the mass of the particles scales as $m\sim 1/N$ this approximation is
valid when $N\gg
1$ \cite{campabook}. We also assume that the
external force
is weak. If we ignore the external force and the fluctuations
of the
potential
due to finite $N$ effects altogether, the collision
term vanishes and Eq. (\ref{n4}) reduces to
the Vlasov
equation
\begin{equation}
\frac{\partial f}{\partial t}+{\bf v}\cdot\frac{\partial f}{\partial {\bf
r}}-\nabla\Phi\cdot \frac{\partial f}{\partial {\bf v}}=0,
\label{n4b}
\end{equation}
where
\begin{eqnarray}
\Phi({\bf r},t)=\int u(|{\bf r}-{\bf r}'|)\rho({\bf r}',t)\, d{\bf
r}'
\label{n2bzero}
\end{eqnarray}
is the potential produced by the mean density of particles $\rho({\bf
r},t)=\int f({\bf r},{\bf v},t)\, d{\bf v}=\langle \rho_{d}({\bf r},t)\rangle$. 
The Vlasov equation describes a collisionless dynamics driven only by
self-consistent mean
field effects. It is valid in the limit $\Phi_e\rightarrow 0$ and in a
proper thermodynamic limit
$N\rightarrow +\infty$ with $m\sim 1/N$. It is also valid for
sufficiently short times.

We now take into account a small correction to the Vlasov equation obtained by
keeping the 
collision term on the right hand side of Eq. (\ref{n4}) but neglecting the
quadratic terms on the right hand side of Eq. (\ref{n5}). We therefore obtain
a set of two coupled equations
\begin{equation}
\frac{\partial f}{\partial t}+{\bf v}\cdot \frac{\partial f}{\partial {\bf
r}}-\nabla\Phi\cdot \frac{\partial f}{\partial {\bf v}}=\frac{\partial}{\partial
{\bf v}}\cdot \left\langle \delta f \nabla(\delta\Phi+\Phi_e)\right\rangle,
\label{n6}
\end{equation}
\begin{equation}
\frac{\partial\delta f}{\partial t}+{\bf v}\cdot \frac{\partial \delta
f}{\partial {\bf r}}-\nabla\Phi\cdot \frac{\partial \delta f}{\partial {\bf
v}}-\nabla(\delta\Phi+\Phi_e)\cdot \frac{\partial f}{\partial {\bf
v}}=0.
\label{n7}
\end{equation}
These equations form the starting point of the quasilinear theory which is
valid in a weak
coupling  approximation ($m\sim 1/N\ll 1$) and for a weak external stochastic
perturbation ($\Phi_e\ll \Phi$). Eq. (\ref{n6})
describes the evolution of the mean distribution function sourced by
the correlations of the fluctuations and Eq. (\ref{n7}) describes the evolution
of the fluctuations due to the
granularities of the system (finite $N$ effects) and the external force.
These equations are valid at the order $1/N$ and to leading
order in  
$\Phi_e$.

For spatially homogeneous systems with $f=f({\bf v},t)$, the foregoing equations
reduce to\footnote{In the following, we assume
that the homogeneous system remains dynamically (Vlasov) stable during the whole
evolution. This
may not always be the case. Even if we start from a Vlasov stable 
distribution function
$f_0({\bf v})$, the ``collision'' term (r.h.s. in Eq. (\ref{i56})) will change
it and induce a temporal evolution of $f({\bf v},t)$. The system may
become dynamically (Vlasov) unstable and undergo a dynamical
phase transition from an homogeneous state to an inhomogeneous state
\cite{ccgm}.
We assume
here
that this transition does not take place or we consider a period of time
preceding this transition.} 
\begin{eqnarray}
\label{i56}
\frac{\partial f}{\partial t}=\frac{\partial}{\partial
{\bf v}}\cdot \left\langle \delta
f \nabla (\delta\Phi+\Phi_e)\right\rangle,
\end{eqnarray}
\begin{eqnarray}
\label{i57}
\frac{\partial \delta f}{\partial t}+{\bf v}\cdot \frac{\partial\delta
f}{\partial {\bf r}}- \nabla (\delta\Phi+\Phi_e)\cdot \frac{\partial
f}{\partial {\bf v}}=0.
\end{eqnarray}

\subsection{Bogoliubov ansatz}
\label{sec_bog2}

In order to solve Eq. (\ref{i57}) for the fluctuations, we resort
to the Bogoliubov
ansatz \cite{bogoliubov}. We  assume that there exists a timescale separation
between a
slow and a
fast dynamics and we regard $f({\bf v})$ in Eq.
(\ref{i57}) as
``frozen'' (independent of time) at the scale of the fast
dynamics. This amounts to neglecting the temporal
variations
of the distribution function when we consider the evolution of the
fluctuations. This is
possible when the mean distribution function evolves on a secular timescale
that is
long compared to the correlation time (the time over which the
correlations of the fluctuations have their essential
support). We can
then introduce Fourier-Laplace
transforms in ${\bf r}$ and $t$ for the
fluctuations, writing
\begin{equation}
\delta \tilde f({\bf k},{\bf v},\omega)=\int \frac{d{\bf
r}}{(2\pi)^d}\int_{0}^{+\infty}dt\, e^{-i({\bf k}\cdot{\bf r}-\omega
t)}\delta
f({\bf r},{\bf v},t)
\label{i58}
\end{equation}
and
\begin{equation}
\delta f({\bf r},{\bf v},t)=\int d{\bf k}\int_{\cal C}\frac{d\omega}{2\pi}\,
e^{i({\bf k}\cdot{\bf r}-\omega t)}\delta\tilde f({\bf k},{\bf v},\omega)
\label{i59}
\end{equation}
with ${\rm Im}\,\omega>0$. Similar expressions hold for the fluctuating
potential $\delta\Phi({\bf
r},t)$ and for the external potential  $\Phi_e({\bf r},t)$.  For
future reference, we recall the Fourier
representation of the Dirac $\delta$-function
\begin{equation}
\delta(t)=\int_{-\infty}^{+\infty} e^{-i \omega t}\, \frac{d\omega}{2\pi},\qquad
\delta({\bf k})=\int e^{-i {\bf k}\cdot {\bf r}}\, \frac{d{\bf
r}}{(2\pi)^d}.
\label{deltac}
\end{equation}

\subsection{Lenard-Balescu equation}
\label{sec_jbl}

We consider an isolated system of particles with long-range interactions and
take $\Phi_e=0$.\footnote{In the present paper we consider systems in which the
fluctuations are due to the internal noise (finite $N$ effects). The quasilinear
evolution of a
self-gravitating system stochastically forced by an external perturbation,
leading to the so-called secular dressed diffusion (SDD) equation, is reviewed
in \cite{sdduniverse}.} We
want to
derive the Lenard-Balescu equation governing the evolution of the mean
distribution function at the
order $1/N$ by using the Klimontovich formalism \cite{klimontovich}. The
derivation follows the one given
in Ref. \cite{epjp1} (see also \cite{angleaction2,kinfd} in the
inhomogeneous case).  We
start from the quasilinear equations
(\ref{i56}) and (\ref{i57}) without the external
potential that we rewrite as
\begin{eqnarray}
\label{j1s}
\frac{\partial f}{\partial t}=\frac{\partial}{\partial
{\bf v}}\cdot \left\langle \delta
f \nabla\delta\Phi\right\rangle,
\end{eqnarray}
\begin{eqnarray}
\label{j2}
\frac{\partial \delta f}{\partial t}+{\bf v}\cdot \frac{\partial\delta
f}{\partial {\bf r}}- \nabla\delta\Phi\cdot \frac{\partial
f}{\partial {\bf v}}=0.
\end{eqnarray}
We make the Bogoliubov ansatz which treats $f$ as
time independent in Eq. (\ref{j2}). Taking the Fourier-Laplace transform of Eq.
(\ref{j2}), we obtain
\begin{eqnarray}
\label{j3s}
\delta{\tilde f}({\bf k},{\bf
v},\omega)=\frac{{\bf k}\cdot \frac{\partial
f}{\partial
{\bf v}}}{{\bf
k}\cdot {\bf v}-\omega}\delta{\tilde\Phi}({\bf k},\omega)+\frac{\delta{\hat
f} ({\bf k},{\bf v},0)}{i(
{\bf k}\cdot {\bf v}-\omega)},
\end{eqnarray}
where $\delta{\hat f}({\bf k},{\bf v},t=0)$ is the Fourier transform of the
initial value of the perturbed distribution function due to finite $N$
effects. 

By using the linear response theory it is shown in Appendix
\ref{sec_inhosrfb} that the Fourier-Laplace transform of the perturbed
potential
is related to the Fourier transform of the initial value of the perturbed
distribution function by
\begin{eqnarray}
\delta{\tilde \Phi}({\bf k},\omega)=
(2\pi)^d \frac{{\hat u}(k)}{\epsilon({\bf k},\omega)}  \int d{\bf
v}\, \frac{\delta{\hat f} ({\bf k},{\bf v},0)}{i(
{\bf k}\cdot {\bf v}-\omega)}.
\label{poto6}
\end{eqnarray}
This equation involves the dressed potential of interaction ${\hat u}_d(k)$ from
Eq. (\ref{udress}) taking into account collective effects. Specifically,  ${\hat
u}(k)$ is the bare potential of interaction -- the Fourier transform of $u(|{\bf
r}-{\bf r}'|)$ -- and $\epsilon({\bf k},\omega)$ is the dielectric function
defined in
Eq. (\ref{lb12}).

Taking the inverse
Laplace transform of Eq. (\ref{poto6}), using the Cauchy residue theorem, and
neglecting the contribution of the damped
modes for sufficiently
late times,\footnote{We only consider the contribution of the
pole $\omega-{\bf k}\cdot {\bf v}$ in Eq. (\ref{poto6}) and ignore the
contribution of the
proper modes $\omega({\bf k})$ of the system which are the solutions of the
dispersion relation 
(\ref{disrel}). This assumes that the system is sufficiently far from the
threshold of instability (see Sec. \ref{sec_nontb} and Appendix \ref{sec_modes}
for results valid close to the critical point). We refer to
\cite{linres} for
general considerations about the linear response theory
of
systems with long-range interactions in the homogeneous case.} we obtain
\begin{eqnarray}
\label{j6s}
\delta{\hat\Phi}({\bf k},t)=(2\pi)^d \int d{\bf v}\, \frac{{\hat
u}(k)}{\epsilon({\bf k},{\bf k}\cdot {\bf v})} \delta{\hat f}({\bf
k},{\bf v},0)e^{-i {\bf k}\cdot {\bf v} t}.
\end{eqnarray}

On the other hand, taking the Fourier transform
of Eq. (\ref{j2}), we find
that
\begin{eqnarray}
\label{j7}
\frac{\partial \delta{\hat f}}{\partial t}+i{\bf k}\cdot{\bf v}\delta{\hat
f}=i{\bf k}\cdot \frac{\partial f}{\partial {\bf v}}\delta{\hat\Phi}.
\end{eqnarray}
This first order differential equation in time can be solved with the method of
the variation of the constant, giving
\begin{eqnarray}
\label{j8}
\delta{\hat f}({\bf k},{\bf v},t)=\delta{\hat
f}({\bf
k},{\bf v},0)e^{-i {\bf k}\cdot
{\bf v} t}+i{\bf k}\cdot
\frac{\partial f}{\partial {\bf v}}\int_0^t dt'\, \delta{\hat \Phi}({\bf
k},t')e^{i{\bf k}\cdot{\bf v}(t'-t)}.
\end{eqnarray}
Substituting Eq. (\ref{j6s}) into Eq. (\ref{j8}), we obtain
\begin{equation}
\label{j9s}
\delta{\hat f}({\bf k},{\bf v},t)=\delta{\hat
f}({\bf
k},{\bf v},0)e^{-i {\bf k}\cdot
{\bf v} t}+i{\bf k}\cdot
\frac{\partial f}{\partial {\bf v}} (2\pi)^d\int d{\bf
v}' \frac{{\hat
u}(k)}{\epsilon({\bf k},{\bf k}\cdot {\bf v}')}
\delta{\hat f}({\bf
k},{\bf v}',0)e^{-i{\bf
k}\cdot{\bf v}t}\int_0^t dt'\, e^{i{\bf
k}\cdot({\bf v}-{\bf v}')t'}.
\end{equation}
Eqs.
(\ref{j6s}) and
(\ref{j9s}) relate $\delta{\hat\Phi}({\bf k},t)$ and
$\delta{\hat f}({\bf k},{\bf v},t)$ to
the
initial fluctuations $\delta{\hat f}({\bf k},{\bf v},0)$. For
sufficiently large
times, we have
\begin{equation}
\label{j9b}
\delta{\hat f}({\bf k},{\bf v},t)=\delta{\hat
f}({\bf
k},{\bf v},0)e^{-i {\bf k}\cdot
{\bf v} t}+{\bf k}\cdot
\frac{\partial f}{\partial {\bf v}} (2\pi)^d\int d{\bf
v}' \frac{{\hat
u}(k)}{\epsilon({\bf k},{\bf k}\cdot {\bf v}')}
\delta{\hat f}({\bf
k}',{\bf v}',0)\frac{e^{-i{\bf k}\cdot {\bf v}'t}}{{\bf
k}\cdot({\bf v}- {\bf v}')}.
\end{equation}

We can now compute the flux
\begin{eqnarray}
\label{j10s}
\left\langle \delta f\nabla\delta\Phi\right\rangle=\int d{\bf
k} d{\bf k}'\, i{\bf k}' e^{i{\bf k}\cdot {\bf r}}e^{i{\bf k}'\cdot {\bf
r}}\langle \delta{\hat f}({\bf k},{\bf v},t)\delta{\hat \Phi}({\bf
k}',t)\rangle.
\end{eqnarray}
From Eqs. (\ref{j6s}) and (\ref{j9s}) we get
\begin{eqnarray}
\label{j11s}
\langle \delta{\hat f}({\bf k},{\bf
v},t)\delta{\hat \Phi}({\bf k}',t)\rangle=(2\pi)^d\int
d{\bf v}' \frac{{\hat
u}(k')}{\epsilon({\bf k}',{\bf k}'\cdot {\bf v}')}
e^{-i {\bf k}'\cdot {\bf v}'t}e^{-i {\bf k}\cdot
{\bf v} t}\langle \delta{\hat f}({\bf
k}',{\bf v}',0)\delta{\hat
f}({\bf
k},{\bf v},0)\rangle\nonumber\\
+(2\pi)^d\int d{\bf
v}' \int d{\bf
v}''\, \frac{{\hat
u}(k')}{\epsilon({\bf k}',{\bf k}'\cdot {\bf v}')}
\langle\delta{\hat f}({\bf
k}',{\bf v}',0)\delta{\hat f}({\bf
k},{\bf v}'',0)\rangle e^{-i {\bf k}'\cdot {\bf v}'t}\nonumber\\
\times i{\bf k}\cdot
\frac{\partial f}{\partial {\bf v}} (2\pi)^d \frac{{\hat
u}(k)}{\epsilon({\bf k},{\bf k}\cdot {\bf v}'')}
e^{-i{\bf
k}\cdot{\bf v}t}\int_0^t dt'\, e^{i{\bf
k}\cdot ({\bf v}-{\bf v}'')t'}.
\end{eqnarray}
We assume that at $t=0$ there are no
correlations among the particles.
Therefore, we have (see Appendix A of \cite{epjp1}) 
\begin{eqnarray}
\label{j12s}
\langle \delta{\hat f}({\bf
k},{\bf v},0)\delta{\hat
f}({\bf
k}',{\bf v}',0)\rangle=\frac{1}{(2\pi)^d}\delta({\bf k}+{\bf k}')\delta({\bf
v}-{\bf v}')mf({\bf v}).
\end{eqnarray}
Eq. (\ref{j11s}) then reduces to
\begin{eqnarray}
\label{j13}
\langle \delta{\hat f}({\bf k},{\bf
v},t)\delta{\hat \Phi}({\bf k}',t)\rangle=\frac{{\hat
u}(k)}{\epsilon(-{\bf k},-{\bf k}\cdot {\bf v})}mf({\bf v})\delta({\bf k}+{\bf
k}')
\nonumber\\
+\int d{\bf
v}' \, \frac{{\hat
u}(k)}{\epsilon(-{\bf k},-{\bf k}\cdot {\bf v}')}
mf({\bf v}')  i{\bf k}\cdot
\frac{\partial f}{\partial {\bf v}} (2\pi)^d \frac{{\hat
u}(k)}{\epsilon({\bf k},{\bf k}\cdot {\bf v}')}
\int_0^t ds\, e^{-i{\bf
k}\cdot ({\bf v}-{\bf v}')s}\delta({\bf k}+{\bf k}'),
\end{eqnarray}
where we have set $s=t-t'$. Using Eq. (\ref{ii2}) we can rewrite this equation
as
\begin{eqnarray}
\label{j13h}
\langle \delta{\hat f}({\bf k},{\bf
v},t)\delta{\hat \Phi}({\bf k}',t)\rangle=\delta({\bf k}+{\bf
k}')\frac{{\hat
u}(k)}{\epsilon({\bf k},{\bf k}\cdot {\bf v})^*}mf({\bf v})
\nonumber\\
+\delta({\bf k}+{\bf k}')\int d{\bf
v}' \, 
mf({\bf v}')  i{\bf k}\cdot
\frac{\partial f}{\partial {\bf v}} (2\pi)^d \frac{{\hat
u}(k)^2}{|\epsilon({\bf k},{\bf k}\cdot {\bf v}')|}
\int_0^t ds\, e^{-i{\bf
k}\cdot ({\bf v}-{\bf v}')s}.
\end{eqnarray}
Substituting this
relation
into Eq. (\ref{j10s}), and taking the limit $t\rightarrow +\infty$, we
obtain
\begin{eqnarray}
\label{j14cbu}
\left\langle \delta f\nabla\delta\Phi\right\rangle=-\int d{\bf
k} \, i{\bf k} \frac{{\hat
u}(k)}{\epsilon({\bf k},{\bf k}\cdot {\bf v})^*}mf({\bf v})\nonumber\\
-\int d{\bf
k}\, i{\bf k} \int d{\bf
v}' \, 
mf({\bf v}')  i{\bf k}\cdot
\frac{\partial f}{\partial {\bf v}} (2\pi)^d \frac{{\hat
u}(k)^2}{|\epsilon({\bf k},{\bf k}\cdot {\bf v}')|^2}
\int_0^{+\infty} ds\, e^{-i{\bf
k}\cdot ({\bf v}-{\bf v}')s}.
\end{eqnarray}
Making the transformations
$s\rightarrow -s$ and ${\bf
k}\rightarrow -{\bf k}$ we see that
we can replace $\int_0^{+\infty} ds$ by $\frac{1}{2}\int_{-\infty}^{+\infty}
ds$. We then get 
\begin{eqnarray}
\label{j15}
\left\langle \delta f\nabla\delta\Phi\right\rangle=-\int d{\bf
k} \, i{\bf k} \frac{{\hat
u}(k)}{\epsilon({\bf k},{\bf k}\cdot {\bf v})^*}mf({\bf v})\nonumber\\
-\int d{\bf
k}\, i{\bf k} \int d{\bf
v}' \, 
mf({\bf v}')  i{\bf k}\cdot
\frac{\partial f}{\partial {\bf v}} (2\pi)^d \frac{{\hat
u}(k)^2}{|\epsilon({\bf k},{\bf k}\cdot {\bf v}')|^2}
\frac{1}{2}\int_{-\infty}^{+\infty} ds\, e^{-i{\bf
k}\cdot ({\bf v}-{\bf v}')s}.
\end{eqnarray}
Using the identity (\ref{deltac}), we arrive at
\begin{eqnarray}
\label{j17}
\left\langle \delta f\nabla\delta\Phi\right\rangle=-\int d{\bf
k} \, i{\bf k} \frac{{\hat
u}(k)}{\epsilon({\bf k},{\bf k}\cdot {\bf v})^*}mf({\bf v})\nonumber\\
-\int d{\bf
k}\, i{\bf k} \int d{\bf
v}' \, 
mf({\bf v}')  i{\bf k}\cdot
\frac{\partial f}{\partial {\bf v}} (2\pi)^d \frac{{\hat
u}(k)^2}{|\epsilon({\bf k},{\bf k}\cdot {\bf v}')|^2}
\pi \delta\lbrack {\bf k}\cdot ({\bf v}-{\bf v}')\rbrack.
\end{eqnarray}
We can rewrite the foregoing equation as
\begin{eqnarray}
\label{j18s}
\left\langle \delta f\nabla\delta\Phi\right\rangle=\int d{\bf
k} \, {\bf k} \, {\rm Im}\left\lbrack \frac{{\hat
u}(k)}{\epsilon({\bf k},{\bf k}\cdot {\bf v})^*}\right\rbrack mf({\bf
v})\nonumber\\
+\pi (2\pi)^d \int d{\bf
k}\, {\bf k} \int d{\bf
v}' \,
 {\bf k}\cdot
\frac{\partial f}{\partial {\bf v}}  \frac{{\hat
u}^2(k)}{|\epsilon({\bf k},{\bf k}\cdot {\bf v}')|^2} \delta\lbrack {\bf k}\cdot
({\bf v}-{\bf v}')\rbrack mf({\bf v}')
\end{eqnarray}
or, equivalently, as
\begin{eqnarray}
\label{j18sb}
\left\langle \delta f\nabla\delta\Phi\right\rangle=\int d{\bf
k} \, {\bf k} \, \frac{{\hat u}(k)}{|\epsilon({\bf k},{\bf k}\cdot {\bf v})|^2}
{\rm Im}\left\lbrack \epsilon({\bf k},{\bf k}\cdot {\bf
v})\right\rbrack m f({\bf
v})\nonumber\\
+\pi (2\pi)^d \int d{\bf
k}\, {\bf k} \int d{\bf
v}' \,
 {\bf k}\cdot
\frac{\partial f}{\partial {\bf v}}  \frac{{\hat
u}^2(k)}{|\epsilon({\bf k},{\bf k}\cdot {\bf v}')|^2} \delta\lbrack {\bf k}\cdot
({\bf v}-{\bf v}')\rbrack mf({\bf v}').
\end{eqnarray}
The first term is the friction by polarization and the second term is the
diffusion (see below).
Using the identity (\ref{ii1}) and substituting the flux from Eq.
(\ref{j18sb}) into Eq. (\ref{j1s}), we obtain
the Lenard-Balescu equation \cite{lenard,balescu}
\begin{equation}
\frac{\partial f}{\partial t}=\pi (2\pi)^{d}m\frac{\partial}{\partial v_i}  \int
d{\bf k} \, d{\bf v}'  \, k_ik_j  \frac{\hat{u}(k)^2}{|\epsilon({\bf k},{\bf
k}\cdot {\bf v})|^2}\delta\lbrack {\bf k}\cdot ({\bf v}-{\bf v}')\rbrack\left
(\frac{\partial}{\partial {v}_{j}}-\frac{\partial}{\partial {v'}_{j}}\right
)f({\bf v},t)f({\bf v}',t).
\label{lb27}
\end{equation}
We can easily extend the derivation
of the Lenard-Balescu equation to the
multispecies case (see, e.g., \cite{landaud,epjp2}).

{\it Remark:} From Eq. (\ref{j6s}) we get
\begin{eqnarray}
\label{int1a}
\langle \delta{\hat\Phi}({\bf k},t)\delta{\hat\Phi}({\bf
k}',t')^*\rangle=(2\pi)^{2d}\int d{\bf
v}d{\bf v}'\, \frac{{\hat u}(k){\hat u}(k')}{\epsilon({\bf k},{\bf k}\cdot
{\bf v})\epsilon({\bf k}',{\bf k}'\cdot
{\bf v}')^*} \langle
\delta{\hat f}({\bf
k},{\bf v},0)\delta{\hat f}({\bf
k}',{\bf v}',0)^*\rangle e^{-i {\bf k}\cdot {\bf v}t} e^{i {\bf
k}'\cdot {\bf v}'t'}.
\end{eqnarray}
Using Eq. (\ref{j12s}) we obtain
\begin{eqnarray}
\label{int1b}
\langle \delta{\hat\Phi}({\bf k},t)\delta{\hat\Phi}({\bf
k}',t')^*\rangle=(2\pi)^{d}\delta({\bf k}-{\bf
k}')\int d{\bf
v}\, \frac{{\hat u}(k)^2}{|\epsilon({\bf k},{\bf k}\cdot
{\bf v})|^2}  e^{-i {\bf k}\cdot {\bf v}(t-t')}mf({\bf v}).
\end{eqnarray}
This is the temporal correlation function  of
the potential fluctuations.  Taking its
Laplace transform, we obtain the power spectrum
produced by a random distribution of particles 
\begin{eqnarray}
\label{int1c}
\langle \delta{\tilde\Phi}({\bf k},\omega)\delta{\tilde\Phi}({\bf
k}',\omega')^*\rangle=(2\pi)^{d+2}m\frac{{\hat u}(k)^2}{|\epsilon({\bf
k},\omega)|^2}\delta({\bf k}-{\bf
k}')\delta(\omega-\omega')\int  \delta({\bf
k}\cdot {\bf v}-\omega) f({\bf v})\, d{\bf
v}.
\end{eqnarray}
This returns Eq. (28) of \cite{epjp1}. Substituting Eq.
(\ref{j8}) into Eq. (\ref{j10s}) we can write the flux as
\begin{eqnarray}
\label{feu1}
\langle \delta f\nabla\delta\Phi\rangle=\int d{\bf k}d{\bf k}' i{\bf k}'
e^{i({\bf k}+{\bf k}')\cdot {\bf r}}\langle \delta{\hat\Phi}({\bf
k}',t)\delta{\hat f}({\bf k},{\bf v},0)\rangle e^{-i{\bf k}\cdot {\bf
v}t}\nonumber\\
+\int d{\bf k}d{\bf k}' i{\bf k}'
e^{i({\bf k}+{\bf k}')\cdot {\bf r}}\left (i{\bf k}\cdot \frac{\partial
f}{\partial {\bf v}}\right )\int_0^t dt' \langle \delta{\hat\Phi}({\bf
k}',t)\delta{\hat \Phi}({\bf k},t')\rangle e^{i{\bf k}\cdot {\bf v}(t'-t)},
\end{eqnarray}
which exhibits the friction term and the diffusion term. This expression shows
that the diffusion coefficient is given by the time integral of the
auto-correlation function of the gravitational force \cite{aa,kinfd,epjp1,hb4}.

\subsection{Landau equation}
\label{sec_le}

If we ignore collective effects, we would be tempted to replace the
dressed potential of interaction ${\hat u}_d={\hat u}/\epsilon$ by
the bare potential of interaction ${\hat u}$ in
Eq. (\ref{j18s}). But, in that case, the friction by polarization would vanish
since
${\hat u}$ is
real. Therefore, we must first compute ${\rm Im}({\hat u}_d)$ by using Eq.
(\ref{ii1}),
{\it then} replace ${\hat u}_d$ by
${\hat u}$ in Eq. (\ref{lb27}). To directly derive the Landau equation from the
quasilinear
equations
(\ref{j1s}) and (\ref{j2}), we can proceed as
follows.\footnote{This approach is related to the iterative method used in
\cite{hb4} to derive the Landau equation in physical space.}

According to Eq. (\ref{bof3bann})  we have
\begin{eqnarray}
\delta{\hat \Phi}({\bf k},t)=(2\pi)^d{\hat u}(k)\int
d{\bf v}\,  \delta{\hat f}({\bf k},{\bf v},t).
\label{p27s}
\end{eqnarray}
Substituting Eq. (\ref{j8}) into Eq. (\ref{p27s}) we obtain
\begin{eqnarray}
\label{p28s}
\delta{\hat\Phi}({\bf k},t)=(2\pi)^d {\hat u}(k)\int d{\bf
v}
\delta{\hat f}({\bf
k},{\bf v},0)e^{-i{\bf k}\cdot {\bf v} t}\nonumber\\
+(2\pi)^d {\hat u}(k)\int d{\bf
v} \, i{\bf k}\cdot
\frac{\partial f}{\partial {\bf v}}\int_0^t dt'\, \delta{\hat \Phi}({\bf
k},t')e^{i{\bf k}\cdot{\bf v}(t'-t)}.
\end{eqnarray}
This is an exact Volterra-Fredholm integral equation equivalent
to Eq.
(\ref{bebe}). In order to relate
$\delta{\hat\Phi}({\bf k},t)$ to $\delta{\hat f}({\bf
k},{\bf v},0)$, instead of using Eq. (\ref{j6s}), we shall use an
iterative
method. We assume that
we can replace $\delta{\hat\Phi}$ in the second line of Eq. (\ref{p28s}) by its
expression
obtained by keeping only the contribution from the first line. This gives
\begin{eqnarray}
\label{p29s}
\delta{\hat\Phi}({\bf k},t)=(2\pi)^d {\hat u}(k)\int d{\bf
v}
\delta{\hat f}({\bf
k},{\bf v},0)e^{-i{\bf k}\cdot {\bf v} t}\nonumber\\
+(2\pi)^d {\hat u}(k)\int d{\bf v}\, i{\bf
k}\cdot
\frac{\partial f}{\partial {\bf v}}\int_0^t dt'\, e^{i{\bf
k}\cdot{\bf v}(t'-t)}(2\pi)^d {\hat u}(k) \int d{\bf
v}' \delta{\hat f}({\bf
k},{\bf v}',0)e^{-i{\bf k}\cdot {\bf v}' t'}.
\end{eqnarray}
We use the same strategy in Eq. (\ref{j8}), thereby obtaining
\begin{eqnarray}
\label{p30s}
\delta{\hat f}({\bf k},{\bf v},t)=\delta{\hat
f}({\bf
k},{\bf v},0)e^{-i {\bf k}\cdot
{\bf v} t}+i{\bf k}\cdot
\frac{\partial f}{\partial {\bf v}}\int_0^t dt'\, e^{i{\bf
k}\cdot{\bf v}(t'-t)} (2\pi)^d {\hat u}(k)\int d{\bf
v}'
\delta{\hat f}({\bf
k},{\bf v}',0)e^{-i{\bf k}\cdot {\bf v}' t'}.
\end{eqnarray}
We see that Eqs. (\ref{p29s}) and (\ref{p30s})
are similar to Eqs.
(\ref{j6s}) and (\ref{j9s}) except that ${\hat u}_d$ is replaced by ${\hat u}$
and there
is a ``new'' term
in Eq.   (\ref{p29s}) in addition to the ``old'' one. This new term
accounts for the friction by polarization. It substitutes itself to the term
proportional to ${\rm Im}({\hat u}_d)$ which vanishes when
${\hat u}_d$ is replaced by ${\hat u}$ (as noted above). On the other hand,
the old term with ${\hat u}_d$ replaced by ${\hat u}$ accounts for the
diffusion. Therefore, repeating
the calculations of Sec. \ref{sec_jbl}, or directly using Eq. (\ref{j18s})
with ${\hat u}_d$ is replaced by ${\hat u}$, we get
\begin{eqnarray}
\label{p26s}
\left\langle \delta f\nabla\delta\Phi\right\rangle_{\rm diffusion}=
\pi (2\pi)^d \int d{\bf
k}\, {\bf k} \int d{\bf
v}' \,
 {\bf k}\cdot
\frac{\partial f}{\partial {\bf v}} {\hat
u}^2(k) \delta\lbrack {\bf k}\cdot
({\bf v}-{\bf v}')\rbrack mf({\bf v}').
\end{eqnarray}
This quantity corresponds to the product of the first term in Eq. (\ref{p29s})
with the two terms in Eq. (\ref{p30s}). Let us now compute the friction term.
To that purpose, we have
to compute
\begin{eqnarray}
\label{aaa3b}
\langle \delta{\hat f}({\bf k},{\bf
v},t)\delta{\hat \Phi}({\bf k}',t)\rangle_{\rm
friction}=(2\pi)^d\int
d{\bf
v}' {\hat u}(k') i{\bf k}'\cdot
\frac{\partial f'}{\partial {\bf v}'}\int_0^t dt'\, e^{i{\bf
k}'\cdot{\bf v}'(t'-t)}\nonumber\\
\times(2\pi)^d\int d{\bf
v}'' {\hat u}(k')
e^{-i{\bf k}'\cdot {\bf v}'' t'}e^{-i {\bf k}\cdot
{\bf v} t}
\langle \delta{\hat
f}({\bf
k},{\bf v},0)\delta{\hat f}({\bf
k}',{\bf v}'',0)\rangle. 
\end{eqnarray}
This quantity corresponds to the product of the second term in Eq. (\ref{p29s})
with the first term in Eq. (\ref{p30s}). In line with our perturbative approach
we neglect the product of the second terms in Eqs. (\ref{p29s}) and
(\ref{p30s}).
Using Eq. (\ref{j12s}) we get 
\begin{eqnarray}
\label{pb28}
\langle \delta{\hat f}({\bf k},{\bf
v},t)\delta{\hat \Phi}({\bf k}',t)\rangle_{\rm
friction}=-(2\pi)^d \delta({\bf k}+{\bf k}')\int
d{\bf
v}' {\hat u}(k) i{\bf k}\cdot
\frac{\partial f'}{\partial {\bf v}'}\int_0^t ds\, {\hat u}(k) e^{-i {\bf
k}\cdot
({\bf v}-{\bf v}')s} m f({\bf v}),
\end{eqnarray}
where we have set $s=t-t'$. Substituting this relation into Eq. (\ref{j10s}), we
obtain
\begin{eqnarray}
\label{aaa1b}
\left\langle \delta f\nabla\delta\Phi\right\rangle_{\rm
friction}=-\int d{\bf k}\, {\bf k} (2\pi)^d\int
d{\bf
v}' {\hat u}(k) {\bf k}\cdot
\frac{\partial f'}{\partial {\bf v}'}\int_0^t ds\, {\hat u}(k) e^{-i {\bf
k}\cdot
({\bf v}-{\bf v}')s} m f({\bf v}).
\end{eqnarray}
If we let $t\rightarrow
+\infty$, we can rewrite the foregoing equation as
\begin{eqnarray}
\label{p31}
\left\langle \delta f\nabla\delta\Phi\right\rangle_{\rm
friction}=-\int d{\bf k}\, {\bf k} (2\pi)^d\int
d{\bf
v}' {\hat u}^2(k) {\bf k}\cdot
\frac{\partial f'}{\partial {\bf v}'}m f({\bf v})\int_0^{+\infty}
ds\,  e^{-i {\bf
k}\cdot
({\bf v}-{\bf v}')s}.
\end{eqnarray}
Making the transformations
$s\rightarrow -s$ and ${\bf
k}\rightarrow -{\bf k}$ we see that
we can replace $\int_0^{+\infty} ds$ by $\frac{1}{2}\int_{-\infty}^{+\infty}
ds$. We then get 
\begin{eqnarray}
\label{p32bu}
\left\langle \delta f\nabla\delta\Phi\right\rangle_{\rm
friction}=-\int d{\bf k}\, {\bf k} (2\pi)^d\int
d{\bf
v}' {\hat u}^2(k) {\bf k}\cdot
\frac{\partial f'}{\partial {\bf v}'}m f({\bf
v})\frac{1}{2}\int_{-\infty}^{+\infty}
ds\,  e^{-i {\bf
k}\cdot
({\bf v}-{\bf v}')s}.
\end{eqnarray}
Using the identity (\ref{deltac}) we arrive
at the expression
\begin{eqnarray}
\label{p35s}
\left\langle \delta f\nabla\delta\Phi\right\rangle_{\rm
friction}=-\pi(2\pi)^d\int d{\bf k}\, {\bf k} \int
d{\bf
v}' {\hat u}^2(k) {\bf k}\cdot
\frac{\partial f'}{\partial {\bf v}'}\delta\lbrack {\bf
k}\cdot
({\bf v}-{\bf v}')\rbrack m f({\bf
v}).
\end{eqnarray}
Adding Eqs. (\ref{p26s}) and (\ref{p35s}), we find that
\begin{eqnarray}
\label{p36s}
\left\langle \delta f\nabla\delta\Phi\right\rangle=
\pi (2\pi)^d \int d{\bf
k}\, {\bf k} \int d{\bf
v}' \,
 {\bf k}\cdot
\frac{\partial f}{\partial {\bf v}} {\hat
u}^2(k) \delta\lbrack {\bf k}\cdot
({\bf v}-{\bf v}')\rbrack mf({\bf v}')\nonumber\\
-\pi(2\pi)^d\int d{\bf k}\, {\bf k} \int
d{\bf
v}' {\hat u}^2(k) {\bf k}\cdot
\frac{\partial f'}{\partial {\bf v}'}\delta\lbrack {\bf
k}\cdot
({\bf v}-{\bf v}')\rbrack m f({\bf
v}).
\end{eqnarray}
Substituting this flux into Eq. (\ref{j1s}), we obtain
the Landau equation \cite{landau}
\begin{equation}
\frac{\partial f}{\partial t}=\pi (2\pi)^{d}m\frac{\partial}{\partial v_i}  \int
d{\bf k} \, d{\bf v}'  \, k_ik_j  \hat{u}(k)^2\delta\lbrack {\bf
k}\cdot ({\bf v}-{\bf v}')\rbrack\left
(\frac{\partial}{\partial {v}_{j}}-\frac{\partial}{\partial {v'}_{j}}\right
)f({\bf v},t)f({\bf v}',t).
\label{j19lands}
\end{equation}
We can easily extend the derivation of the Landau equation to the
multispecies case (see, e.g., \cite{landaud,aa,kinquant}).

{\it Remark:} The integral over ${\bf k}$ can be performed explicitly (see,
e.g.,
\cite{kindetail}) and we recover the original form of the Landau equation
\cite{landau}
\begin{equation}
\frac{\partial f}{\partial t}=K_d\frac{\partial}{\partial v_i}\int d{\bf v}'
\frac{w^2\delta_{ij}-w_iw_j}{w^3}\left (\frac{\partial}{\partial
{v}_{j}}-\frac{\partial}{\partial {v'}_{j}}\right )f({\bf v},t)f({\bf v}',t),
\label{lb29}
\end{equation}
where ${\bf w}={\bf v}-{\bf v}'$ is the relative velocity and $K_d$ is a
constant with value $K_3=8\pi^5m\int_0^{+\infty}k^3\hat{u}(k)^2\, dk$ in $d=3$
and  $K_2=8\pi^3 m\int_0^{+\infty}k^2\hat{u}(k)^2\, dk$ in $d=2$. We note that
the structure of this kinetic equation does not depend on the specific form of
the potential of interaction. The potential of interaction only appears in
factor of
the integral in a constant $K_d$ that determines the relaxation time (see,
e.g., \cite{epjp1,epjp2,epjp3} for more details).  For a 3D self-gravitating
system,\footnote{The
case of Coulombian plasmas is treated in \cite{epjp1,epjp2,epjp3} and in
standard textbooks of plasma physics.} using
$(2\pi)^3\hat{u}(k)=-4\pi
G/k^2$, we get $K_3=2\pi mG^2\ln\Lambda$ where $\ln\Lambda=\int_0^{+\infty}
dk/k$ is the Coulomb factor that has to be regularized with appropriate
cut-offs. The large-scale cut-off is the Jeans length $\lambda_J^2\sim k_B T/G
m^2 n$ (which is of the order of the system's size $R$ according to the virial
theorem) and the small-scale
cut-off is the gravitational Landau length $\lambda_{L}\sim Gm/v_m^2\sim G
m^2/k_B T\sim 1/n\lambda_J^2$ which corresponds to the distance at which the
particles are deflected by about  $90^{\rm o}$. Therefore $\Lambda\sim
\lambda_J/\lambda_{L}\sim
n\lambda_J^3\sim N \gg 1$ represents the number of stars  in the cluster. The
Chandrasekhar relaxation time then scales as $t_{\rm relax}\sim (N/\ln N)t_D$.

\subsection{Fokker-Planck equation}
\label{sec_fphom}

The Lenard-Balescu equation (\ref{lb27}) can be written under the form of a
Fokker-Planck equation
\begin{equation}
\label{fp8}
\frac{\partial f}{\partial t}=\frac{\partial}{\partial v_{i}} \left
(D_{ij}\frac{\partial f}{\partial
v_{j}}- f F_i^{\rm pol}\right )
\end{equation}
with a diffusion tensor and a friction by polarization\footnote{The difference
between the friction by polarization and the total friction is discussed in
\cite{hb4,epjp1,angleaction2,kinfd}.} given by
\begin{equation}
D_{ij}=\pi(2\pi)^d m \int d{\bf k}\, d{\bf v}' \, k_i k_j
\frac{\hat{u}(k)^2}{|\epsilon({\bf k},{\bf k}\cdot {\bf v})|^2}\delta\lbrack
{\bf k}\cdot ({\bf v}-{\bf v}')\rbrack f({\bf v}',t),
\label{diff7again}
\end{equation}
\begin{equation}
{F}_i^{\rm pol}=\pi (2\pi)^d m\int d{\bf k}\, d{\bf v}'\, k_i k_j
\frac{\hat{u}(k)^2}{|\epsilon({\bf k},{\bf k}\cdot {\bf v})|^2}\delta\lbrack
{\bf k}\cdot ({\bf v}- {\bf v}')\rbrack \frac{\partial f}{\partial {v'_j}}({\bf
v}',t).
\label{ess14again}
\end{equation}
The  Lenard-Balescu equation (\ref{lb27}) can be directly derived from the
Fokker-Planck approach by calculating the diffusion and friction coefficients
(first and second moments of the velocity increment) as shown in Sec. 3
of \cite{epjp1}. The Lenard-Balescu  equation (\ref{lb27}) is valid at the order
$1/N$ so it describes the ``collisional'' evolution of the system  due to
finite $N$ effects on a timescale
$\sim N t_D$, where $t_D$ is the dynamical time. The Lenard-Balescu equation
conserves the mass $M$ and the energy $E$ which
reduces to the kinetic energy for spatially homogeneous systems. It also
monotonically increases the Boltzmann entropy $S$ ($H$-theorem)
\cite{klimontovich}. The collisional evolution of the system is due to a
condition of resonance. This condition of resonance, encapsulated in the
$\delta$-function, corresponds to ${\bf k}\cdot {\bf v}'={\bf k}\cdot {\bf v}$.
When $d>1$, this condition of
resonance can always
be satisfied and the Maxwell-Boltzmann
distribution  $f_e({\bf v})=M(\beta m/2\pi)^{d/2}e^{-\beta m v^2/2}$ is the
unique steady state of the Lenard-Balescu
equation. Therefore, under the effect of ``collisions'', the system reaches the
Maxwell-Boltzmann distribution on a relaxation time $t_{R}\sim
Nt_D$.\footnote{For 3D plasmas and 3D self-gravitating systems there is a
logarithmic correction due to the effect of strong
collisions at small scales. As a result, the relaxation time scales as $t_R\sim
(N/\ln N)
t_D$. For neutral plasmas $N$ represents the number of
charges in the Debye sphere.} The dimension $d=1$ leads to a
situation of kinetic blocking which is specifically discussed in
Sec. \ref{sec_kbh}.

In the thermal bath approximation, where the field particles have the
Maxwell-Boltzmann distribution $f({\bf v}')\sim e^{-\beta m {v'}^2/2}$, the
friction by polarization is linked to the diffusion tensor by 
the Einstein relation ${F}_i^{\rm pol}=-D_{ij}({\bf v})\beta m v_j$
\cite{epjp1,epjp2}
and the Fokker-Planck equation takes the
form 
\begin{equation}
\label{tb6}\frac{\partial P}{\partial t}=\frac{\partial}{\partial v_{i}} \left
\lbrack D_{ij}({\bf v})\left (\frac{\partial P}{\partial
v_{j}}+\beta m  P v_j\right )\right\rbrack.
\end{equation}
This equation is similar to
the Kramers equation in Brownian theory \cite{risken}. The diffusion
tensor depends on the velocity. It can be calculated analytically when
collective effects are neglected (see, e.g., \cite{landaud,epjp1,epjp2,epjp3}).
The distribution
function of the test particles relaxes towards the Maxwell-Boltzmann
distribution $P_e({\bf v})=(\beta m/2\pi)^{d/2}e^{-\beta m v^2/2}$ on a
timescale $t_{R}^{\rm bath}\sim N\,  t_D$ in all dimensions of space $d$
including  $d=1$ (for 3D plasmas and  3D homogeneous self-gravitating systems
the relaxation time scales as $t_{R}^{\rm
bath}\sim (N/\ln N)\, t_D$).

\subsection{The case $d=1$: kinetic blocking}
\label{sec_kbh}

For spatially homogeneous systems in $d=1$ dimension, there is no resonance and
the Lenard-Balescu collision term vanishes:
 \begin{equation}
\frac{\partial f}{\partial t}=2\pi^2 m\frac{\partial}{\partial v}  \int d{k} \,
d{v}'  \, |k|  \frac{\hat{u}(k)^2}{|\epsilon({k},k v)|^2}\delta ({v}-{v}')\left
(\frac{\partial}{\partial {v}}-\frac{\partial}{\partial {v'}}\right
)f({v},t)f({v}',t)=0.
\label{rw2}
\end{equation}
Therefore, the distribution function $f(v)$ does not evolve at all on the
timescale $\sim N t_D$. This is a situation of kinetic blocking due to the
absence of resonances at the order $1/N$. The
kinetic theory predicts no thermalization to a Maxwellian at first order in
$1/N$. The maxwellization is a second order effect in $1/N$, and consequently a
very slow process. A kinetic equation has been obtained at the order $1/N^2$
when collective effects are neglected \cite{fbcn2,fcpn2}. It relaxes
towards the
Maxwell-Boltzmann distribution on a timescale $t_{R}\sim N^2t_D$. In that case,
the relaxation is caused by three-body correlations.

In the bath approximation, where the field particles can have any
distribution $f_0(v')$,\footnote{As we have seen, such a distribution function
does not evolve on
a timescale $Nt_D$ in $d=1$. By contrast, in $d>1$, only the Maxwell-Boltzmann
distribution function is steady on this timescale and can therefore form a
bath.} the friction by
polarization satisfies the
generalized Einstein relation $F_{\rm
pol}=D\frac{d\ln f_0}{dv}$ and the Fokker-Planck equation
takes the
form
\begin{equation}
\label{genn1}  {\partial P\over\partial t}={\partial\over\partial
v}\biggl\lbrack D(v)\biggl ({\partial P\over\partial v}-P {d\ln f_0\over
dv}\biggr )\biggr\rbrack,
\end{equation}
with a diffusion coefficient
\begin{equation}
\label{genn2} D(v)=m f_0(v)\int_{0}^{+\infty} dk {4\pi^{2}\hat{u}(k)^{2}k\over
|\epsilon(k,kv)|^{2}}.
\end{equation}
Equation (\ref{genn1}) is similar to the Smoluchowski equation
describing the motion of a Brownian particle in a potential $U({v})=-\ln
f_0({v})$. The distribution function of the test particle $P(v,t)$
relaxes towards the distribution of the bath $f_0(v)$ on a timescale
$t_R^{\rm bath}\sim N t_D$. In the case of a thermal bath, Eq.
(\ref{genn1}) returns Eq. (\ref{tb6}).

\subsection{Approaching the critical point of marginal stability}
\label{sec_nontb}

The previous results are valid for neutral plasmas which are spatially
homogeneous and stable. They are also valid for systems with long-range
interactions
such as the HMF model which are spatially homogeneous and stable above a
critical energy
$E_c$. In general, self-gravitating systems are spatially inhomogeneous (see
Sec. \ref{sec_avp}). Furthermore, spatially homogeneous self-gravitating
systems suffer from the Jeans instability. However, we can consider a toy model
consisting of a spatially homogeneous self-gravitating system
confined within a domain of size $R$. This system is stable if
the size $R$ of the domain is smaller than the Jeans length $\lambda_J$ (see
Appendix \ref{sec_drs}).\footnote{The case of a self-gravitating system in a
periodic
cube has been considered in Refs. \cite{weinberglb,epjp3,aa,magorrian}. In
order to have a stable homogeneous phase above a critical temperature $T_c$, one
can also work in an infinite
domain and modify the potential of interaction, e.g., by replacing the Poisson
equation by a screened Poisson equation like in \cite{ksjeans,cd1,cd2}. This
type of potential occurs in the Keller-Segel model of chemotaxis
\cite{ks}.} Let us consider what happens as we approach the
critical point of marginal stability, so that the homogeneous phase tends to
become unstable. This corresponds to the limit $E\rightarrow E_c^+$ in the HMF
model or the limit
$R\rightarrow \lambda_J^-$ for self-gravitating systems. This is the generic
case for self-gravitating systems since the size of the system is precisely of
the order of the Jeans length according to the virial theorem. We first recall
and extend the discussion given in former works
\cite{weinberglb,epjp3,aa,magorrian}, then consider more recent suggestions
\cite{hh}.

The normal modes of the system $\omega_{\alpha}({\bf k})$ are the solutions of
the
dispersion relation $\epsilon({\bf k},\omega)=0$ (see Appendix \ref{sec_drs}).
Let us write them as $\omega=\Omega+i\gamma$,
where $\Omega({\bf k})$ is the real pulsation and $\gamma({\bf k})$ the
exponential rate. The mode ${\bf k}$ is stable when $\gamma({\bf k})<0$ and
unstable
when $\gamma({\bf k})>0$. The condition of marginal stability is $\gamma({\bf
k})=0$. If we place ourselves at the point of marginal stability $\gamma=0$, we
have 
$\epsilon({\bf k},\Omega)=0$. Therefore, the quantity $\epsilon({\bf
k},{\bf k}\cdot{\bf v})=0$ which occurs in the denominator of the dressed
potential of interaction (\ref{udress}) appearing in the Lenard-Balescu equation
(\ref{lb27}) vanishes for the velocity ${\bf v}$ satisfying  ${\bf
k}\cdot{\bf v}=\Omega$. This suggests that $D_{ij}$ and ${\bf F}_{\rm pol}$
(or the total flux) diverge at that point, implying an acceleration of the
relaxation (the relaxation
time decreases). To be more specific, let us assume that $\Omega=0$ at the point
of marginal stability (i.e. marginal stability corresponds to $\omega=0$). In
that case $\epsilon({\bf k},{\bf k}\cdot{\bf v})$ vanishes when ${\bf
v}={\bf 0}$ (or more generally when ${\bf k}\cdot{\bf v}=0$). Let us then
consider
the ``Debye-H\"uckel'' approximation  \cite{epjp1,epjp2,epjp3} where we replace 
$\epsilon({\bf k},{\bf
k}\cdot{\bf v})$ by its static value $\epsilon({k},0)$ [see Eq. (\ref{disrel2})]
in the 
Lenard-Balescu equation (\ref{lb27}). In that case, we obtain an
equation
equivalent to
the Landau equation (\ref{j19lands}) with the ``Debye-H\"uckel'' potential
of
interaction
\begin{eqnarray}
{\hat u}_{\rm DH}({k},\omega)=\frac{{\hat
u}(k)}{\epsilon({k},0)}.
\label{no1}
\end{eqnarray}
Let us consider specific examples:

(i) For self-gravitating systems, using the results of Appendix
\ref{sec_drs}, the ``Debye-H\"uckel'' potential reads
\begin{eqnarray}
(2\pi)^d{\hat u}_{\rm DH}({k},\omega)=-\frac{S_d G}{k^2-k_J^2}.
\label{no2}
\end{eqnarray}
It leads to a situation of anti-shielding where a star tends to concentrate
other stars around it -- by forming a polarization cloud -- because of the
gravitational attraction (see Appendix E of
\cite{aa} for details). This has the effect of increasing the effective mass
of the star. The integral over the wavenumber $k$ that
occurs in the 3D Landau equation 
(\ref{lb29}) with the  ``Debye-H\"uckel'' potential is given by
\begin{eqnarray}
K_3^{\rm DH}=2\pi G^2 m \int_{k_R}^{k_{L}} 
\frac{1}{\left (1-\frac{k_J^2}{k^2}\right )^2}\, \frac{dk}{k},
\label{no4}
\end{eqnarray}
where we have introduced a small scale cut-off $k_L=2\pi/\lambda_L$ at the
Landau length and a large scale cut-off $k_R=2\pi/R$ at the box size (see Sec.
\ref{sec_le}). We
assume that $R<\lambda_J$ (i.e. $k_R>k_J$) so that the system is stable. The
integral (\ref{no4}) can be calculated analytically,
yielding
\begin{eqnarray}
K_3^{\rm DH}=\pi G^2 m \left\lbrack
\ln\left(\frac{k_L^2-k_J^2}{k_R^2-k_J^2}\right )+\frac{k_J^2(k_L^2-k_R^2)}{
(k_L^2-k_J^2) (k_R^2-k_J^2)}\right\rbrack.
\label{no5}
\end{eqnarray}
We now consider its behavior as we approach the critical point. For 
$R\rightarrow \lambda_J$ (i.e. $k_R\rightarrow k_J$) we get
\begin{eqnarray}
K_3^{\rm DH}\sim \pi G^2 m \frac{k_J^2}{k_R^2-k_J^2}\sim \frac{\pi}{2} G^2 m
\frac{k_J}{k_R-k_J}.
\label{no6}
\end{eqnarray}
Therefore, $K_3^{\rm DH}\sim (\lambda_J-R)^{-1}$ diverges algebraically as
we approach the critical point (see Appendix E of
\cite{aa}) instead of diverging logarithmically when collective effects are
neglected (see Sec.
\ref{sec_le}). As a result, collective effects are expected to
speed up the
relaxation \cite{weinberglb,epjp3,aa,magorrian}.

(ii) For the HMF model, using the results of
Appendix \ref{sec_drs}, the ``Debye-H\"uckel'' potential reads
\begin{eqnarray}
{\hat u}_n^{\rm DH}=-\frac{k}{4\pi}\frac{1}{1-B}\delta_{n,\pm 1}.
\label{no7}
\end{eqnarray}
For a thermal bath, the parameter $B$  is given by
$B=k\rho/2T=T_c/T$. It represents the normalized inverse temperature. The
diffusion
coefficient from Eq. (\ref{genn2}) then becomes
\begin{equation}
\label{no8} 
D(v)=\frac{k^2}{4}f_0(v)\frac{1}{(1-B)^2}.
\end{equation}
It diverges as $(1-B)^{-2}$ at the critical point $B_c=1$. Correspondingly, the
relaxation time in the bath approximation\footnote{Recall that the relaxation
of the system as a whole is blocked at the order $1/N$ because of the
absence of resonances (see Sec. \ref{sec_kbh}). Therefore, we only consider
this system in the bath approximation where the Fokker-Planck current is
nonzero (except at equilibrium).} decreases
as $t_{\rm relax}\sim (1-B)^{2}$ (see Sec. 6 of \cite{epjp3}).

The foregoing results are based on the standard Lenard-Balescu equation
(\ref{lb27}) that ignores the contribution of the damped modes (see footnote
11). However, these damped modes decay more and more slowly as one approaches
the critical point. Therefore, they have to be taken into account close to the
critical point.
Recently, Hamilton and Heinemann \cite{hh} made a calculation suggesting that no
divergence occurs at the critical point when the contribution of the Landau
modes are taken into account. We had also made a similar calculation in a former
paper \cite{linres} leading to the same result without realizing at that time
all its implications. In Appendix \ref{sec_modes} we extend our original
calculation (valid
for the Cauchy distribution) and compute the friction by polarization close to
the critical point. We show that the friction by polarization does not
diverge at the critical point (at fixed time $t$). This is qualitatively
consistent with the numerical simulations of Weinberg \cite{weinberglb} which
show that
the divergence at the critical point (if any) is much weaker than naively
expected on the basis of the Lenard-Balescu equation (without taking the Landau
modes into account). The computation of the
diffusion coefficient is more complicated and is left for a future work. It is
not obvious if this quantity remains finite at the critical point or if it
diverges. We note that correlation functions usually diverge at the critical
point, due to the enhancement of fluctuations. This critical behavior is related
to the phenomenon of  critical opalescence \cite{sdduniverse}. This is the case
in particular for the systems of Brownian particles with long-range interactions
(corresponding to a canonical
ensemble description) considered in \cite{hb5}. We note in this respect that the
Lenard-Balescu equation, which is an equation for the averaged distribution
function, does not take
fluctuations into account.\footnote{As we approach the
critical point, the number of particles $N$ must be larger and larger to
reduce the fluctuations \cite{hb5,bmf}.} When the fluctuations
are important (i.e.
close to the critical point) it may be necessary to consider more
general {\it stochastic} kinetic
equations \cite{bouchetld,jb,kinfd} which are similar to the stochastic
Fokker-Planck equation
introduced in the context of
Brownian particles with long-range interactions \cite{hb5,cd2,bmf,entropy,gsse}.
These stochastic
equations could be used to describe random transitions between different
equilibrium states \cite{cd2,bmf,gsse,nardini2,bouchetsimmonet,rbs,brs}.

\section{Kinetic theory of inhomogeneous systems with long-range interactions}
\label{sec_avp}

Systems with long-range interactions can be spatially inhomogenous. This is the
case of the  HMF model below the critical energy $E_c$. On the other hand, 
self-gravitating systems (that are not in periodic boxes) are
generically spatially inhomogeneous. Therefore, it is important to generalize
the kinetic theory to spatially inhomogeneous systems. This can be done most
conveniently by working with angle-action variables.

\subsection{Inhomogeneous Lenard-Balescu equation}

The inhomogeneous Lenard-Balescu equation written in terms of angle-action
variables reads \cite{heyvaerts,angleaction2}:
\begin{equation}
\frac{\partial f}{\partial t}=\pi (2\pi)^d m \frac{\partial}{\partial {\bf
J}}\cdot \sum_{{\bf k},{\bf k}'}\int d{\bf J}' \, {\bf k}\,
|A^d_{{\bf k}{\bf k}'}({\bf J},{\bf J}',{\bf k}\cdot {\bf
\Omega})|^2\delta({\bf k}\cdot {\bf \Omega}-{\bf k}'\cdot {\bf \Omega}') \left
(f' {\bf k}\cdot \frac{\partial f}{\partial {\bf J}}-f {\bf
k}'\cdot \frac{\partial f'}{\partial {\bf J}'}\right ),
\label{ilb6}
\end{equation}
where $f$ stands for $f({\bf J},t)$, $f'$ stands for $f({\bf J}',t)$,
${\bf\Omega}$ stands for
${\bf\Omega}({\bf J},t)$ and ${\bf\Omega}'$ stands for
${\bf\Omega}({\bf J}',t)$. The pulsation ${\bf\Omega}({\bf J},t)=\partial
H/\partial {\bf J}$ is determined by $f({\bf J},t)$.  The Fourier-Laplace
transform of the dressed potential of
interaction 
$A^d_{{\bf k}{\bf k}'}({\bf J},{\bf J}',\omega)$ is defined by the integral
equation (see
Appendix A of \cite{angleaction2} and Appendix H of \cite{kinfd})
\begin{eqnarray}
\label{nmat3}
A^d_{{\bf k}{\bf k}'}({\bf J},{\bf J}',\omega)-(2\pi)^d\sum_{{\bf
k}''} \int d{\bf
J}''\, 
A_{{\bf k}{\bf k}''}({\bf J},{\bf
J}'') \frac{{\bf k}''\cdot \frac{\partial
f''}{\partial
{\bf J}''}}{{\bf
k}''\cdot {\bf
\Omega}''-\omega}A^d_{{\bf k}''{\bf k}'}({\bf J}'',{\bf J}',\omega)
=A_{{\bf k}{\bf k}'}({\bf J},{\bf
J}'),
\end{eqnarray}
where $A_{{\bf k}{\bf k}'}({\bf J},{\bf J}')$ is the Fourier
transform of the bare potential of
interaction. $A^d_{{\bf k}{\bf k}'}({\bf J},{\bf J}',\omega)$ can be interpreted
 as a Green
function [see Eq. (H6) of \cite{kinfd} where $\delta\tilde\Phi({\bf k},{\bf
J},\omega)$ satisfies the Volterra-Fredholm equation (H5)]. We can decompose the
dressed potential of interaction $A^d_{{\bf k}{\bf k}'}({\bf J},{\bf
J}',\omega)$  in angle-action variables on a
complete biorthogonal basis where the density-potential pairs
$\rho_{\alpha}({\bf r})$ and
$\Phi_{\alpha}({\bf r})$
satisfy \cite{kalnajs}\footnote{We consider an attractive self-interaction, like
gravity, hence
the sign $-$
in the second term of Eq. (\ref{i63}).} 
\begin{equation}
\Phi_{\alpha}({\bf r})=\int u(|{\bf r}-{\bf
r}'|) \rho_{\alpha}({\bf
r}')\,
d{\bf r}',\qquad \int \rho_{\alpha}({\bf r})\Phi_{\alpha'}({\bf r})^*\, d{\bf
r}=-\delta_{\alpha\alpha'}.
\label{i63}
\end{equation}
We then find that \cite{heyvaerts,angleaction2}
\begin{eqnarray}
A^d_{{\bf k}{\bf k}'}({\bf J},{\bf
J'},\omega)=-\sum_{\alpha\alpha'}{\hat\Phi}_{\alpha}({\bf k},{\bf
J})(\epsilon^{-1})_{\alpha\alpha'}(\omega){\hat\Phi}^*_{\alpha'}({\bf k}',{\bf
J}')
\label{can1b}
\end{eqnarray}
with the dielectric tensor
\begin{eqnarray}
\epsilon_{\alpha\alpha'}(\omega)=\delta_{\alpha\alpha'}+(2\pi)^d\int d{\bf
J}\sum_{{\bf
k}}\, \frac{{\bf k}\cdot \frac{\partial
f}{\partial {\bf J}}}{{\bf k}\cdot {\bf
\Omega}-\omega} \hat\Phi_{\alpha'}({\bf
k},{\bf J})\hat\Phi_{\alpha}^*({\bf
k},{\bf J}).
\label{i70}
\end{eqnarray}
Eqs. (\ref{nmat3}) and (\ref{i70}) are defined for any $\omega$ by performing
the
integration along the Landau contour. Eq. (\ref{can1b}) is the
proper generalization of the expression of the dressed potential of
interaction  ${\hat u}_{d}={\hat u}(k)/|\epsilon({\bf k},{\bf
k}\cdot {\bf v})|$ in the homogeneous case (see Appendix 
\ref{sec_gfn}). Eq.
(\ref{nmat3}) can be used as a definition of $A^d_{{\bf k}{\bf k}'}({\bf J},{\bf
J}',\omega)$ which does not require to introduce a biorthogonal basis (see
Appendix A of \cite{angleaction2}).  The two manners to define $A^d_{{\bf k}{\bf
k}'}({\bf J},{\bf
J}',\omega)$, either by Eq. (\ref{nmat3}) or by Eq. (\ref{can1b}),  are
equivalent.  If we
neglect collective effects by taking $\epsilon=1$, or by
replacing the dressed potential  $A^d_{{\bf k}{\bf
k}'}({\bf J},{\bf J}',{\bf k}\cdot {\bf
\Omega})$  by the bare potential $A_{{\bf k}{\bf k}'}({\bf J},{\bf J}')$, the
inhomogeneous Lenard-Balescu equation (\ref{ilb6}) reduces to the inhomogeneous
Landau
equation \cite{aa}. These equations can be extended to the multi-species case
\cite{hfcp,kinfd}.

{\it Remark:} The dispersion relation determining the proper complex pulsations
$\omega$ of an inhomogeneous system with long-range interactions reads 
\begin{eqnarray}
{\rm det}[\epsilon(\omega)]=0.
\label{disrelinh}
\end{eqnarray}
It can be used to study the linear dynamical stability of the system with
respect to the Vlasov equation. A
zero of the dispersion relation ${\rm det}[\epsilon(\omega)]=0$ for
$\omega_i>0$ signifies an unstable growing mode. The condition of marginal
stability corresponds to $\omega_i=0$. In the following, we
assume that the system is stable so that $\omega_i<0$ for all modes. This
implies that $\epsilon(\omega)$ does not vanish for any real $\omega$. This
assumption is necessary to make the inhomogeneous Lenard-Balescu
equation well-defined.

\subsection{Fokker-Planck equation}

The Lenard-Balescu equation (\ref{ilb6}) can be written under the form of a
Fokker-Planck equation
\begin{equation}
\label{ilb10}
\frac{\partial f}{\partial t}=\frac{\partial}{\partial J_{i}} \left
(D_{ij}\frac{\partial f}{\partial
J_{j}}- f F^{\rm pol}_i\right )
\end{equation} 
with a diffusion tensor and a friction by polarization given by
\begin{eqnarray}
D_{ij}&=&\pi(2\pi)^{d}m \sum_{{\bf k},{\bf k}'}\int d{\bf
J}'\, k_ik_j
|A^d_{{\bf k}{\bf k}'}({\bf J},{\bf
J'},{\bf k}\cdot {\bf\Omega})|^2 \delta({\bf
k}\cdot {\bf\Omega}-{\bf k}'\cdot
{\bf\Omega}') f({\bf J}'),
\label{ilb11}
\end{eqnarray}
\begin{eqnarray}
{\bf F}_{\rm pol}=\pi(2\pi)^{d}m\sum_{{\bf k},{\bf k}'}\int d{\bf
J}'\, {\bf k}
|A^d_{{\bf k}{\bf k}'}({\bf J},{\bf
J'},{\bf k}\cdot {\bf\Omega})|^2 \delta({\bf
k}\cdot {\bf\Omega}-{\bf k}'\cdot
{\bf\Omega}')\left ({\bf k}'\cdot \frac{\partial f'}{\partial {\bf J}'}\right
).
\label{ilb12}
\end{eqnarray}
The  Lenard-Balescu equation (\ref{ilb6}) can be directly derived
from the Fokker-Planck approach by calculating the diffusion and friction
coefficients (first and second moments of the action increment) as shown in Sec.
3 of \cite{angleaction2} and in \cite{kinfd}. The Lenard-Balescu  equation
(\ref{ilb6}) is valid at the
order $1/N$ so it describes the ``collisional'' evolution of the system due to
finite $N$ effects on a
timescale $\sim N t_D$, where $t_D$ is the dynamical time. The Lenard-Balescu
equation (\ref{ilb6}) conserves
the mass $M=\int f\, d{\bf J}$ and the energy $E=\int f H\, d{\bf J}$. The
Boltzmann entropy  $S=-\int
\frac{f}{m}\ln \frac{f}{m}\, d{\bf J}$ increases monotonically:
$\dot S\ge 0$ ($H$-theorem) \cite{hfcp}. The collisional evolution of the system
is due to a
condition of resonance. This condition of resonance, encapsulated in the
$\delta$-function, corresponds to ${\bf k}'\cdot {\bf \Omega}'={\bf k}\cdot {\bf
\Omega}$. In
most cases,
this condition of resonance can always be satisfied and the Boltzmann
distribution   $f_{\rm eq}({\bf J})=A e^{-\beta m H({\bf J})}$ is the
unique steady state of the Lenard-Balescu
equation. Therefore, under the effect of ``collisions'', the system reaches the
Boltzmann distribution on a relaxation time $t_{R}\sim Nt_D$.\footnote{For 3D
stellar systems, the relaxation time scales as $t_R\sim (N/\ln N) t_D$.
There is a logarithmic correction due to
the effect of strong collisions at small scales (the divergence at large scales
is regularized by the inhomogeneity of the system and by its finite
extension). Furthermore, in an infinite domain, there is no statistical
equilibrium state: the system evaporates and undergoes a
gravothermal catastrophe leading to core collapse (see, e.g., \cite{aa} for
more details).}
There is a case where the foregoing results are not valid
because of a
situation of kinetic blocking at the order $1/N$. This situation is
specifically discussed in Sec. \ref{sec_mono}.

In the thermal bath approximation, where the field particles have the
Boltzmann distribution  $f({\bf J}')\sim e^{-\beta m H({\bf J}')}$,
the friction by polarization is linked to the diffusion tensor by the Einstein
relation $F_i^{\rm
pol}=-\beta m D_{ij}({\bf J})\Omega_j({\bf J})$ \cite{angleaction2} and the 
Fokker-Planck equation takes the form
\begin{eqnarray}
\frac{\partial P}{\partial t}=\frac{\partial}{\partial J_i}\left
\lbrack D_{ij}\left
(\frac{\partial P}{\partial J_j}+\beta m P \Omega_j\right )\right\rbrack.
\label{ilb19}
\end{eqnarray}
This equation is similar to
the Kramers equation or to the Smoluchowski equation describing the motion of a
Brownian particle in a potential $H({\bf J})$. The diffusion
tensor depends on the action.  The distribution
function of the test particles relaxes towards the Boltzmann
distribution $P_{\rm eq}({\bf J})=A\, e^{-\beta m H({\bf J})}$ on a
timescale $t_{R}^{\rm bath}\sim N\,  t_D$ in any case.\footnote{For
3D self-gravitating systems the relaxation time scales as
$t_{R}^{\rm bath}\sim (N/\ln N)\, t_D$ and the system evaporates unless
artificially confined.}

\subsection{One dimensional systems with a monotonic frequency
profile: kinetic blocking}
\label{sec_mono}

In this section, we consider an inhomogeneous 1D system with long-range
interactions and assume that the binary potential of interaction is of the form
$u=u(|w-w'|,J,J')$. This is the case for spins with long-range interactions
moving on a sphere in relation to the process of vector resonant
relaxation (VRR) in galactic nuclei \cite{fbc}. In that case only
$1:1$ resonances are permitted and the Lenard-Balescu equation (\ref{ilb6}) 
reduces to 
\begin{equation}
\frac{\partial f}{\partial t}=m\frac{\partial}{\partial J}
\int dJ' \, \chi(J,J',\Omega(J))\delta(\Omega-\Omega') \left
(f' \frac{\partial f}{\partial J}-f \frac{\partial
f'}{\partial J'}\right ),
\label{ham106}
\end{equation}
where we have introduced the convenient notation 
$\chi(J,J',\Omega(J))=2\pi^2\sum_{k} |k|\,
|A^d_{kk}(J,J',k \Omega)|^2$. Using the identity
\begin{eqnarray}
\delta[g(x)]=\sum_j \frac{1}{|g'(x_j)|}\delta(x-x_j),
\label{gei7d}
\end{eqnarray}
where the $x_j$ are the simple zeros of the function $g(x)$ (i.e. $g(x_j)=0$ and
$g'(x_j)\neq 0$), we can rewrite Eq. (\ref{ham106}) as
\begin{equation}
\frac{\partial f}{\partial t}= m\frac{\partial}{\partial J}
\sum_r \frac{\chi(J,J_r,\Omega(J))}{|\Omega'(J_r)|}
\left
(f_r \frac{\partial f}{\partial J}- f \frac{\partial
f_r}{\partial J_r}\right ),
\label{aham105h}
\end{equation}
where the $J_r$ are the points that resonate with $J$, i.e., the points that
satisfy $\Omega(J_r)=\Omega(J)$.

If the frequency profile $\Omega(J)$ is
monotonic, we find that
\begin{eqnarray}
\frac{\partial f}{\partial t}=m\frac{\partial}{\partial J} \int dJ'\,
\chi(J,J',\Omega(J))   \frac{1}{|\Omega'(J)|}\delta(J'-J)\left (f'
\frac{\partial f}{\partial J}-f\frac{\partial f'}{\partial J'}\right
)=0.
\label{ham107}
\end{eqnarray}
Therefore, the distribution function $f(J)$ does not evolve at all on the
timescale $\sim N t_D$. This is a situation of kinetic
blocking due to the absence of resonance at the order $1/N$. This ``kinetic
blocking'' is
illustrated in \cite{fbc}. A kinetic equation has been obtained at the order
$1/N^2$
when collective effects are neglected \cite{fouvry}. For
generic potentials (see \cite{jbr} for more
details), this equation does not display kinetic blocking and
relaxes
towards the Boltzmann distribution on a timescale $t_{\rm R}\sim {N}^2
t_D$. In that case,
the relaxation is caused by three-body correlations.

In the bath approximation, where the field particles can have any
distribution $f_0(J')$ with a monotonic frequency profile, the friction by
polarization satisfies the
generalized Einstein relation $F_{\rm pol}=D\frac{d\ln f_0}{dJ}$
and the Fokker-Planck
equation takes the
form 
\begin{eqnarray}
\frac{\partial P}{\partial t}=\frac{\partial}{\partial J}\left\lbrack D(J)\left
(\frac{\partial P}{\partial
J}-P\frac{d\ln f_0}{dJ}\right )\right\rbrack
\label{ham98}
\end{eqnarray}
with a diffusion coefficient
\begin{eqnarray}
D=\frac{\chi(J,J,\Omega(J))}{|\Omega'(J)|}m f_0(J).
\label{tb3c}
\end{eqnarray}
Equation (\ref{tb3c}) is similar to the Smoluchowski equation
describing the motion of a Brownian particle in a potential $U(J)=-\ln
f_0(J)$. The distribution function of the test particle $P(J,t)$
relaxes towards the distribution of the bath $f_0(J)$ on a timescale
$t_R^{\rm bath}\sim N t_D$. In the case of a thermal bath, we
recover Eq. (\ref{ilb19}).

\section{Kinetic theory of 2D point vortices}
\label{sec_inhos3}

\subsection{Basic equations}
\label{sec_inhos4}

We consider a system of 2D point vortices of individual
circulation
$\gamma$ on the plane (see, e.g., Appendix A of \cite{kinfdvortex} for a
derivation of
this
model). The vortices may be subjected to an external
incompressible velocity field (exterior
perturbation) arising from a stream function $\psi_e({\bf r},t)$.
The equations of motion of the
point vortices   are  
\begin{eqnarray}
\frac{d{\bf
r}_{i}}{dt}=-{\bf z}\times \nabla\psi_d({\bf r}_i)-{\bf
z}\times \nabla\psi_e({\bf r}_i,t),
\label{n1zerob}
\end{eqnarray}
where $\psi_d({\bf r})=-(1/2\pi)\sum_j \gamma \ln|{\bf r}-{\bf r}_j|$ is the
exact stream function
produced by the point vortices. These equations can be written in
Hamiltonian form as $\gamma d{\bf r}_i/dt=-{\bf z}\times
\nabla(H_d+H_e)$, where $H_d=-(1/2\pi)\sum_{i<j} \gamma^2 \ln|{\bf
r}_i-{\bf r}_j|$ is the Hamiltonian of the point vortices and
$H_e=\sum_i\gamma \psi_e({\bf r}_i,t)$ is the Hamiltonian associated with the
external flow. 
The discrete (or singular exact)
vorticity
$\omega_d({\bf r},t)=\sum_i \gamma\delta({\bf r}-{\bf r}_i(t))$ of the point
vortex gas
satisfies the equations
\begin{eqnarray}
\frac{\partial\omega_d}{\partial t}+({\bf u}_d+{\bf u}_e)\cdot \nabla\omega_d=0,
\label{ham1}
\end{eqnarray}
\begin{eqnarray}
{\bf u}_d=-{\bf z}\times\nabla\psi_d,\qquad \omega_d=-\Delta\psi_d,
\label{ham2}
\end{eqnarray}
\begin{eqnarray}
{\bf u}_e=-{\bf z}\times\nabla\psi_e,\qquad \omega_e=-\Delta\psi_e,
\label{ham2b}
\end{eqnarray}
where $\psi_d({\bf r},t)$ is the 
stream function produced by the point vortices. These equations are similar to
the 2D
Euler-Poisson equations for an incompressible continuous flow but they apply
here to a singular vorticity field which is a sum of Dirac distributions. The
2D Euler-Poisson equations for an incompressible continuous flow are the
counterparts of the Vlasov-Poisson equations  in  stellar dynamics
\cite{jeans} and plasma physics \cite{vlasov} and the 2D Euler-Poisson equations
for a
singular system of point vortices are the counterparts of the Klimontovich
equations \cite{klimontovich} in plasma physics.

We introduce the mean vorticity $\omega({\bf r},t)=\langle
\omega_{d}({\bf r},t)\rangle$ corresponding to an ensemble
average of
$\omega_{d}({\bf r},t)$. We then write
$\omega_d({\bf r},t)=\omega({\bf r},t)+\delta \omega({\bf r},t)$ where
$\delta \omega({\bf r},t)$ denotes the fluctuations about the mean vorticity.
Similarly, we write $\psi_d({\bf r},t)=\psi({\bf
r},t)+\delta\psi({\bf r},t)$ where $\delta\psi({\bf
r},t)$ denotes the fluctuations about the mean stream function $\psi({\bf
r},t)=\langle \psi_d({\bf
r},t)\rangle$. Substituting this decomposition into Eq. (\ref{ham1}), we get
\begin{eqnarray}
\frac{\partial\omega}{\partial t}+\frac{\partial\delta\omega}{\partial t}+({\bf
u}+\delta{\bf u}+{\bf u}_e)\cdot \nabla(\omega+\delta\omega)=0,
\label{ham3}
\end{eqnarray}
\begin{eqnarray}
{\bf u}=-{\bf z}\times\nabla\psi,\qquad \omega=-\Delta\psi,
\label{ham2bb}
\end{eqnarray}\begin{eqnarray}
\delta{\bf u}=-{\bf z}\times\nabla\delta\psi,\qquad
\delta\omega=-\Delta\delta\psi.
\label{ham4}
\end{eqnarray}
If we introduce the total fluctuations $\delta {\bf u}_{\rm tot}=\delta{\bf
u}+{\bf u}_e$, $\delta\psi_{\rm tot}=\delta\psi+\psi_e$ and $\delta\omega_{\rm
tot}=\delta\omega+\omega_e$, which include the contribution of
the external perturbation, we can rewrite the foregoing
equations as
\begin{eqnarray}
\frac{\partial\omega}{\partial t}+\frac{\partial\delta\omega}{\partial t}+({\bf
u}+\delta{\bf u}_{\rm tot})\cdot \nabla(\omega+\delta\omega)=0,
\label{ham5}
\end{eqnarray}
\begin{eqnarray}
{\delta{\bf u}}_{\rm tot}=-{\bf z}\times\nabla\delta\psi_{\rm tot},\qquad
\delta\omega_{\rm tot}=-\Delta\delta\psi_{\rm tot}.
\label{ham5b}
\end{eqnarray}
Expanding the advection term in Eq. (\ref{ham5}) we obtain
\begin{eqnarray}
\frac{\partial\omega}{\partial t}+\frac{\partial\delta\omega}{\partial t}+{\bf
u}\cdot \nabla \omega+{\bf u}\cdot\nabla\delta\omega+\delta{\bf u}_{\rm
tot}\cdot \nabla\omega+\delta{\bf u}_{\rm tot}\cdot \nabla\delta\omega=0.
\label{ham6}
\end{eqnarray}
Taking the ensemble average of Eq. (\ref{ham6}) and subtracting the resulting
equation from Eq. (\ref{ham6}) we obtain the two coupled equations
\begin{eqnarray}
\frac{\partial\omega}{\partial t}+{\bf u}\cdot \nabla \omega=-\nabla\cdot
\langle \delta\omega \delta{\bf u}_{\rm tot}\rangle,
\label{ham7}
\end{eqnarray}
\begin{eqnarray}
\frac{\partial\delta\omega}{\partial t}+{\bf
u}\cdot\nabla\delta\omega+\delta{\bf u}_{\rm tot}\cdot \nabla\omega=-\nabla\cdot
(\delta\omega \delta{\bf u}_{\rm tot})+\nabla\cdot \langle \delta\omega
\delta{\bf u}_{\rm tot}\rangle,
\label{ham8}
\end{eqnarray}
which govern the evolution of the mean flow  and the fluctuations. 
To get the right hand side of Eqs. (\ref{ham7}) and (\ref{ham8}) we have used
the incompressibility of the flow $\nabla\cdot \delta{\bf u}_{\rm tot}=0$.
Equations (\ref{ham7}) and (\ref{ham8})  are exact
in the sense that no approximation has been made for the moment. The right hand
side of Eq.
(\ref{ham7}) can be interpreted as
a ``collision'' or  ``correlational'' term arising from the granularity of the
system
(finite $N$ effects).

We now assume that the
fluctuations $\delta\psi({\bf r},t)$  of the stream function created by the
point
vortices are weak. Since the circulation of the point vortices scales as
$\gamma\sim 1/N$ this approximation is valid when $N\gg
1$. We also assume that the external velocity
is weak. If we
ignore the external velocity and the fluctuations of
the stream function due to finite $N$ effects altogether, the
collision term vanishes and Eq. (\ref{ham7}) reduces to the 2D Euler
equation
\begin{eqnarray}
\frac{\partial\omega}{\partial t}+{\bf u}\cdot \nabla \omega=0.
\label{ham7ep}
\end{eqnarray}
The 2D Euler-Poisson equations (\ref{ham2bb}) and (\ref{ham7ep})  describe
a
self-consistent mean field (collisionless) dynamics. It is valid in the limit
$\psi_e\rightarrow 0$ and  in a proper thermodynamic limit
$N\rightarrow +\infty$ with $\gamma\sim 1/N$. It is also valid for
sufficiently short times.

We now take into account a small correction to the 2D Euler equation obtained by
keeping the 
collision term on the right hand side of Eq. (\ref{ham7}) but neglecting the
quadratic terms on the right hand side of Eq. (\ref{ham8}). We therefore obtain
a set of two coupled equations
\begin{eqnarray}
\frac{\partial\omega}{\partial t}+{\bf u}\cdot \nabla \omega=-\nabla\cdot
\langle \delta\omega \delta{\bf u}_{\rm tot}\rangle,
\label{ham7b}
\end{eqnarray}
\begin{eqnarray}
\frac{\partial\delta\omega}{\partial t}+{\bf
u}\cdot\nabla\delta\omega+\delta{\bf u}_{\rm tot}\cdot \nabla\omega=0.
\label{ham8b}
\end{eqnarray}
These equations form the starting point of the quasilinear theory of 2D
point vortices which is
valid in a weak
coupling  approximation ($\gamma\sim 1/N\ll 1$) and for a weak external
stochastic perturbation ($\psi_e\ll \psi$). Eq. (\ref{ham7b})
describes the evolution of the mean vorticity sourced by
the correlations of the fluctuations and Eq. (\ref{ham8b}) describes the
evolution
of the fluctuations due to  the
granularities of the system (finite $N$ effects) and the external flow. These
equations are valid at the order $1/N$ and to leading order in  
$\psi_e$.

If we restrict ourselves to axisymmetric mean
flows\footnote{In
the following, we
assume
that the axisymmetric flow remains dynamically (Euler) stable during the
whole evolution. This
may not always be the case. Even if we start from an Euler stable vorticity
field $\omega_0(r)$, 
the ``collision'' term (r.h.s. in Eq. (\ref{ham7b})) will change it and induce a
temporal 
evolution of the vorticity field $\omega(r,t)$. The system may
become
dynamically (Euler) unstable and undergo a dynamical phase
transition from an axisymmetric flow to a more complicated flow.
We assume here that this transition does not take place or we consider a period
of time preceding this transition.} and introduce a polar system of
coordinates, we have
\begin{eqnarray}
{\bf u}=u(r,t){\bf e}_{\theta},\qquad \psi=\psi(r,t),\qquad \omega=\omega(r,t),
\label{ham9}
\end{eqnarray}
\begin{eqnarray}
u=-\frac{\partial\psi}{\partial r}, \qquad
\omega=-\frac{1}{r}\frac{\partial}{\partial r}\left
(r\frac{\partial\psi}{\partial r}\right ),\qquad 
\omega=\frac{1}{r}\frac{\partial}{\partial r}
(r u).
\label{ham10}
\end{eqnarray}
On the other hand,  the two components of the  fluctuating velocity field read
\begin{eqnarray}
(\delta{\bf u}_{\rm tot})_r=\frac{1}{r}\frac{\partial\delta\psi_{\rm
tot}}{\partial \theta},\qquad (\delta{\bf u}_{\rm
tot})_\theta=-\frac{\partial\delta\psi_{\rm tot}}{\partial
r}.
\label{ham11}
\end{eqnarray}
As a result, Eqs. (\ref{ham7b}) and (\ref{ham8b}) become 
\begin{eqnarray}
\frac{\partial\omega}{\partial t}=-\frac{1}{r}\frac{\partial}{\partial r}
\left\langle
\delta\omega \frac{\partial\delta\psi_{\rm tot}}{\partial \theta}\right\rangle,
\label{ham12}
\end{eqnarray}
\begin{eqnarray}
\frac{\partial\delta\omega}{\partial
t}+\Omega\frac{\partial\delta\omega}{\partial
\theta}+\frac{1}{r} \frac{\partial\delta\psi_{\rm tot}}{\partial \theta}
\frac{\partial\omega}{\partial r}=0,
\label{ham13}
\end{eqnarray}
where we have introduced the angular velocity $\Omega(r,t)=u(r,t)/r$. It is
related to the vorticity by $\Omega(r,t)=\frac{1}{r^2}\int_0^r\omega(r',t)r'\,
dr'$.

\subsection{Bogoliubov ansatz}
\label{sec_bog}

In order to solve Eq. (\ref{ham13})  for the fluctuations, we resort to
the Bogoliubov ansatz. We  assume that there exists a timescale separation
between a slow and a fast dynamics and we regard $\Omega(r)$ and $\omega(r)$ in
Eq.
(\ref{ham13}) as ``frozen'' (independent of time) at the scale of the fast
dynamics. This amounts to neglecting
the temporal variation of the mean flow when we consider the evolution of the
fluctuations. This is possible when the mean
vorticity field evolves on
a
secular timescale that is long compared to the correlation time (the time
over which the
correlations of the fluctuations have their essential
support).  We can then
introduce Fourier-Laplace transforms in $r$ and $t$ for the vorticity
fluctuations,
writing 
\begin{eqnarray}
\delta{\tilde\omega}(n,r,\sigma)=\int_{0}^{2\pi}\frac{d\theta}{2\pi}
\int_{0}^{+\infty} dt\, e^{-i(n\theta-\sigma
t)}\delta\omega(\theta,r,t)
\label{ham14}
\end{eqnarray}
and
\begin{eqnarray}
\delta{\omega}(\theta,r,t)=\sum_{n=-\infty}^{+\infty}\int_{\cal
C}\frac{d\sigma}{ 2\pi } \,
e^{i(n\theta-\sigma
t)}\delta{\tilde \omega}(n,r,\sigma)
\label{ham15}
\end{eqnarray}
with ${\rm Im}\, \sigma>0$. Similar expressions hold  for the fluctuating stream
function $\delta\psi(r,\theta,t)$
and for the external stream function $\psi_e(r,\theta,t)$.  For future
reference, we recall the Fourier
representation of the Dirac $\delta$-function
\begin{equation}
\delta(x)=\int_{-\infty}^{+\infty} e^{i x t}\, \frac{dt}{2\pi}.
\label{delta}
\end{equation} 
We also recall the identity
\begin{eqnarray}
\delta(\lambda x)=\frac{1}{|\lambda|}\delta(x).
\label{gei7b}
\end{eqnarray}

\subsection{Lenard-Balescu-like equation}
\label{sec_lblike}

We consider an isolated system of point vortices and take
$\psi_e=0$.\footnote{The case of a system of point vortices stochastically
forced by an external perturbation, leading to the so-called SDD equation, is
considered in \cite{kinfdvortex}.} We want to
derive the kinetic equation governing the evolution of the mean vorticity at the
order $1/N$ by using the Klimontovich formalism \cite{klimontovich}. The
derivation follows the one given in Refs.
\cite{dubin,klim} (see also \cite{kinfdvortex} for unidirectional flows). We
start
from the
quasilinear equations
(\ref{ham12}) and (\ref{ham13}) without the external stream function that we
rewrite as
\begin{eqnarray}
\frac{\partial\omega}{\partial t}=-\frac{1}{r}\frac{\partial}{\partial r}
\left\langle
\delta\omega \frac{\partial\delta\psi}{\partial \theta}\right\rangle,
\label{j1}
\end{eqnarray}
\begin{eqnarray}
\frac{\partial\delta\omega}{\partial
t}+\Omega\frac{\partial\delta\omega}{\partial
\theta}+\frac{1}{r}\frac{\partial\delta\psi}{\partial \theta}
\frac{\partial\omega}{\partial r}=0.
\label{j2v}
\end{eqnarray}
We make the Bogoliubov ansatz which treats $\omega$ and $\Omega$ as
time independent in Eq. (\ref{j2v}). Taking the Fourier-Laplace transform of Eq.
(\ref{j2v}), we obtain
\begin{eqnarray}
\label{j3}
\delta{\tilde\omega}(n,r,\sigma)=-\frac{n\frac{1}{r}\frac{\partial\omega}{
\partial
r}}{n\Omega-\sigma}\delta{\tilde\psi}(n,r,\sigma)+\frac{\delta{\hat\omega}(n,r
,0)}{i(
n\Omega-\sigma)},
\end{eqnarray}
where $\delta{\hat\omega}(n,r,t=0)$ is the Fourier transform of the initial
vorticity fluctuation due to finite $N$ effects. 

By using the linear response theory it is shown in Appendix
\ref{sec_dpicv} that the Fourier-Laplace transform of the perturbed
stream function
is related to the Fourier transform of the initial
vorticity fluctuation by
\begin{eqnarray}
\delta{\tilde\psi}(n,r,\sigma)=\int_0^{+\infty} 2\pi r'  dr'\,
G(n,r,r',\sigma)\frac{\delta{\hat\omega}(n,r',0)}{i(
n\Omega'-\sigma)},
\label{j5v}
\end{eqnarray}
where the Green function $G(n,r,r',\sigma)$ is defined by Eq.
(\ref{pot11ht}) in the general case and
by
Eq. (\ref{defg}) for the usual potential of interaction between point
vortices. This is the dressed potential
of interaction taking into account collective effects [see Eq.
(\ref{pot6blues2})]. We have also introduced
the notation $\Omega'$ for $\Omega(r',t)$.

Taking the inverse
Laplace transform of Eq. (\ref{j5v}), using the Cauchy residue theorem, and
neglecting the contribution of the damped
modes for sufficiently
late times,\footnote{We only consider the contribution of the
pole $\sigma-n\Omega'$ in Eq. (\ref{j5v}) and ignore the contribution of the
proper modes $\sigma(n)$ of the
flow which are the solutions of the Rayleigh equation (\ref{raynw}).} we obtain
\begin{eqnarray}
\label{j6}
\delta{\hat\psi}(n,r,t)=\int_0^{+\infty} 2\pi r' dr'\,
G(n,r,r',n\Omega') 
\delta{\hat \omega}(n,r',0)e^{-i n \Omega't}.
\end{eqnarray}

On the other hand, taking the Fourier transform of Eq. (\ref{j2v}), we find that
\begin{eqnarray}
\label{j7bu}
\frac{\partial \delta{\hat \omega}}{\partial t}+in\Omega\delta{\hat
\omega}=-\frac{in}{r}\frac{\partial \omega}{\partial r}\delta{\hat\psi}.
\end{eqnarray}
This first order differential equation in time can be solved with the method of
the variation of the constant, giving
\begin{eqnarray}
\label{j8v}
\delta{\hat \omega}(n,r,t)=\delta{\hat\omega}(n,r,0)e^{-i n\Omega
t}-\frac{in}{r}
\frac{\partial \omega}{\partial r}\int_0^t dt'\, \delta{\hat
\psi}(n,r,t')e^{in\Omega (t'-t)}.
\end{eqnarray}
Substituting Eq. (\ref{j6}) into Eq. (\ref{j8v}), we obtain
\begin{eqnarray}
\label{j9}
\delta{\hat \omega}(n,r,t)=\delta{\hat\omega}(n,r,0)e^{-i n\Omega
t}-\frac{in}{r}\frac{\partial \omega}{\partial r}\int_0^{+\infty} 2\pi r' dr'\,
G(n,r,r',n\Omega') \delta{\hat \omega}(n,r',0) e^{-i n\Omega t}\int_0^t dt'\,
e^{in(\Omega-\Omega')t'}.
\end{eqnarray}
Eqs.
(\ref{j6}) and
(\ref{j9}) relate $\delta{\hat\psi}(n,r,t)$ and $\delta{\hat \omega}(n,r,t)$ to
the
initial fluctuation $\delta{\hat \omega}(n,r,0)$.

We can now compute the flux
\begin{eqnarray}
\label{j10}
\left\langle \delta
\omega\frac{\partial\delta\psi}{\partial\theta}\right\rangle=\sum_{n,n'}
i n' e^{in\theta}e^{in'\theta}\langle \delta{\hat \omega}(n,r,t)\delta{\hat
\psi}(n',r,t)\rangle.
\end{eqnarray}
From Eqs. (\ref{j6}) and (\ref{j9}) we get
\begin{eqnarray}
\label{j11}
\langle \delta{\hat \omega}(n,r,t)\delta{\hat
\psi}(n',r,t)\rangle=\int_0^{+\infty}2\pi r'dr'\, G(n',r,r',n'\Omega')
e^{-i n'\Omega't}e^{-i n\Omega t}\langle \delta{\hat \omega}(n,r,0)\delta{\hat
\omega}(n',r',0)\rangle\nonumber\\
-\int_0^{+\infty}2\pi r'dr'\, G(n',r,r',n'\Omega')
e^{-i n'\Omega't}\frac{in}{r}\frac{\partial\omega}{\partial r}
\int_0^{+\infty}2\pi r''dr''\, G(n,r,r'',n\Omega'')
\nonumber\\
\times \langle\delta{\hat \omega}(n,r'',0)\delta{\hat
\omega}(n',r',0)\rangle e^{-i n\Omega t}\int_0^t dt'\,
e^{in(\Omega-\Omega'')t'}. 
\end{eqnarray}
We assume that at $t=0$ there are no
correlations among the point vortices.
Therefore, we have (see Appendix D of \cite{klim}) 
\begin{eqnarray}
\label{j12}
\langle \delta{\hat
\omega}(n,r,0)\delta{\hat\omega}(n',r',0)\rangle=\delta_{n,-n'}
\frac{\delta(r-r')}{2\pi r}\gamma\omega(r).
\end{eqnarray}
Equation (\ref{j11}) then reduces to 
\begin{eqnarray}
\label{j13bu}
\langle \delta{\hat \omega}(n,r,t)\delta{\hat
\psi}(n',r,t)\rangle=G(-n,r,r,-n\Omega)\gamma\omega(r)\delta_{n,-n'}
\nonumber\\
-\delta_{n,-n'}\int_0^{+\infty}2\pi r'dr'\, G(-n,r,r',-n\Omega')
\frac{in}{r}\frac{\partial\omega}{\partial r}
G(n,r,r',n\Omega')\gamma\omega(r')
\int_0^t ds\, e^{-in(\Omega-\Omega')s}, 
\end{eqnarray}
where we have set $s=t-t'$. Using Eq. (\ref{obv}), we can rewrite
this equation as
\begin{eqnarray}
\label{j13cu}
\langle \delta{\hat \omega}(n,r,t)\delta{\hat
\psi}(n',r,t)\rangle=G(n,r,r,n\Omega)^*\gamma\omega(r)\delta_{n,-n'}
\nonumber\\
-\delta_{n,-n'}\int_0^{+\infty}2\pi r'dr'\, 
\frac{in}{r}\frac{\partial\omega}{\partial r}
|G(n,r,r',n\Omega')|^2\gamma\omega(r')
\int_0^t ds\, e^{-in(\Omega-\Omega')s}. 
\end{eqnarray}
Substituting this relation into Eq. (\ref{j10}),
and taking the limit $t\rightarrow +\infty$, we obtain
\begin{eqnarray}
\label{j14}
\left\langle\delta\omega\frac{\partial\delta\psi}{\partial\theta}
\right\rangle=-\sum_n
i n G(n,r,r,n\Omega)^*\gamma\omega(r)
\nonumber\\
-\sum_n  n^2 \int_0^{+\infty}2\pi r'dr'\,
\frac{1}{r}\frac{\partial\omega}{\partial r}
|G(n,r,r',n\Omega')|^2\gamma\omega(r')
\int_0^{+\infty} ds\, e^{-in(\Omega-\Omega')s}. 
\end{eqnarray}
Making the transformations $s\rightarrow -s$ and $n\rightarrow -n$ we see
that
we can replace $\int_0^{+\infty} ds$ by $\frac{1}{2}\int_{-\infty}^{+\infty}
ds$. We then get 
\begin{eqnarray}
\label{j15bu}
\left\langle\delta\omega\frac{\partial\delta\psi}{\partial\theta}
\right\rangle=-\sum_n i
n G(n,r,r,n\Omega)^*\gamma\omega(r)
\nonumber\\
-\sum_n  n^2 \int_0^{+\infty}2\pi r'dr'\, 
\frac{1}{r}\frac{\partial\omega}{\partial r}
|G(n,r,r',n\Omega')|^2\gamma\omega(r')
\frac{1}{2}\int_{-\infty}^{+\infty} ds\, e^{in(\Omega-\Omega')s}. 
\end{eqnarray}
Using the identity (\ref{delta}), we arrive at
\begin{eqnarray}
\label{j17add}
\left\langle\delta\omega\frac{\partial\delta\psi}{\partial\theta}
\right\rangle=-\sum_n i
n G(n,r,r,n\Omega)^*\gamma\omega(r)
\nonumber\\
-\pi\sum_n  n^2 \int_0^{+\infty}2\pi r'dr'\, 
\frac{1}{r}\frac{\partial\omega}{\partial r}
|G(n,r,r',n\Omega')|^2\gamma\omega(r')\delta\lbrack n(\Omega-\Omega')\rbrack.
\end{eqnarray}
Then, using the identity (\ref{gei7b}), we obtain
\begin{eqnarray}
\label{j17h}
\left\langle\delta\omega\frac{\partial\delta\psi}{\partial\theta}
\right\rangle=-\sum_n i
n G(n,r,r,n\Omega)^*\gamma\omega(r)
\nonumber\\
-\pi\sum_n  |n| \int_0^{+\infty}2\pi r'dr'\, 
\frac{1}{r}\frac{\partial\omega}{\partial r}
|G(n,r,r',n\Omega')|^2\gamma\omega(r')\delta(\Omega-\Omega').
\end{eqnarray}
We can rewrite the foregoing
equation as
\begin{eqnarray}
\label{j18}
\left\langle\delta\omega\frac{\partial\delta\psi}{\partial\theta}
\right\rangle=-\sum_n 
n\, {\rm Im} G(n,r,r,n\Omega)\gamma\omega(r)
-\pi\sum_n  |n| \int_0^{+\infty}2\pi r'dr'\, 
\frac{1}{r}\frac{\partial\omega}{\partial r}
|G(n,r,r',n\Omega')|^2\gamma\omega(r')\delta(\Omega-\Omega').
\end{eqnarray}
The first term is the drift by polarization term and the second term is the
diffusion term (see below).
Using the identity (\ref{hami4}) and substituting
the flux from Eq.
(\ref{j18}) into Eq. (\ref{j1}), we obtain
the Lenard-Balescu-like equation \cite{dubin,dubin2,klim,onsagerkin}
\begin{eqnarray}
\frac{\partial\omega}{\partial
t}=2\pi^2\gamma\frac{1}{r}\frac{\partial}{\partial r} \sum_n
\int_0^{+\infty} r' dr'\, |n| |G(n,r,r',n\Omega)|^2  
\delta(\Omega-\Omega')\left
(\frac{1}{r}\omega' \frac{\partial \omega}{\partial
r}-\frac{1}{r'}\omega\frac{\partial \omega'}{\partial r'}\right ).
\label{j19}
\end{eqnarray}
We can easily extend the derivation of the Lenard-Balescu equation to the
multispecies case \cite{dubin2,kinfdvortex}.

{\it Remark:} From Eq. (\ref{j6}) we get
\begin{eqnarray}
\label{po1}
\langle
\delta{\hat\psi}(n,r,t)\delta{\hat\psi}(n',r,t')\rangle=\int_0^{+\infty}
2\pi r'\, dr' \int_0^{+\infty} 2\pi r''\, dr'' G(n,r,r',n\Omega')
G(n',r,r'',n'\Omega'')\nonumber\\
\times  \langle \delta{\hat
\omega}(n,r',0)\delta{\hat \omega}(n',r'',0)\rangle e^{-i n\Omega' t} e^{-i
n'\Omega'' t'}.
\end{eqnarray}
Using Eq. (\ref{j12}) we obtain
\begin{eqnarray}
\label{po2}
\langle \delta{\hat\psi}(n,r,t)\delta{\hat\psi}(n',r,t')\rangle=\delta_{n,-n'}
\int_0^ { +\infty} 2\pi r'\, dr' |G(n,r,r',n\Omega')|^2  e^{-i
n\Omega'(t-t')}\gamma \omega(r').
\end{eqnarray}
This is the temporal correlation function  of
the stream function fluctuations.  Taking its
Laplace transform, we obtain the power spectrum
produced by a random distribution of point vortices 
\begin{eqnarray}
\label{po3}
\langle\delta{\tilde\psi}(n,r,\sigma)\delta{\tilde\psi}(n',r,
\sigma')\rangle=(2\pi)^3 \gamma \delta_{n, -n' }\delta(\sigma+\sigma')
\int_0^{+\infty} r'\, dr' |G(n,r,r',\sigma)|^2 \delta(\sigma-n\Omega')
\omega(r').
\end{eqnarray}
This returns Eq. (34) of \cite{klim}. Substituting Eq.
(\ref{j8v}) into Eq. (\ref{j10}) we can write the flux as
\begin{eqnarray}
\label{feu2}
\left\langle \delta
\omega\frac{\partial\delta\psi}{\partial\theta}\right\rangle=\sum_{n,n'} in'
e^{i(n+n')\theta}\langle \delta{\hat\psi}(n',r,t)\delta{\hat
\omega}(n,r,0)\rangle e^{-in\Omega t}\nonumber\\
+\sum_{n,n'} in'e^{i(n+n')\theta}
\left (-\frac{in}{r}\frac{\partial\omega}{\partial r}\right )\int_0^t dt'
\langle \delta{\hat\psi}(n',r,t)\delta{\hat \psi}(n,r,t')\rangle
e^{in\Omega(t'-t)},
\end{eqnarray}
which exhibits the drift term and the diffusion term. This expression shows
that the diffusion coefficient is given by the time integral of the
auto-correlation function of the velocity
\cite{bbgky,klim,kinfdvortex}.

\subsection{Landau-like equation}

If we ignore collective effects, we would be tempted to replace $G$ by $G_{\rm
bare}$ in
Eq. (\ref{j18}). But, in that case, the drift by polarization would vanish
since
$G_{\rm bare}$ is
real (see Appendix \ref{sec_swift}). Therefore, we must first compute ${\rm
Im}G$ by using Eq.
(\ref{hami4}),
{\it then} replace $G$
by $G_{\rm bare}$. To directly derive the Landau equation from the quasilinear
equations
(\ref{j1}) and (\ref{j2v}), we can proceed as
follows.\footnote{This approach is related to the iterative
method used in
\cite{bbgky} to derive the Landau equation in physical space.}

According to Eq. (\ref{pot10})  we have
\begin{eqnarray}
\delta{\hat\psi}(n,r,t)=\int_0^{+\infty} 2\pi r' dr'\,
G_{\rm bare}(n,r,r') 
\delta{\hat \omega}(n,r',t).
\label{p27}
\end{eqnarray}
Substituting Eq. (\ref{j8v}) into Eq. (\ref{p27}) we obtain
\begin{eqnarray}
\label{p28}
\delta{\hat\psi}(n,r,t)=\int_0^{+\infty} 2\pi r' dr'\,
G_{\rm bare}(n,r,r') 
\delta{\hat \omega}(n,r',0)e^{-in\Omega' t}\nonumber\\
-\int_0^{+\infty} 2\pi r' dr'\, G_{\rm bare}(n,r,r') \frac{in}{r'}
\frac{\partial \omega'}{\partial r'}\int_0^t dt'\, \delta{\hat
\psi}(n,r',t')e^{in\Omega'(t'-t)}.
\end{eqnarray}
This is an exact Volterra-Fredholm integral equation equivalent
to Eq. (\ref{dq3}). In order to relate
$\delta{\hat\psi}(n,r,t)$ to $\delta{\hat \omega}(n,r,0)$, instead of using Eq.
(\ref{j6}), we shall use an
iterative
method. We assume that
we can replace $\delta{\hat\psi}$ in the second line of Eq. (\ref{p28}) by its
expression
obtained by keeping only the contribution from the first line. This gives
\begin{eqnarray}
\label{p29}
\delta{\hat\psi}(n,r,t)=\int_0^{+\infty} 2\pi r' dr'\,
G_{\rm bare}(n,r,r') 
\delta{\hat \omega}(n,r',0)e^{-in\Omega' t}\nonumber\\
-\int_0^{+\infty} 2\pi r' dr'\, G_{\rm bare}(n,r,r') \frac{in}{r'}
\frac{\partial \omega'}{\partial r'}\int_0^t dt'\int_0^{+\infty} 2\pi r'' dr''\,
G_{\rm bare}(n,r',r'') 
\delta{\hat \omega}(n,r'',0)e^{-in\Omega'' t'}
e^{in\Omega'(t'-t)}.
\end{eqnarray}
We use the same strategy in Eq. (\ref{j8v}), thereby obtaining
\begin{eqnarray}
\label{p30}
\delta{\hat \omega}(n,r,t)=\delta{\hat\omega}(n,r,0)e^{-i n\Omega
t}-\frac{in}{r}
\frac{\partial \omega}{\partial r}
\int_0^{+\infty} 2\pi r' dr'\,
G_{\rm bare}(n,r,r') 
\delta{\hat \omega}(n,r',0)e^{-in\Omega t}
\int_0^t dt'\, e^{in(\Omega-\Omega') t'}.
\end{eqnarray}
We see that Eqs. (\ref{p29}) and (\ref{p30})
are similar to Eqs.
(\ref{j6}) and (\ref{j9}) except that $G$ is replaced by $G_{\rm bare}$ and
there
is a ``new'' term
in Eq.   (\ref{p29}) in addition to the ``old'' one. This new term
accounts for the drift by polarization. It substitutes itself to the term
proportional to ${\rm Im}G$ which vanishes when
$G$ is replaced by $G_{\rm bare}$ (as noted above). On the other hand, the old
term with
$G$ replaced by $G_{\rm bare}$ accounts for the diffusion. Therefore, repeating
the calculations of Sec.  \ref{sec_lblike}, or directly using Eq.
(\ref{j18})
with $G$
replaced by $G_{\rm bare}$, we get
\begin{eqnarray}
\label{p26}
\left\langle\delta\omega\frac{\partial\delta\psi}{\partial\theta}
\right\rangle_{\rm diffusion}=
-\pi\sum_n  |n| \int_0^{+\infty}2\pi r'dr'\, 
\frac{1}{r}\frac{\partial\omega}{\partial r}
G_{\rm bare}(n,r,r')^2\gamma\omega(r')\delta(\Omega-\Omega').
\end{eqnarray}
This quantity corresponds to the product of the first term in Eq. (\ref{p29})
with the two terms in Eq. (\ref{p30}). Let us now compute the drift term.
To that purpose, we have
to compute
\begin{eqnarray}
\label{aaa3}
\langle \delta{\hat \omega}(n,r,t)\delta{\hat
\psi}(n',r,t)\rangle_{\rm drift}=-\int_0^{+\infty}2\pi r'dr'\,
G_{\rm bare}(n',r,r')\frac{in'}{r'}\frac{\partial\omega'}{\partial r'}
\int_0^tdt'\, e^{in'\Omega'(t'-t)}\nonumber\\
\times\int_0^{+\infty}2\pi r''dr''\, G_{\rm bare}(n',r',r'')
e^{-i n'\Omega'' t'}e^{-in\Omega t} \langle\delta{\hat \omega}(n,r,0)\delta{\hat
\omega}(n',r'',0)\rangle.
\end{eqnarray}
This quantity corresponds to the product of the second term in Eq. (\ref{p29})
with the first term in Eq. (\ref{p30}). In line with our perturbative approach
we neglect the product of the second terms in Eqs. (\ref{p29}) and (\ref{p30}).
Using Eq. (\ref{j12}) we get
\begin{eqnarray}
\label{pb28bu}
\langle \delta{\hat \omega}(n,r,t)\delta{\hat
\psi}(n',r,t)\rangle_{\rm drift}=\delta_{n,-n'}\int_0^{+\infty}2\pi r'dr'\,
G_{\rm bare}(-n,r,r')\frac{in}{r'}\frac{\partial\omega'}{\partial r'}
\int_0^tds\, G_{\rm bare}(-n,r',r) e^{-in(\Omega-\Omega')s}
  \gamma\omega(r),
\end{eqnarray}
where we have set $s=t-t'$. Substituting this relation into Eq. (\ref{j10}), we
obtain
\begin{eqnarray}
\label{aaa1}
\left\langle\delta\omega\frac{\partial\delta\psi}{\partial\theta}
\right\rangle_{\rm drift}=\sum_n n^2 \int_0^{+\infty}2\pi r'dr'\,
G_{\rm bare}(-n,r,r')\frac{1}{r'}\frac{\partial\omega'}{\partial r'}
\int_0^tds\, G_{\rm bare}(-n,r',r) e^{-in(\Omega-\Omega')s}
  \gamma\omega(r).
\end{eqnarray}
If we let $t\rightarrow
+\infty$, we can rewrite the foregoing equation as
\begin{eqnarray}
\label{p31bu}
\left\langle\delta\omega\frac{\partial\delta\psi}{\partial\theta}
\right\rangle_{\rm drift}=\sum_n n^2 \int_0^{+\infty}2\pi r'dr'\,
G_{\rm bare}(-n,r,r')\frac{1}{r'}\frac{\partial\omega'}{\partial r'}G_{\rm
bare}(-n,r',r)\gamma\omega(r)
\int_0^{+\infty}ds\,  e^{-in(\Omega-\Omega')s}.
 \end{eqnarray}
Making the transformations
$s\rightarrow -s$ and $n\rightarrow -n$ we see that
we can replace $\int_0^{+\infty} ds$ by $\frac{1}{2}\int_{-\infty}^{+\infty}
ds$. We then get 
\begin{eqnarray}
\label{p32}
\left\langle\delta\omega\frac{\partial\delta\psi}{\partial\theta}
\right\rangle_{\rm drift}=\sum_n n^2 \int_0^{+\infty}2\pi r'dr'\,
G_{\rm bare}(-n,r,r')\frac{1}{r'}\frac{\partial\omega'}{\partial r'}G_{\rm
bare}(-n,r',r)\gamma\omega(r)\frac{1}{2}
\int_{-\infty}^{+\infty}ds\,  e^{-in(\Omega-\Omega')s}.
\end{eqnarray}
Using the identity (\ref{delta}) we arrive at the expression
\begin{eqnarray}
\label{p34}
\left\langle\delta\omega\frac{\partial\delta\psi}{\partial\theta}
\right\rangle_{\rm drift}=2\pi^2\sum_n n^2 \int_0^{+\infty} r'dr'\,
G_{\rm bare}(-n,r,r')\frac{1}{r'}\frac{\partial\omega'}{\partial r'}G_{\rm
bare}(-n,r',r)\gamma\omega(r)
\delta\left\lbrack n(\Omega-\Omega')\right\rbrack.
\end{eqnarray}
Then, using the identity (\ref{gei7b}), we obtain
\begin{eqnarray}
\label{p35}
\left\langle\delta\omega\frac{\partial\delta\psi}{\partial\theta}
\right\rangle_{\rm drift}=2\pi^2\sum_n |n| \int_0^{+\infty} r'dr'\,
G_{\rm bare}(n,r,r')^2 \frac{1}{r'}\frac{\partial\omega'}{\partial r'}
\delta (\Omega-\Omega')\gamma\omega(r).
\end{eqnarray}
Adding Eqs. (\ref{p26}) and (\ref{p35}), we find that
\begin{eqnarray}
\label{p36}
\left\langle\delta\omega\frac{\partial\delta\psi}{\partial\theta}
\right\rangle=
-\pi\sum_n  |n| \int_0^{+\infty}2\pi r'dr'\, 
\frac{1}{r}\frac{\partial\omega}{\partial r}
G_{\rm bare}(n,r,r')^2\delta(\Omega-\Omega')\gamma\omega(r')\nonumber\\
+2\pi^2\sum_n |n| \int_0^{+\infty} r'dr'\,
G_{\rm bare}(n,r,r')^2 \frac{1}{r'}\frac{\partial\omega'}{\partial r'}
\delta (\Omega-\Omega')\gamma\omega(r).
\end{eqnarray}
Substituting this flux into Eq. (\ref{j1}), we obtain
the Landau-like equation \cite{pre,cl,bbgky,kindetail}
\begin{eqnarray}
\frac{\partial\omega}{\partial
t}=2\pi^2\gamma\frac{1}{r}\frac{\partial}{\partial r} \sum_n
\int_0^{+\infty} r' dr'\, |n| G_{\rm bare}(n,r,r')^2  
\delta(\Omega-\Omega')\left
(\omega' \frac{1}{r}\frac{\partial \omega}{\partial
r}-\omega\frac{1}{r'}\frac{\partial \omega'}{\partial r'}\right ).
\label{j19land}
\end{eqnarray}
We can easily extend the derivation of the Landau equation to the
multispecies case \cite{cl}.

{\it Remark:} The sum over $n$ can be performed explicitly (see Appendix
\ref{sec_nce}) and the Landau equation
(\ref{j19land}) can be written as \cite{pre,cl,bbgky,kindetail}
\begin{eqnarray}
\frac{\partial\omega}{\partial
t}=-\frac{1}{4}\gamma\frac{1}{r}\frac{\partial}{\partial r}
\int_0^{+\infty} r' dr'\, \ln\left\lbrack 1-\left (\frac{r_{<}}{r_{>}}\right
)^2\right\rbrack 
\delta(\Omega-\Omega')\left
(\omega' \frac{1}{r}\frac{\partial \omega}{\partial
r}-\omega\frac{1}{r'}\frac{\partial \omega'}{\partial r'}\right ).
\label{j19landexp}
\end{eqnarray}
We note that this
equation is perfectly well-defined without any divergence (see footnote 28
below), contrary to the usual Landau equation (\ref{lb29}) for
plasmas and self-gravitating
systems in $d=3$ which presents a logarithmic divergence.

\subsection{Fokker-Planck equation}

The Lenard-Balescu equation (\ref{j19}) can be written under the
form of a
Fokker-Planck equation
\begin{equation}
\label{tp1}
\frac{\partial \omega}{\partial t}=\frac{1}{r}\frac{\partial}{\partial r}
\left\lbrack r\left (D\frac{\partial \omega}{\partial r}- \omega V_{\rm
pol}\right )\right\rbrack
\end{equation} 
with a diffusion coefficient and a drift by polarization\footnote{The difference
between the drift by polarization and the total drift is discussed in
\cite{bbgky,klim,kinfdvortex}.} given by
\begin{eqnarray}
D=2\pi^2\gamma \frac{1}{r^2}\sum_n
\int_0^{+\infty} r' dr'\, |n| |G(n,r,r',n\Omega)|^2  
\delta(\Omega-\Omega')\omega(r'),
\label{tp2}
\end{eqnarray}
\begin{eqnarray}
V_{\rm pol}=2\pi^2\gamma \frac{1}{r}\sum_n
\int_0^{+\infty} r' dr'\, |n| |G(n,r,r',n\Omega)|^2  
\delta(\Omega-\Omega')\frac{1}{r'}\frac{\partial
\omega'}{\partial r'}.
\label{tp3}
\end{eqnarray}
The  Lenard-Balescu equation (\ref{j19}) can be directly derived
from the Fokker-Planck approach by calculating the diffusion and drift
coefficients (first and second moments of the position increment) as shown in
Sec. 4 of \cite{klim}. The Lenard-Balescu  equation (\ref{j19}) is valid at the
order $1/N$ so it describes the ``collisional'' evolution of the system  due to
finite $N$ effects on a
timescale $\sim N t_D$, where $t_D$ is the dynamical time. It
conserves
the circulation $\Gamma=\int \omega\, d{\bf r}$ and the energy
$E=\frac{1}{2}\int \omega \psi\, d{\bf r}$. The
Boltzmann entropy  $S=-\int
\frac{\omega}{\gamma}\ln \frac{\omega}{\gamma}\, d{\bf r}$ increases
monotonically:
$\dot S\ge 0$ ($H$-theorem) \cite{cl}. The collisional evolution of the system
is due to a
condition of resonance. This condition of resonance, encapsulated in the
$\delta$-function, corresponds to $\Omega(r')=\Omega(r)$. Only the field
vortices which rotate with the same angular velocity as the test vortex have a
contribution to the collision term. The Boltzmann
distribution   $\omega_{\rm eq}(r)=A e^{-\beta \gamma \psi(r)}$ is always
a steady state of the Lenard-Balescu equation. However, it is not the only
one \cite{cl}. 
Any distribution with a monotonic profile of angular velocity is a steady state
of the Lenard-Balescu equation. In that case there is no resonance. This leads
to a situation of kinetic blocking at the order $1/N$ which is detailed in Sec.
\ref{sec_monov}.

In the thermal bath approximation, where the field vortices have the
Boltzmann distribution  $\omega(r')\sim e^{-\beta \gamma\psi(r')}$,
the drift by polarization is linked to the diffusion coefficient by the Einstein
relation $V_{\rm
pol}=-D(r)\beta \gamma
d\psi/dr$ \cite{preR,pre,klim,onsagerkin,cl,bbgky,kindetail} and the 
Fokker-Planck equation takes the form
\begin{equation}
\label{tp4}
\frac{\partial P}{\partial t}=\frac{1}{r}\frac{\partial}{\partial r}
\left\lbrack rD(r)\left (\frac{\partial P}{\partial r}+\beta
\gamma P\frac{d\psi}{dr} \right )\right\rbrack.
\end{equation}
This equation is similar to the Smoluchowski equation describing the evolution
of the probability density of a
Brownian particle in a potential $\psi(r)$. The diffusion coefficient depends on
the position. The probability density of the test vortex relaxes towards the
Boltzmann
distribution $P_{\rm eq}(r)=A\, e^{-\beta\gamma\psi(r)}$ on a
timescale $t_{R}^{\rm bath}\sim (N/\ln N)\,  t_D$.\footnote{The
sum over $n$ in the Lenard-Balescu equation (\ref{j19}) diverges logarithmically
when $r'=r$ (see Appendix \ref{sec_nce}). Since the term in
parenthesis in this equation
precisely vanishes when
$r'=r$ the Lenard-Balescu equation (\ref{j19}) as a whole is perfectly
well-defined without any divergence. This is why it is valid on a timescale
$\sim Nt_D$, not $\sim (N/\ln N)\,  t_D$. However, when we make the bath
approximation, the term in parenthesis does not vanish anymore when
$r'=r$ and we get a
logarithmic divergence. This is why the relaxation timescale of the system in a
bath described by the Fokker-Planck equation (\ref{tp4}) is
$t_{R}^{\rm bath}\sim (N/\ln N)\,  t_D$.}

\subsection{Kinetic equation with a monotonic velocity profile: Kinetic
blocking}
\label{sec_monov}

The Lenard-Balescu equation (\ref{j19}) can be rewritten as
\begin{eqnarray}
\frac{\partial\omega}{\partial
t}=2\pi^2\gamma\frac{1}{r}\frac{\partial}{\partial r} 
\int_0^{+\infty} r' dr'\,  \chi(r,r',\Omega(r))
\delta(\Omega-\Omega')\left
(\frac{1}{r}\omega' \frac{\partial \omega}{\partial
r}-\frac{1}{r'}\omega\frac{\partial \omega'}{\partial r'}\right ),
\label{tp5}
\end{eqnarray}
where we have introduced the convenient notation
$\chi(r,r',\Omega(r))=\sum_n |n| |G(n,r,r',n\Omega(r))|^2$ (see Appendix
\ref{sec_nce}). Using
the identity (\ref{gei7d}) we can rewrite Eq. (\ref{tp5}) as
\begin{equation}
\frac{\partial \omega}{\partial t}=2\pi^2\gamma
\frac{1}{r}\frac{\partial}{\partial r} 
\sum_a r_a\,  \frac{\chi(r,r_a,\Omega(r))}{|\Omega'(r_a)|}\left
(\frac{1}{r}\omega_a \frac{\partial \omega}{\partial
r}-\frac{1}{r_a}\omega\frac{\partial \omega_a}{\partial r_a}\right ),
\label{tp6}
\end{equation}
where the $r_a$ are the points that resonate with $r$, i.e., the points that
satisfy $\Omega(r_a)=\Omega(r)$.

If the profile of angular velocity $\Omega(r)$ is monotonic, there is no
distant resonance ($r_a\neq r$) and the Lenard-Balescu collision term vanishes:
\begin{eqnarray}
\frac{\partial\omega}{\partial
t}=2\pi^2\gamma\frac{1}{r}\frac{\partial}{\partial r} 
\int_0^{+\infty} r' dr'\,  \chi(r,r',\Omega(r))\frac{1}{|\Omega'(r)|}
\delta(r-r')\left
(\frac{1}{r}\omega' \frac{\partial \omega}{\partial
r}-\frac{1}{r'}\omega\frac{\partial \omega'}{\partial r'}\right )=0.
\label{tp7}
\end{eqnarray}
We recall that the Lenard-Balescu equation (\ref{j19})  is 
valid at the order $1/N$ so it describes the evolution of the average vorticity
on a timescale $N t_D$ under the effect of two-body correlations. Equation
(\ref{tp7}) shows that the vorticity does not change on this timescale (the
current vanishes at the order $1/N$). This is a situation of kinetic
blocking due to the absence of resonance at the order
$1/N$.\footnote{It is possible that, initially, the profile of angular velocity
is nonmonotonic but that it becomes monotonic during the evolution. As long as
the profile of angular velocity is nonmonotonic, there are resonances leading to
a
nonzero current ($J\neq
0$) and to the  evolution of the vorticity.  However, the relaxation stops
($J=0$) on a timescale $N t_D$ when the profile of angular velocity
becomes monotonic even if the system has not reached the Boltzmann distribution
of statistical equilibrium. This ``kinetic blocking'' at the order $1/N$ for
axisymmetric flows is
illustrated in \cite{cl}.} The vorticity may evolve on a longer timescale due
to higher order correlations  between the point vortices. For example,
three-body correlations which are of order $1/N^2$ are expected to induce an
evolution of
the
vorticity on a timescale $N^2\, t_D$.

In the bath approximation, where the field vortices can
have any
distribution $\omega_0(r')$ with a monotonic profile of angular
velocity,\footnote{As we have seen, such distributions do not evolve on
a timescale $Nt_D$.} the
drift by polarization satisfies the
generalized Einstein relation $V_{\rm pol}=D\frac{d\ln
\omega_0}{dr}$ \cite{pre,klim,onsagerkin,cl,bbgky,kindetail}
and the Fokker-Planck
equation takes the
form
\begin{eqnarray}
\frac{\partial P}{\partial t}=\frac{1}{r}\frac{\partial}{\partial r}\left\lbrack
rD(r)\left
(\frac{\partial
P}{\partial r}-P\frac{d\ln|\omega_0|}{dr}\right
)\right\rbrack
\label{tp8}
\end{eqnarray}
with a diffusion coefficient 
\begin{eqnarray}
D(r)=2\pi^2\gamma
\frac{\chi(r,r,\Omega(r))}{r|\Omega'(r)|}\omega_0(r)=2\pi^2\gamma
\frac{\chi(r,r,\Omega(r))}{|\Sigma(r)|}\omega_0(r),
\label{tp9}
\end{eqnarray}
where $\Sigma(r)=r\Omega'(r)$ is the local shear created by the field
vortices. Equation (\ref{tp8}) is similar to the Smoluchowski equation
describing the
evolution of the probability density of a Brownian particle in a potential 
$U(r)=-\ln |\omega_0(r)|$.
The probability density  of the test vortex $P(r,t)$
relaxes towards the distribution of the bath $\omega_0(r)$ on a timescale
$t_R^{\rm bath}\sim (N/\ln N) t_D$. In the case of a thermal bath with a
monotonic profile of angular velocity, we recover Eq. (\ref{tp4}).

If we neglect collective effects, we just have to replace
$\chi(r,r',\Omega(r))$ by $\chi_{\rm bare}(r,r')$ [see Eq. (\ref{pot15})]
in
the foregoing equations. This leads to the Landau equation (\ref{j19landexp}).
On the other
hand, when the profile of angular velocity is monotonic, the diffusion
coefficient (\ref{tp9}) is
given by \cite{preR,pre,bbgky,klim}
\begin{eqnarray}
D(r)=\frac{1}{4}\gamma
\frac{\ln\Lambda}{r|\Omega'(r)|}\omega_0(r)=\frac{1}{4}\gamma
\frac{\ln\Lambda}{|\Sigma(r)|}\omega_0(r),
\label{tp10}
\end{eqnarray}
where $\ln\Lambda$ is the Coulomb logarithm.

\section{Conclusion}

In this paper, we have developed the analogy between the kinetic theory of
stellar systems and the kinetic theory of 2D point
vortices \cite{houchesPH}. The evolution of the system as a
whole is governed  at the order $1/N$ by the Lenard-Balescu equation which
reduces to the Landau equation when collective effects are neglected. The
relaxation is due to a condition of resonance encapsulated in a
$\delta$-function. In the thermal bath approximation, we obtain a Fokker-Planck
equation which has the form of a Kramers equation for stellar systems and the
form of a Smoluchowski equation for 2D point vortices. Indeed,
point vortices have no inertia\footnote{This corresponds to the
conception of motion according to Aristote and Descartes (velocity proportional
to ``force'') as compared to Newton (acceleration proportional to
force).} so their ``collisional'' dynamics is similar to that of Brownian
particles in the overdamped limit. The relaxation to the Boltzmann distribution
is due to a competition between diffusion and friction in the case of stellar
systems and between diffusion and drift in the case of 2D point vortices. The
systematic drift experienced by a point vortex \cite{preR,pre,bbgky,klim} is the
counterpart of Chandrasekhar's dynamical friction experienced by a star
\cite{chandra1}. We have also mentioned the situation of ``kinetic blocking'' (a
term introduced in \cite{cl}) in certain cases. This peculiarity was originally
discovered in
the case of 1D homogeneous plasmas \cite{ef}, then in the case of axisymmetric
distributions of 2D point vortices with a monotonic profile and in the case of
unidirectional flows of point vortices
\cite{pre,cl}.\footnote{For more general flows that are neither
unidirectional
nor axisymmetric, the kinetic equation of 2D point vortices  is given by Eq.
(128) or Eq. (137) of \cite{pre} (see also Eq. (54) in \cite{bbgky} and Eq.
(16) in \cite{cl}) and the collision
term does not necessarily vanish. This is
because there are potentially more resonances at the order $1/N$ for
complicated flows than for unidirectional and axisymmetric flows. Similarly,
resonances appear at the order $1/N$ for inhomogeneous 1D systems with
long-range interactions (such as 1D self-gravitating systems)  that are not
present for homogeneous 1D systems \cite{ktaav,kindetail}.} Kinetic blocking
also
occurs in the homogeneous phase of the HMF model \cite{bd,cvb} and for spins
with long-range interactions moving on a sphere in
relation to the process of vector resonant relaxation (VRR) in galactic nuclei
\cite{fbc}. In that case, the Lenard-Balescu collision term vanishes at the
order $1/N$ and it is necessary to develop the kinetic theory at the order
$1/N^2$ \cite{fbcn2,fcpn2,fouvry}. Finally, we have investigated the kinetic
theory close to a critical point where the system becomes unstable. We have
considered the simple case of homogeneous systems with long-range
interactions described by the Cauchy distribution which can be treated
analytically \cite{linres}. We have shown that when the contribution of the
normal (Landau) modes are properly taken into account \cite{linres,hh} the
friction by polarization does not diverge at the critical point. This result
seems to confirm the claim of Hamilton and Heinemann \cite{hh}
that the kinetic equation does not diverge at the critical point. However, to
be conclusive, it
remains to determine the behavior of the diffusion coefficient close to the
critical point and to consider the case of more general distribution functions.
On the other hand, it is expected that fluctuations become important close to
the critical point like in other areas of physics (e.g. the phenomenon of
critical opalescence mentioned in \cite{sdduniverse}) so that kinetic equations
for the mean distribution function (like the Landau and Lenard-Balescu
equations) may not be relevant in that case and should
be replaced by stochastic kinetic equations (see footnote 19). This type of
stochastic kinetic
equations have been introduced a long time ago in the case of Brownian particles
with long-range interctions described by the canonical ensemble (see,
e.g., \cite{hb5,cd2,bmf,entropy,gsse}). The enhancement of
fluctuations close to the critical point has been demonstrated explicitly
\cite{hb5} as well as the random transitions between metastable states and their
relation to the Kramers problem and the instanton theory
\cite{cd2,bmf,gsse,nardini2,bouchetsimmonet,rbs,brs}. The generalization of
these equations to the case of Hamiltonian particles with
long-range interactions described by the microcanonical
ensemble \cite{bouchetld,jb,kinfd} is clearly an important topic for future
works.

\appendix

\section{Kinetic theory close to the critical point}
\label{sec_modes}

In the kinetic theory presented in Sec. \ref{sec_jbl}, we have ignored the
contribution of the normal (Landau) modes. This approximation is valid if the
system is strongly
stable so that these modes decay sufficiently rapidly. However, if we are
close to a critical point, the decay rate $\gamma\lesssim 0$ is small (it
vanishes at the critical point and becomes positive above the critical point
leading to an instability). In that case,
it is not
possible to ignore the contribution of the normal modes anymore. Since
$\epsilon({\bf k},0)\rightarrow 0$ when $\gamma\rightarrow 0$ (assuming
$\Omega=0$ -- see below) the kinetic theory of Sec. \ref{sec_jbl}
leads to a divergence of the diffusion and friction coefficients for ${\bf
v}={\bf 0}$ (see Sec. \ref{sec_nontb}). Recently, Hamilton and Heinemann
\cite{hh} showed that no divergence occurs in the response function when
the normal modes are properly taken into
account. Actually, we had obtained a similar result in a former paper
\cite{linres} without realizing at that time all of its implications. We
computed the response  of a homogeneous system with long-range
interactions described by the Vlasov equation to a step function. For the
Cauchy distribution function, the response function can be calculated exactly
analytically. We obtained an explicit formula showing how the
contribution of the normal modes regularizes the divergence  of
the response function at
the critical point. In this Appendix, we recall and extend our previous
results \cite{linres} and use a similar approach to compute the friction by
polarization close to the critical point. We show that it
does not diverge at the critical point (for a fixed time $t$). The calculation
of the diffusion coefficient is more complicated and left for a future work.

\subsection{Cauchy distribution}

We assume that the reduced distribution function (see Appendix \ref{sec_drs})
is the Cauchy
distribution
\begin{eqnarray}
\label{cd1}
f(v)=\frac{\rho}{\pi u_0}\frac{1}{1+\frac{v^2}{u_0^2}}.
\end{eqnarray}
In that case, the dielectric function (\ref{disrel1}) has the explicit
expression (see Eq. (94) of \cite{linres}):
\begin{eqnarray}
\epsilon(k,\omega)=\frac{(iku_0+\omega)^2-(2\pi)^d{\hat u}(k)\rho
k^2}{(iku_0+\omega)^2}.
\label{cd2}
\end{eqnarray}
In the following, we consider an attractive potential ${\hat u}(k)<0$ and define
$\Delta=(2\pi)^d|{\hat u}(k)|\rho k^2$. The dispersion relation
$\epsilon(k,\omega)=0$ gives $\omega=i(-ku_0\pm\sqrt{\Delta})$.
Therefore, the normal modes of the Cauchy distribution are purely imaginary
($\Omega=0$). The
mode
$\omega=i(-ku_0-\sqrt{\Delta})$ is always stable. The mode
$\omega=i(-ku_0+\sqrt{\Delta})$ may become unstable. We write it as
$\omega=i\gamma$ with $\gamma=\sqrt{\Delta}-ku_0$. The system is stable when
$\gamma<0$, i.e., $\Delta<(ku_0)^2$ and unstable when
$\gamma>0$, i.e., $\Delta>(ku_0)^2$. The
critical point $\gamma=0$ (marginal stability) corresponds to
$\Delta=(ku_0)^2$. This condition determines the critical wavenumber $k_c$
for a given $u_0$ or the critical velocity $u_0$ when the potential of
interaction is
restricted to a few Fourier modes (like in the HMF model). We will be
particularly interested in the limit $\gamma\rightarrow 0^-$. Let us consider
specific examples:

(i) For self-gravitating systems, using the results of Appendix
\ref{sec_drs}, we have 
\begin{eqnarray}
\label{cd3}
\gamma=u_0(k_J-k),
\end{eqnarray}
where $k_J=(S_d G\rho/u_0^2)^{1/2}$ is the Jeans wavenumber.
The system is stable when $k>k_J$, marginally stable when $k=k_J$, and
unstable when $k<k_J$.

(ii) For the HMF model, using the results of Appendix
\ref{sec_drs}, we have
\begin{eqnarray}
\label{cd4}
\gamma=u_c-u_0<0,
\end{eqnarray}
where $u_c=\sqrt{k\rho/2}$ is the critical velocity. The homogeneous phase is
stable when $u_0>u_c$, marginally stable when $u_0=u_c$, and unstable when
$u_0<u_c$.

\subsection{Response to a step function}

The response $\delta{\hat\Phi}({\bf k},t)=J({\bf k},t){\hat\Phi}_e({\bf
k})$ of a stable homogeneous system with long-range interactions to a step
function has
been determined in \cite{linres}. It is measured by the quantity (see Eq. (23)
of
\cite{linres})
\begin{eqnarray}
J({\bf k},t)=\int\frac{d\omega}{2\pi}e^{-i\omega
t}\frac{1-\epsilon({\bf k},\omega)}{\epsilon({\bf k},\omega)}\frac{i}{\omega}.
\label{sf0}
\end{eqnarray}
Using the Cauchy residue theorem, it can be written as (see Eq. (24) of
\cite{linres})
\begin{eqnarray}
J({\bf k},t)=\frac{1-\epsilon({\bf k},0)}{\epsilon({\bf k},0)}+\sum_{\alpha}
e^{-i\omega_{\alpha}({\bf
k})t}\left\lbrack
{\rm Res} \frac{1-\epsilon({\bf
k},\omega)}{\epsilon({\bf
k},\omega)\omega}\right\rbrack_{\omega=\omega_{\alpha}({\bf
k})}.
\label{sf0b}
\end{eqnarray}
The first term is the ``usual'' (or ``naive'') term and the second term,
which takes the normal (Landau) modes into account, is the ``additional''
term. When $t\rightarrow
+\infty$ the ``additional'' term tends to zero and only the
``usual'' term remains. For the Cauchy distribution function (\ref{cd1}),
using the expression
(\ref{cd2})
of the dielectric function, we get (see Eq. (105) of
\cite{linres})
\begin{eqnarray}
J({\bf k},t)=\frac{\Delta}{(u_0 k)^2-\Delta}\left\lbrace
1-e^{-u_0 k t}\left\lbrack\frac{u_0k}{\sqrt{\Delta}}\sinh(\sqrt{\Delta}t)+
\cosh(\sqrt{\Delta}t)\right\rbrack\right\rbrace.
\label{sf1}
\end{eqnarray}
Introducing the notation $\gamma$ defined above, the foregoing equation can be
rewritten as 
\begin{eqnarray}
J(t,\gamma)=-\frac{(\gamma+ku_0)^2}{\gamma(\gamma+2u_0k)}\left\lbrace
1-e^{-u_0 k
t}\left\lbrack\frac{u_0k}{\gamma+u_0k}\sinh\lbrack (\gamma+u_0k)t\rbrack+
\cosh\lbrack (\gamma+u_0k)t\rbrack \right\rbrack\right\rbrace.
\label{sf2}
\end{eqnarray}
In the following it will be convenient to rescale $\gamma$ by $ku_0$ and $t$ by
$(ku_0)^{-1}$ in which
case we obtain
\begin{eqnarray}
J(t,\gamma)=-\frac{(\gamma+1)^2}{\gamma(\gamma+2)}\left\lbrace
1-e^{-t}\left\lbrack\frac{1}{\gamma+1}\sinh\lbrack
(\gamma+1)t\rbrack+
\cosh\lbrack (\gamma+1)t\rbrack \right\rbrack\right\rbrace.
\label{sf3}
\end{eqnarray}
Let us study this function close to the critical point ($\gamma=0$):

(i) If we fix the time $t$ and let $\gamma\rightarrow 0$ we can make the
expansion
\begin{eqnarray}
J(t,\gamma)=\frac{1}{2}e^{-t}\left\lbrack
t\cosh(t)+(t-1)\sinh(t)\right\rbrack+\frac{\gamma}{4}e^{-t}\left\lbrack
t(t+1)\cosh(t)+(t^2+3t-1)\sinh(t)\right\rbrack+...
\label{sf4}
\end{eqnarray}
We see that $J(t,\gamma)$ converges for  $\gamma\rightarrow 0$ towards the
function
\begin{eqnarray}
J_0(t)=\frac{1}{2}e^{-t}\left\lbrack
t\cosh(t)+(t-1)\sinh(t)\right\rbrack.
\label{sf5}
\end{eqnarray}
For $t\rightarrow 0$, this function behaves as
$J_0(t)=\frac{t^2}{2}-\frac{t^3}{3}+...$.
For $t\rightarrow +\infty$, we
get $J_0(t)\sim t/2$ which increases linearly with time.

(ii) If we fix $\gamma<0$ and let $t\rightarrow +\infty$ the function
$J(t,\gamma)$
tends to the limit
\begin{eqnarray}
J_{\infty}(\gamma)=-\frac{(\gamma+1)^2}{\gamma(\gamma+2)},
\label{sf6}
\end{eqnarray}
which corresponds to the ``usual'' term in Eq. (\ref{sf0b}). If we now let
$\gamma\rightarrow 0$ we can make the
expansion $J_{\infty}(\gamma)=-\frac{1}{2\gamma}-\frac{3}{4}-\frac{\gamma}{ 8}
+...$. We see that $J_{\infty}(\gamma)$ diverges as $-1/(2\gamma)$ at
the critical point. In the unstable case ($\gamma<0$), the response function
increases exponentially rapidly with time as $J(t,\gamma)\sim
\frac{\gamma+1}{2\gamma} e^{\gamma t}$ for $t\rightarrow +\infty$.

(iii) If we fix $\gamma$ and let $t\rightarrow 0$ we can make the expansion
\begin{eqnarray}
J(t,\gamma)= \frac{1}{2}(1+\gamma)^2t^2-\frac{1}{3}(1+\gamma)^2
t^3+...
\label{sf8}
\end{eqnarray}
If we now let $\gamma\rightarrow 0$ we find
that $J(t,\gamma)=\frac{1}{2}t^2-\frac{1}{3}
t^3+...$ in agreement with the behavior of $J_0(t)$ for $t\rightarrow 0$.

\begin{figure}[!h]
\begin{center}
\includegraphics[clip,scale=0.3]{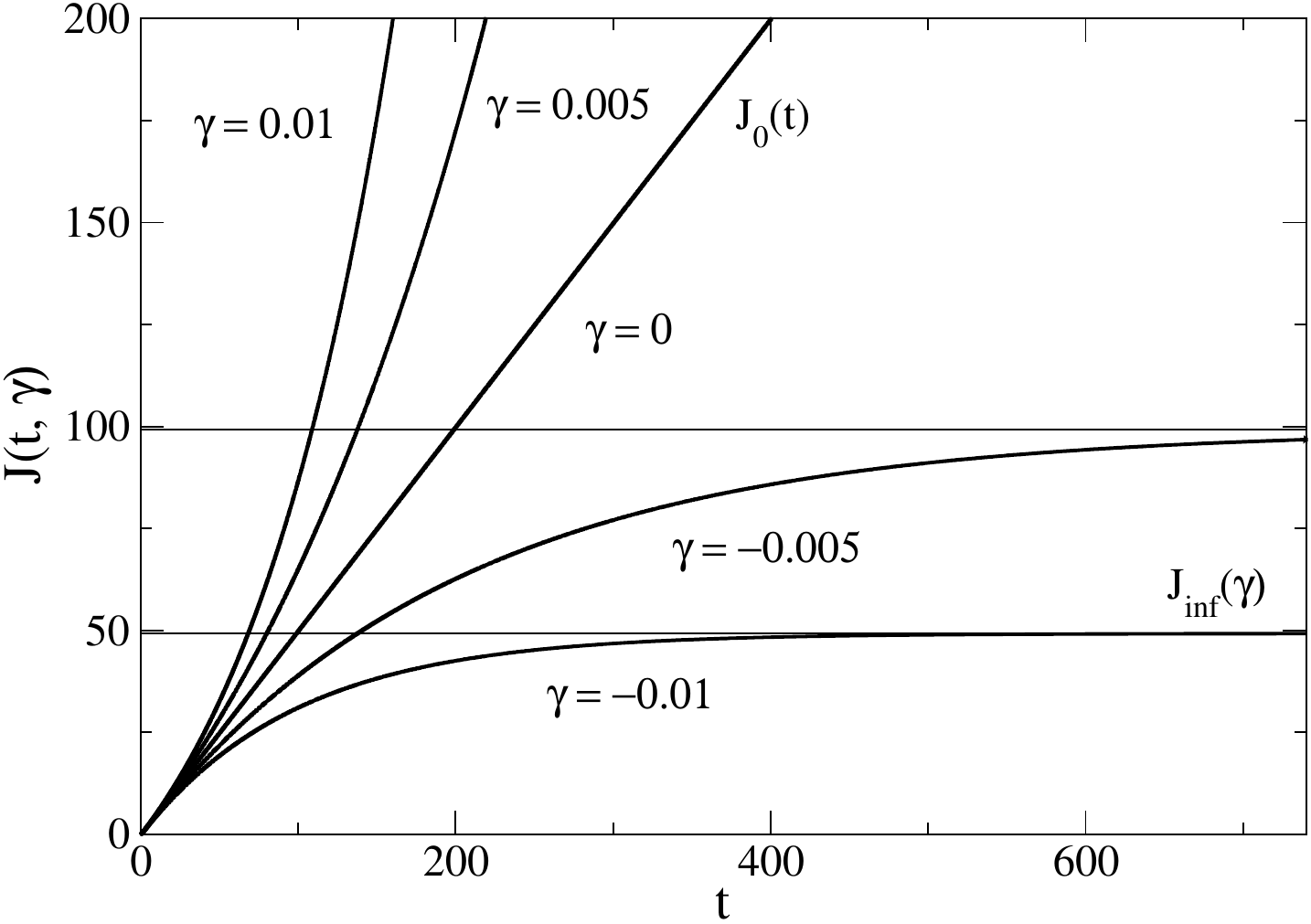}
\caption{Temporal evolution of the response function $J(t,\gamma)$ for
different values of $\gamma$ close to the critical point ($\gamma=0$).}
\label{Jt}
\end{center}
\end{figure}

\begin{figure}[!h]
\begin{center}
\includegraphics[clip,scale=0.3]{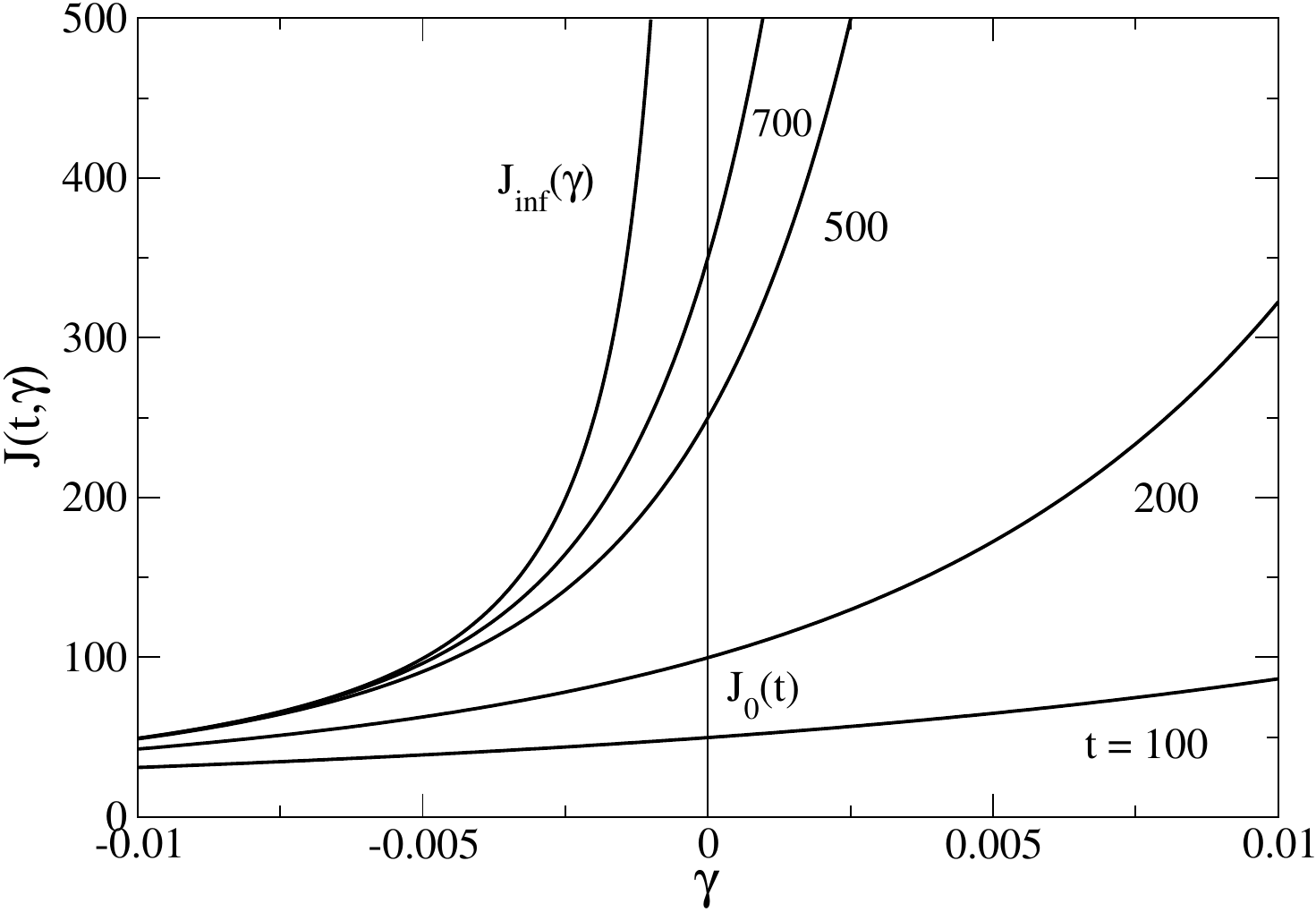}
\caption{Response function $J(t,\gamma)$ as a function of $\gamma$ at
different times.}
\label{Jg}
\end{center}
\end{figure}

In conclusion, if we first take the limit  $t\rightarrow +\infty$ (which
corresponds to the
situation where all the modes have decayed) we obtain the response function from
Eq. (\ref{sf6}), which diverges at the critical point $\gamma\rightarrow 0^{-}$.
However, if we take into account the fact that observations are made on a
finite timescale, and if we  take the limit
$\gamma\rightarrow 0^{-}$ at fixed time $t$, we obtain the response
function from Eq. (\ref{sf5}) which is finite at the critical point
($\gamma=0$). More
generally, if we fix the time $t$ and plot the function $J(t,\gamma)$ as a
function of $\gamma$ there is no divergence at any value of $\gamma$ (negative,
zero, or positive). This
is in agreement with the results of Hamilton and Heinemann \cite{hh}. The
function $J(t,\gamma)$ is plotted in Figs. \ref{Jt} and  \ref{Jg} for
illustration.

{\it Remark:} The case of a waterbag distribution function has also been
considered in
\cite{linres}.
In that case, the normal modes are purely real in the stable
case (there is no Landau damping) and purely imaginary in the unstable case. The
response to a step function is given by Eq. (53) of \cite{linres} in the stable
case.\footnote{It is  given by Eq. (58) of \cite{linres} in the unstable
case.} It is purely oscillating. It does not diverge at the
critical point. Since the normal modes
are not damped in the stable case this implies that the assumptions made to
derive the
Lenard-Balescu equation in Sec. \ref{sec_jbl} are not fulfilled. In
particular, we need to take into account all the modes. The resulting kinetic
equation may be very pathological.

\subsection{Response to an initial fluctuation}

In Sec. \ref{sec_jbl} we have seen that
the fluctuations of the potential are related to the initial fluctuations by
Eq. (\ref{poto6}). Taking the inverse
Laplace transform of this equation, we
obtain
\begin{eqnarray}
\delta{\tilde \Phi}({\bf k},t)=(2\pi)^d {\hat u}(k)\int d{\bf
v}\, \delta{\hat f} ({\bf k},{\bf v},0)\int_{\cal C}
\frac{d\omega}{2\pi} e^{-i\omega t} \frac{1}{\epsilon({\bf
k},\omega)} \frac{1}{i(
{\bf k}\cdot {\bf v}-\omega)}.
\label{rf1}
\end{eqnarray}
If we use the Cauchy residue theorem and neglect the contribution of the normal
modes, we recover Eq. (\ref{j6s}).
However, in the case of the Cauchy distribution, we can compute the contribution
of the normal modes  exactly analytically by using a method similar to
\cite{linres}. Since
the
dielectric function is given by  Eq.
(\ref{cd2}), we have to compute the integral
\begin{eqnarray}
I({\bf k},{\bf v},t)=\int_{\cal C}
\frac{d\omega}{2\pi} e^{-i\omega t}
\frac{(iku_0+\omega)^2}{\left\lbrack (iku_0+\omega)-\sqrt{(2\pi)^d{\hat
u}(k)\rho
k^2}\right\rbrack \left\lbrack (iku_0+\omega)+\sqrt{(2\pi)^d{\hat
u}(k)\rho
k^2}\right\rbrack}
\frac{1}{i(
{\bf k}\cdot {\bf v}-\omega)}.
\label{rf2}
\end{eqnarray}
It can be written as
\begin{eqnarray}
I({\bf k},{\bf v},t)=e^{-i{\bf k}\cdot {\bf v}t}\frac{1}{\epsilon({\bf k},{\bf
k}\cdot {\bf
v})}+N({\bf k},{\bf v},t),
\label{rf3}
\end{eqnarray}
where the first term is the ``usual'' term neglecting the normal
modes (see Sec. \ref{sec_jbl}) and the second term is the ``additional'' term
taking
into account the contribution of
the normal modes. In the attractive case, we find that
\begin{eqnarray}
N({\bf k},{\bf v},t)=\frac{i}{2}e^{-k u_0 t}\sqrt{\Delta} \left\lbrace
\frac{e^{-\sqrt{\Delta}t}\left\lbrack {\bf k}\cdot {\bf
v}-i(\sqrt{\Delta}+k u_0)\right\rbrack}{({\bf k}\cdot {\bf
v})^2+(\sqrt{\Delta}+ku_0)^2}-\frac{e^{\sqrt{\Delta}t}\left\lbrack {\bf k}\cdot
{\bf
v}+i(\sqrt{\Delta}-k u_0)\right\rbrack}{({\bf k}\cdot {\bf
v})^2+(\sqrt{\Delta}-ku_0)^2}\right\rbrace.
\label{rf4}
\end{eqnarray}
Introducing the notation $\gamma$ defined above, the foregoing equation can be
rewritten as
\begin{eqnarray}
N({\bf k},{\bf v},t)=\frac{i}{2}e^{-k u_0 t}(\gamma+u_0 k) \left\lbrace
\frac{e^{-(\gamma+u_0 k) t}\left\lbrack {\bf k}\cdot {\bf
v}-i(\gamma+2u_0 k)\right\rbrack}{({\bf k}\cdot {\bf
v})^2+(\gamma+2u_0 k)^2}
-\frac{e^{(\gamma+u_0 k) t}({\bf k}\cdot
{\bf
v}+i\gamma)}{({\bf k}\cdot {\bf
v})^2+\gamma^2}\right\rbrace.
\label{rf5}
\end{eqnarray}
Then, the temporal fluctuations of the potential are given by
\begin{eqnarray}
\label{rf6}
\delta{\hat\Phi}({\bf k},t)=(2\pi)^d {\hat u}(k)\int d{\bf
v}\, \delta{\hat f} ({\bf k},{\bf v},0)\left\lbrack e^{-i{\bf k}\cdot {\bf
v}t}\frac{1}{\epsilon({\bf k},{\bf
k}\cdot {\bf
v})}+N({\bf k},{\bf v},t)\right \rbrack.
\end{eqnarray}

\subsection{Usual friction by polarization}

Let us first compute the usual
friction by
polarization for a Cauchy distribution. From the results of Sec. \ref{sec_jbl}
[see in
particular the first term in Eq. (\ref{j17})] the usual friction by
polarization  is given by
\begin{eqnarray}
\label{uf1}
{\bf F}_{\rm pol}=m\int d{\bf
k} \, i{\bf k} \frac{{\hat
u}(k)}{\epsilon(-{\bf k},-{\bf k}\cdot {\bf v})}.
\end{eqnarray}
This equation can also be obtained by the method detailed in Sec. 10 of
\cite{kinfd}. Using the expression
of the dielectric
function from Eq. (\ref{cd2}) we get 
\begin{eqnarray}
\label{uf2}
{\bf F}_{\rm pol}=m\int d{\bf
k} \, i{\bf k} {\hat
u}(k)\frac{(iku_0-{\bf k}\cdot {\bf v})^2}{(iku_0-{\bf k}\cdot {\bf
v})^2-(2\pi)^d{\hat u}(k)\rho
k^2}.
\end{eqnarray}
Multiplying the numerator of the integrand by the complex conjugate of the
denominator, we obtain 
\begin{eqnarray}
\label{uf3}
{\bf F}_{\rm pol}=m\int d{\bf
k} \, i{\bf k} {\hat
u}(k)(iku_0-{\bf k}\cdot {\bf v})^2\frac{({\bf k}\cdot {\bf
v})^2-(ku_0)^2+\Delta+2iku_0({\bf k}\cdot {\bf v})}{\left\lbrack ({\bf k}\cdot
{\bf v})^2-(ku_0)^2+\Delta\right\rbrack^2+4 (ku_0)^2({\bf k}\cdot {\bf v})^2}.
\end{eqnarray}
After simplification, we find that
\begin{eqnarray}
\label{uf4}
{\bf F}_{\rm pol}=2m\int d{\bf
k} \, {\bf k} {\hat
u}(k)ku_0 ({\bf k}\cdot {\bf
v}) \frac{\Delta}{\left\lbrack
({\bf k}\cdot
{\bf v})^2-(ku_0)^2+\Delta\right\rbrack^2+4 (ku_0)^2({\bf k}\cdot {\bf v})^2}.
\end{eqnarray}
We note that this term does not diverge for ${\bf v}\neq {\bf 0}$ when
$\Delta\rightarrow (ku_0)^2$. This is
because $\epsilon({\bf k},{\bf k}\cdot {\bf v})$ tends to zero when
$\Delta\rightarrow (ku_0)^2$ only for ${\bf v}={\bf 0}$. Let us therefore
consider the case of small velocities. For small
$|{\bf v}|$ at fixed $\Delta$ the foregoing expression reduces to
\begin{eqnarray}
\label{uf5}
{\bf F}_{\rm pol}=2m\int d{\bf
k} \, {\bf k} {\hat
u}(k)k u_0 ({\bf k}\cdot {\bf v}) \frac{\Delta}
{\left\lbrack \Delta-(ku_0)^2\right\rbrack^2}.
\end{eqnarray}
Introducing the notation $\gamma$ defined above, this quantity can be
rewritten as
\begin{eqnarray}
\label{uf6}
{\bf F}_{\rm pol}=2m\int d{\bf
k} \, {\bf k} {\hat
u}(k)  k u_0 ({\bf k}\cdot {\bf v})       
\frac{(\gamma+u_0k)^2}
{\gamma^2(\gamma+2u_0 k)^2}.
\end{eqnarray}
For
$\gamma\rightarrow 0$ we get 
\begin{eqnarray}
\label{uf7}
{\bf F}_{\rm pol}=m\int_{|{\bf k}|\sim k_c} d{\bf
k} \, {\bf k} {\hat
u}(k)  \frac{1}{k u_0} ({\bf k}\cdot {\bf v}) \left
\lbrack\frac{(u_0k)^2}{2\gamma^2}+\frac{u_0 k}{2\gamma}-\frac{1}{8}+
\frac{\gamma^2}{32(u_0k)^2}+...\right \rbrack .    
\end{eqnarray}
We see that
the ``usual'' friction by polarization for $|{\bf v}|\rightarrow 0$ diverges as
$\gamma^{-2}$ at the critical point (see Sec. \ref{sec_nontb}).

{\it Remark:} We can also calculate the usual friction by polarization from
Eq. (\ref{ess14again}). In $d=1$, this expression reduces to
\begin{eqnarray}
\label{uf8}
F_{\rm pol}=2\pi^2 m f'(v)\int dk\, |k|\frac{{\hat u}(k)^2}{|\epsilon(k,kv)|^2}.
\end{eqnarray}
For the Cauchy distribution (\ref{cd1}) we have
\begin{eqnarray}
\label{uf9}
f'(v)=-\frac{2\rho}{\pi u_0^3}\frac{v}{\left (1+\frac{v^2}{u_0^2}\right
)^2}.
\end{eqnarray}
Therefore, for small $v$, we obtain 
\begin{eqnarray}
\label{uf10}
F_{\rm pol}=-\frac{4\pi m\rho}{u_0^3} v \int dk\, |k|\frac{{\hat
u}(k)^2}{|\epsilon(k,0)|^2}.
\end{eqnarray}
With the expression of the dielectric function from Eq. (\ref{cd2}) we get
\begin{eqnarray}
\label{uf11}
F_{\rm pol}=-4\pi m\rho u_0 v \int dk\, |k|^5 \frac{{\hat
u}(k)^2}{\left\lbrack \Delta-(ku_0)^2\right\rbrack^2},
\end{eqnarray}
which coincides with Eq. (\ref{uf5}).

\subsection{Additional term in the friction by polarization}

Repeating the calculations of Sec. \ref{sec_jbl} while keeping the contribution
of
the normal modes in Eq. (\ref{rf6}) we find that the additional term in the 
friction
by polarization is
\begin{eqnarray}
\label{new5}
{\bf F}_{\rm pol}^{\rm add}=\int d{\bf
k} \, i{\bf k} {\hat
u}(k)  
e^{-i {\bf k}\cdot
{\bf v} t}mN(-{\bf k},{\bf v},t),
\end{eqnarray}
i.e.
\begin{eqnarray}
\label{new6}
{\bf F}_{\rm pol}^{\rm add}=-\frac{m}{2}\int d{\bf
k} \, {\bf k} {\hat
u}(k)  
e^{-i {\bf k}\cdot
{\bf v} t}e^{-k u_0 t}(\gamma+u_0 k)\nonumber\\
\times
\left\lbrace
\frac{e^{-(\gamma+u_0 k) t}\left\lbrack -{\bf k}\cdot {\bf
v}-i(\gamma+2u_0 k)\right\rbrack}{(-{\bf k}\cdot {\bf
v})^2+(\gamma+2u_0 k)^2}
-\frac{e^{(\gamma+u_0 k) t}(-{\bf k}\cdot
{\bf
v}+i\gamma)}{(-{\bf k}\cdot {\bf
v})^2+{\gamma}^2}\right\rbrace .
\end{eqnarray}
We obtain the same expression of  ${\bf F}_{\rm pol}^{\rm add}$ by extending 
the method
detailed in Sec. 10 of \cite{kinfd}. We note that this term does not diverge for
${\bf v}\neq 0$
when $\gamma\rightarrow 0$  and that it
tends to zero when $t\rightarrow +\infty$. Let us therefore focus on the case of
small
velocities. For $|{\bf v}|\rightarrow 0$ at fixed $\gamma$ and $t$ the foregoing
equation reduces to
\begin{eqnarray}
\label{new7}
{\bf F}_{\rm pol}^{\rm add}=-\frac{m}{2}\int d{\bf
k} \, {\bf k} {\hat
u}(k)  
(1-i {\bf k}\cdot
{\bf v} t) (\gamma+u_0 k)\nonumber\\
\times
\left\lbrace
\frac{e^{-(\gamma+2u_0 k) t}\left\lbrack -{\bf k}\cdot {\bf
v}-i(\gamma+2u_0 k)\right\rbrack}{(\gamma+2u_0 k)^2}
-\frac{e^{\gamma t}(-{\bf k}\cdot
{\bf
v}+i\gamma)}{{\gamma}^2}\right\rbrace. 
\end{eqnarray}
This is the sum of two terms. The first term is
\begin{eqnarray}
\label{new8}
({\bf F}_{\rm pol}^{\rm add})_{\rm I}=\frac{m}{2}\int d{\bf
k} \, {\bf k} {\hat
u}(k) ({\bf k}\cdot
{\bf
v})(\gamma+u_0 k)
\left\lbrack
\frac{e^{-(\gamma+2u_0 k) t}}{(\gamma+2u_0 k)^2}
-\frac{e^{\gamma t}}{{\gamma}^2}\right\rbrack
\end{eqnarray}
and the second term is
\begin{eqnarray}
\label{new10}
({\bf F}_{\rm pol}^{\rm add})_{\rm II}=\frac{m}{2}\int d{\bf
k} \, {\bf k} {\hat
u}(k)  ({\bf k}\cdot
{\bf v}) t (\gamma+u_0 k)
\left\lbrack
\frac{e^{-(\gamma+2u_0 k) t}}{\gamma+2u_0 k}
+\frac{e^{\gamma t}}{\gamma}\right\rbrack.
\end{eqnarray}
For $\gamma\rightarrow 0$ we find that
\begin{eqnarray}
\label{new11}
{\bf F}_{\rm pol}^{\rm add}=m\int_{|{\bf k}|\sim k_c} d{\bf
k} \, {\bf k} {\hat
u}(k)  ({\bf k}\cdot
{\bf v}) \frac{1}{u_0 k}
\left\lbrack
-\frac{(u_0k)^2}{2\gamma^2}-\frac{u_0k}{2\gamma}+\frac{1}{8}e^{-2u_0kt}
(1+2u_0kt)+\frac{1}{4}(u_0kt)^2+...\right\rbrack,
\end{eqnarray}
where the terms (...) go to zero when
$\gamma\rightarrow 0$.  We see that
the ``additional'' friction by polarization for $|{\bf v}|\rightarrow 0$
diverges as
$-\gamma^{-2}$ at the critical point.

Summing the contributions of Eqs. (\ref{uf7}) and (\ref{new11}) we see that the
diverging terms cancel each other. We are then left with
\begin{eqnarray}
\label{new12}
{\bf F}_{\rm pol}^{\rm total}=m\int_{|{\bf k}|\sim k_c} d{\bf
k} \, {\bf k} {\hat
u}(k)  \frac{1}{k u_0} ({\bf k}\cdot {\bf v}) \left
\lbrack  -\frac{1}{8}+\frac{1}{8}e^{-2u_0kt}
(1+2u_0kt)+\frac{1}{4}(u_0kt)^2\right \rbrack,    
\end{eqnarray}
to which we have to add the contribution of the other
modes ${\bf k}\neq {\bf k}_c$. Therefore, the total friction
by polarization does not diverge when
$t$ is fixed and $\gamma\rightarrow 0$. However, we note that this asymptotic
expression increases like $t^2$.

\subsection{Diffusion term}

Repeating the calculations of Sec. \ref{sec_jbl} while keeping the contribution
of
the normal modes in Eq. (\ref{rf6}) we find that the total diffusion term reads
\begin{eqnarray}
\label{j14c}
\left\langle \delta f\nabla\delta\Phi\right\rangle_{\rm diff}=-\int d{\bf
k} \, i{\bf k} {\hat u}(k)\int
d{\bf
v}'\, mf({\bf v}') \left\lbrack e^{i{\bf
k}\cdot {\bf
v}'t}\frac{1}{\epsilon(-{\bf k},-{\bf
k}\cdot {\bf
v}')}+N(-{\bf k},{\bf v}',t)\right \rbrack\nonumber\\
\times
i{\bf k}\cdot
\frac{\partial f}{\partial {\bf v}}\int_0^t dt'\, e^{i{\bf
k}\cdot{\bf v}(t'-t)}(2\pi)^d {\hat u}(k) \left\lbrack e^{-i{\bf k}\cdot {\bf
v}'t'}\frac{1}{\epsilon({\bf k},{\bf
k}\cdot {\bf
v}')}+N({\bf k},{\bf v}',t')\right \rbrack. 
\end{eqnarray}
The explicit calculation of this term is left for a future work.

\subsection{Application to the HMF model}

The previous expressions can be simplified for the HMF model where the
Fourier transform of the potential of interaction is restricted to the
modes $n=\pm 1$ [see Eq.
(\ref{poig5})]. In $d=1$, the usual friction by polarization (\ref{uf4}) can be
written for
a general potential of interaction as
\begin{eqnarray}
\label{rev1}
F_{\rm pol}=2m u_0 v\int dk \, k^2 {\hat
u}(k)|k| \frac{(\gamma+ku_0)^2}{\left\lbrack
(kv)^2+\gamma^2+2\gamma k u_0\right\rbrack^2+4 (ku_0)^2(kv)^2}.
\end{eqnarray}
For the HMF model, we get
\begin{eqnarray}
\label{rev2}
F_{\rm pol}=-\frac{k}{\pi}u_0 v \frac{(\gamma+u_0)^2}{\left (
v^2+\gamma^2+2\gamma u_0\right )^2+4 u_0^2 v^2}.
\end{eqnarray}
Let us consider particular limits:

(i) For $v\neq 0$ and $\gamma\rightarrow 0$, we find that
\begin{eqnarray}
\label{rev3}
F_{\rm pol}=-\frac{k}{\pi} \frac{u_0^3}{v^3+4 u_0^2 v}.
\end{eqnarray}
In the limit $v\rightarrow 0$, the usual friction by polarization behaves as
$F_{\rm pol}\sim -\frac{k}{4\pi} \frac{u_0}{v}$.

(ii) For $\gamma<0$ and
$v\rightarrow 0$, we find that
\begin{eqnarray}
\label{rev4}
F_{\rm pol}=-\frac{k}{\pi}u_0 v \frac{(\gamma+u_0)^2}{\gamma^2\left (
\gamma+2u_0\right )^2}.
\end{eqnarray}
In the limit $\gamma\rightarrow 0$, the usual friction by polarization behaves
as $F_{\rm pol}\sim -\frac{k}{4\pi}u_0 v \frac{1}{\gamma^2}$. 

As indicated previously, the usual friction by polarization diverges for
$v\rightarrow 0$ at the critical point $\gamma\rightarrow 0$. Furthermore, the
limits $v\rightarrow 0$ and
$\gamma\rightarrow 0$ do not commute. By contrast, the total friction
force which takes  the Landau modes into account is finite at
the critical point $u_0\rightarrow u_c$ and its expression 
for the HMF model when $v\rightarrow 0$ is [see Eq. (\ref{new12})]
\begin{eqnarray}
\label{rev6}
F_{\rm pol}=-\frac{k}{2\pi}  \frac{v}{u_c} \left
\lbrack  -\frac{1}{8}+\frac{1}{8}e^{-2u_ct}
(1+2u_ct)+\frac{1}{4}(u_ct)^2\right \rbrack.    
\end{eqnarray}
For $t\rightarrow 0$ we get $F_{\rm pol}\sim
-\frac{k}{6\pi}u_c^2 t^3 v$ and for $t\rightarrow +\infty$ we get $F_{\rm
pol}\sim -\frac{k}{8\pi} 
u_c t^2 v$.

{\it Remark:} We know that for 1D homogeneous systems with long-range
interactions (like the HMF model above the critical energy) the diffusion term
and the friction term cancel each other, resulting in a vanishing Lenard-Balescu
flux. It would be interesting to know if this property persists at the critical
point when the Landau modes are properly taken into account. This requires to
calculate the diffusion term explicitly.

\section{General results for homogeneous systems with long-range interactions}
\label{sec_gfn}

\subsection{Bare potential of interaction}
\label{sec_gfnbare}

We consider a system of particles interacting via a long-range
binary
potential $u(|{\bf r}-{\bf r}'|)$. The potential is
related to the distribution function (or to the density $\rho=\int f\, d{\bf
v}$) by
\begin{eqnarray}
\Phi({\bf r},t)=\int u(|{\bf r}-{\bf r}'|) f({\bf r}',{\bf v}',t)\, d{\bf
r}'d{\bf v}'.
\label{pot1}
\end{eqnarray}
The potential $u(|{\bf r}-{\bf r}'|)$ is called the bare
potential of interaction. Taking the Fourier-Laplace
transform of
this equation, and using the fact that the integral is a product of
convolution, we get 
\begin{eqnarray}
{\tilde \Phi}({\bf k},\omega)=(2\pi)^d{\hat u}(k)\int
d{\bf v}\, {\tilde f}({\bf k},{\bf v},\omega),
\label{bof3bann}
\end{eqnarray}
where ${\hat u}(k)$ is the Fourier transform of $u(|{\bf r}-{\bf r}'|)$. We note
that ${\hat u}({\bf k})$ is real and
that ${\hat u}(-{\bf k})={\hat u}({\bf k})$.  Eq. (\ref{bof3bann}) is
the
counterpart of Eq. (A5) of \cite{angleaction2} or Eq. (C4) of \cite{kinfd}
for inhomogeneous stellar
systems.

For self-gravitating systems \cite{bt}, the potential $\Phi$
produced by the
density  $\rho$ is
determined by the Poisson equation
\begin{eqnarray}
\Delta\Phi=S_d G\rho,
\label{poig1}
\end{eqnarray}
where $S_d$ is the surface of a unit sphere in $d$ dimensions. In that
case, the potential
of interaction is explicitly  given by
\begin{eqnarray}
u(|{\bf r}-{\bf r}'|)=-\frac{G}{d-2}\frac{1}{|{\bf r}-{\bf
r}'|^{d-2}} \qquad (d\neq 2),
\label{vi1}
\end{eqnarray}
\begin{eqnarray}
u(|{\bf r}-{\bf r}'|)=G \ln |{\bf r}-{\bf r}'| \qquad (d=2).
\label{vi2}
\end{eqnarray}
Taking the
Fourier transform of Eq. (\ref{poig1}), we find that 
\begin{eqnarray}
(2\pi)^d {\hat u}(k)=-\frac{S_d G}{k^2}.
\label{poig3}
\end{eqnarray}
Therefore, we can decompose the gravitational potential of interaction as
\begin{eqnarray}
u(|{\bf r}-{\bf r}'|)=-S_d G\int \frac{d{\bf k}}{(2\pi)^d}\frac{e^{i{\bf
k}\cdot
({\bf r}-{\bf r}')}}{k^2}.
\label{alld}
\end{eqnarray}

For the HMF model \cite{ar,cvb}, the potential of interaction reads
\begin{eqnarray}
u(\theta-\theta')=-\frac{k}{2\pi}\cos(\theta-\theta')
\label{poig4}
\end{eqnarray}
with $k>0$ in the attractive case and $k<0$ in the repulsive case. Its Fourier
transform is
\begin{eqnarray}
\hat{u}_n=-\frac{k}{4\pi}\delta_{n,\pm 1}.
\label{poig5}
\end{eqnarray}
It involves only the modes $n=\pm 1$.

\subsection{Dressed potential of interaction from an external perturbation}
\label{sec_inhosrf}

Let us determine the linear response of a spatially homogeneous
system with long-range interactions to a weak external perturbation $\Phi_e({\bf
r},t)$ produced by a distribution function $f_e({\bf r},{\bf v},t)$. Since the
perturbation is small, we can use the linearized Vlasov equation
(\ref{i57}). Taking its Fourier-Laplace
transform and
ignoring the initial condition (or assuming that there is no initial
fluctuation) we
obtain
\begin{eqnarray}
\label{j3z}
\delta{\tilde f}({\bf k},{\bf
v},\omega)=\frac{{\bf k}\cdot \frac{\partial
f}{\partial
{\bf v}}}{{\bf
k}\cdot {\bf v}-\omega}\delta{\tilde\Phi}_{\rm tot}({\bf k},\omega),
\end{eqnarray}
where $\delta{\Phi}_{\rm tot}=\Phi_e+\delta\Phi$ is the total fluctuation of
the potential acting on a particle.
On the other hand, Eq. (\ref{bof3bann}) with $\delta f_{\rm tot}=f_e+\delta f$
gives
\begin{eqnarray}
\delta {\tilde \Phi}_{\rm tot}({\bf k},\omega)=(2\pi)^d{\hat u}(k)\int
d{\bf v}\, \left\lbrack {\tilde f}_e({\bf k},{\bf v},\omega)+\delta{\tilde
f}({\bf k},{\bf v},\omega)\right\rbrack.
\label{ping}
\end{eqnarray}
Combining these relations we get
\begin{eqnarray}
\left\lbrack 1-(2\pi)^d {\hat u}(k)  \int d{\bf
v}\, \frac{{\bf k}\cdot \frac{\partial
f}{\partial
{\bf v}}}{{\bf
k}\cdot {\bf v}-\omega}\right\rbrack\delta{\tilde\Phi}_{\rm tot}({\bf k},\omega)
=(2\pi)^d {\hat u}(k) \int
d{\bf v}\, {\tilde f}_e({\bf k},{\bf v},\omega).
\label{mat1}
\end{eqnarray}
This is a degenerate Volterra-Fredholm equation relating the Fourier-Laplace
transform of the total potential $\delta
{\tilde
\Phi}_{\rm
tot}({\bf k},\omega)$ to the  Fourier-Laplace
transform of the external distribution function ${\tilde f}_e({\bf k},{\bf
v},\omega)$. Its solution is
\begin{eqnarray}
\delta{\tilde \Phi}_{\rm tot}({\bf k},\omega)=
(2\pi)^d \frac{{\hat u}(k)}{\epsilon({\bf k},\omega)}  \int
d{\bf v}\, {\tilde f}_e({\bf k},{\bf v},\omega),
\label{pot6s}
\end{eqnarray}
where  
\begin{equation}
\epsilon({\bf k},\omega)=1-(2\pi)^d\hat{u}(k)\int  \frac{{\bf k}\cdot
\frac{\partial f}{\partial {\bf v}}}{{\bf k}\cdot {\bf v}-\omega}\, d{\bf v}
\label{lb12}
\end{equation}
is the ``dielectric'' function. It is defined for any $\omega$ by performing
the integration along the Landau contour (see,
e.g., \cite{nyquisthmf,nyquistgrav}).  As
a result, $\epsilon({\bf k},\omega)$ is a complex function.
It determines the
response of the system $\delta{\tilde \Phi}_{\rm tot}({\bf k},\omega)$ to an
external
perturbation ${\tilde f}_e({\bf k},{\bf v},\omega)$ through Eq. (\ref{pot6s}). 
We can rewrite Eq. (\ref{pot6s}) as 
\begin{eqnarray}
\delta{\tilde \Phi}_{\rm tot}({\bf k},\omega)=
(2\pi)^d {\hat u}_d({\bf k},\omega)  \int
d{\bf v}\, {\tilde f}_e({\bf k},{\bf v},\omega),
\label{pot6sb}
\end{eqnarray}
where 
\begin{eqnarray}
{\hat u}_d({\bf k},\omega)=\frac{{\hat u}(k)}{\epsilon({\bf k},\omega)}.
\label{udress}
\end{eqnarray}
is the
dressed potential of interaction. It can be interpreted as a
degenerate Green function. 
If we neglect collective effects in Eq. (\ref{ping}), we obtain
\begin{eqnarray}
\delta{\tilde \Phi}_{\rm tot}({\bf k},\omega)={\tilde \Phi}_{e}({\bf
k},\omega)=
(2\pi)^d {\hat u}(k)  \int
d{\bf v}\, {\tilde f}_e({\bf k},{\bf v},\omega).
\label{pot6sc}
\end{eqnarray}
This amounts  to replacing the dressed potential of interaction ${\hat
u}_d(k,\omega)$ by the bare potential of interaction ${\hat
u}(k)$ in Eq. (\ref{pot6sb}). 
Eqs. (\ref{mat1}), (\ref{lb12}) and (\ref{pot6sb}) are the counterparts
of Eqs.
(C17), (C18) and (C19) of \cite{kinfd} for inhomogeneous stellar
systems.

{\it Remark:} From Eqs. (\ref{bof3bann}) and (\ref{pot6s}) we get
\begin{eqnarray}
\delta{\tilde \Phi}_{\rm tot}({\bf k},\omega)=\frac{1}{\epsilon({\bf
k},\omega)} {\tilde \Phi}_e({\bf k},\omega)
\label{pot6sh}
\end{eqnarray}
and
\begin{eqnarray}
\delta{\tilde \Phi}({\bf k},\omega)=\frac{1-\epsilon({\bf
k},\omega)}{\epsilon({\bf
k},\omega)} {\tilde \Phi}_e({\bf k},\omega).
\label{pot6shb}
\end{eqnarray}
These equations show how the external perturbation is modified (dressed) by
collective effects.

\subsection{Dressed potential of interaction from an initial perturbation}
\label{sec_inhosrfb}

In Appendix \ref{sec_inhosrf}, we have studied the response of a
collisionless
system described by the Vlasov equation to a weak external potential
$\Phi_e({\bf r},t)$ by using the linear response
theory. Here, we compare these
results to those obtained when the system is isolated (i.e., $\Phi_e({\bf
r},t)=0$) but the distribution function is slightly perturbed at $t=0$. Taking
the Fourier-Laplace transform of the linearized
Vlasov equation (\ref{i57}), and assuming now that $\delta f({\bf r},{\bf
v},0)\neq 0$, we obtain 
\begin{eqnarray}
\label{j3sann}
\delta{\tilde f}({\bf k},{\bf
v},\omega)=\frac{{\bf k}\cdot \frac{\partial
f}{\partial
{\bf v}}}{{\bf
k}\cdot {\bf v}-\omega}\delta{\tilde\Phi}({\bf k},\omega)+\frac{\delta{\hat
f} ({\bf k},{\bf v},0)}{i(
{\bf k}\cdot {\bf v}-\omega)},
\end{eqnarray}
where $\delta\hat{f}({\bf k},{\bf v},0)$ is the Fourier transform of the initial
perturbation $\delta f({\bf r},{\bf v},0)$. According to Eq.
(\ref{bof3bann}) the fluctuations of the potential are related to the
fluctuations of
the distribution function by 
\begin{eqnarray}
\delta{\tilde \Phi}({\bf k},\omega)=(2\pi)^d{\hat u}(k)\int
d{\bf v}\,  \delta{\tilde f}({\bf k},{\bf v},\omega).
\label{poip}
\end{eqnarray}
Substituting Eq. (\ref{j3sann}) into Eq. (\ref{poip}), we obtain
\begin{eqnarray}
\left\lbrack 1-(2\pi)^d {\hat u}(k)  \int d{\bf
v}\, \frac{{\bf k}\cdot \frac{\partial
f}{\partial
{\bf v}}}{{\bf
k}\cdot {\bf v}-\omega}\right\rbrack\delta{\tilde\Phi}({\bf k},\omega)
=(2\pi)^d {\hat u}(k)  \int d{\bf
v}\, \frac{\delta{\hat f} ({\bf k},{\bf v},0)}{i(
{\bf k}\cdot {\bf v}-\omega)}.
\label{mat1b}
\end{eqnarray}
This is a degenerate Volterra-Fredholm integral equation relating the
Fourier-Laplace transform of the fluctuations of the potential
$\delta{\tilde\Phi}({\bf k},\omega)$  to the Fourier transform of the initial
fluctuations of the
distribution function $\delta{\hat f} ({\bf k},{\bf v},0)$. It is similar to
Eq. (\ref{mat1}). Its solution is
\begin{eqnarray}
\delta{\tilde \Phi}({\bf k},\omega)=
(2\pi)^d \frac{{\hat u}(k)}{\epsilon({\bf k},\omega)}  \int d{\bf
v}\, \frac{\delta{\hat f} ({\bf k},{\bf v},0)}{i(
{\bf k}\cdot {\bf v}-\omega)},
\label{bebe}
\end{eqnarray}
where the dielectric function is defined by Eq. (\ref{lb12}). We can also
rewrite Eq. (\ref{bebe}) as
\begin{eqnarray}
\delta{\tilde \Phi}({\bf k},\omega)=
(2\pi)^d {\hat u}_d({\bf k},\omega)  \int d{\bf
v}\, \frac{\delta{\hat f} ({\bf k},{\bf v},0)}{i(
{\bf k}\cdot {\bf v}-\omega)},
\label{poto6dri}
\end{eqnarray}
with the dressed potential of interaction from Eq. (\ref{udress}).  If we
neglect
collective effects in Eq. (\ref{j3sann}), the foregoing
equation reduces to
\begin{eqnarray}
\delta{\tilde \Phi}({\bf k},\omega)=
(2\pi)^d {\hat u}(k)  \int d{\bf
v}\, \frac{\delta{\hat f} ({\bf k},{\bf v},0)}{i(
{\bf k}\cdot {\bf v}-\omega)}.
\label{poto6dr}
\end{eqnarray}
This amounts to replacing the dressed potential of interaction
${\hat
u}_d({\bf k},\omega)$ by the bare potential of interaction ${\hat
u}(k)$ in Eq. (\ref{poto6dri}). Eqs.
(\ref{lb12}), (\ref{poip}), (\ref{mat1b}) and (\ref{poto6dri}) are the
counterparts
of Eqs. (37), (A5), (A6) and (A8) of \cite{angleaction2} or Eqs. (C19), (H4),
(H5) and (H6) of \cite{kinfd} for inhomogeneous stellar
systems.

\subsection{Dispersion relation}
\label{sec_drs}

In the absence of external ($f_e=0$) or initial
($\delta f(t=0)=0$) perturbation, Eqs. (\ref{mat1}) and (\ref{mat1b}) lead to
the dispersion relation
\begin{equation}
\epsilon({\bf k},\omega)=0,
\label{disrel}
\end{equation}
which determines the proper pulsations $\omega_{\alpha}({\bf k})$ of the system
as a
function of the
wavevector ${\bf k}$. This relation can be used to study the linear dynamical
stability of a spatially homogeneous distribution with respect to the Vlasov
equation (\ref{n4b}) \cite{nicholson,bt}. Some illustrations are given
in \cite{nyquisthmf,nyquistgrav}. Let us write the normal
modes $\omega({\bf k})$ as
$\omega=\Omega+i\gamma$,
where $\Omega({\bf k})$ (pulsation) and $\gamma({\bf k})$ (exponential rate)
are real so that $\delta f\sim e^{-i\Omega t}e^{\gamma t}$. The mode ${\bf k}$
is
stable when $\gamma({\bf k})<0$ and unstable
when $\gamma({\bf k})>0$. The condition of marginal stability is $\gamma({\bf
k})=0$. In that case, the perturbation oscillates with a pulsation $\Omega$.

Assuming that $f({\bf v})$ is isotropic, we can rewrite the dielectric function
(\ref{lb12}) as
\begin{eqnarray}
\epsilon(k,\omega)=1-(2\pi)^d{\hat u}(k)\int\frac{f'(v)}{v-\frac{\omega}{k}}\,
dv,
\label{disrel1}
\end{eqnarray}
where we have taken the $z$-axis in the direction of ${\bf k}$, noted $v$ for
$v_z$, and introduced the reduced distribution function $f(v)=\int f({\bf
v})\, dv_xdv_y$
\cite{nyquisthmf,nyquistgrav}. For $\omega=0$ we have
\begin{equation}
\epsilon(k,0)=1-(2\pi)^d{\hat
u}(k)\int_{-\infty}^{+\infty}\frac{f'(v)}{v}\, dv,
\label{disrel2}
\end{equation}
where we have used $f'(0)=0$. If the reduced distribution function $f(v)$ has
a single maximum at $v=0$ then, according to the
Nyquist criterion \cite{nyquisthmf,nyquistgrav}, the system is stable if 
\begin{equation}
\epsilon({k},0)>0
\label{disrel2b}
\end{equation}
and unstable otherwise. Furthermore, there is
only one unstable mode. Therefore, if an unstable mode exists with a purely
imaginary pulsation $\omega=i\gamma>0$, then this is the only
one. By contrast, there may exist several stable modes.  We also recall the
identity
\begin{equation}
c_s^2=-\frac{\rho}{\int_{-\infty}^{+\infty}\frac{f'(v)}{v}\,
dv},
\label{disrel3}
\end{equation}
where $c_s^2=P'(\rho)$ is the squared speed of sound in the corresponding
barotropic gas (see \cite{nyquisthmf,nyquistgrav} for details). Let us consider
specific examples:

(i) The Fourier transform of the gravitational potential is given by Eq.
(\ref{poig3}). In that case, Eq. (\ref{disrel2}) becomes
\begin{eqnarray}
\epsilon({k},0)=1+\frac{S_d G}{k^2}\int\frac{f'(v)}{v}\,
dv.
\label{frigo2}
\end{eqnarray}
It can be written as
\begin{eqnarray}
\epsilon({k},0)=1-\frac{k_J^2}{k^2},
\label{frigo3}
\end{eqnarray}
where $k_J$ is the Jeans wavenumber defined by
\begin{equation}
k_J^2=-S_dG\int_{-\infty}^{+\infty}\frac{f'(v)}{v}\,
dv.
\label{disrel8}
\end{equation}
The stability criterion (\ref{disrel2b}) then reads
\begin{equation}
\epsilon({k},0)=1-\frac{k_J^2}{k^2}>0,\qquad {\rm i.e.}\qquad k>k_J.
\label{disrel7}
\end{equation}
The system is stable if, and only if, the wavelength $\lambda$ of the
perturbation is smaller
than the Jeans length $\lambda_J$. For
the Maxwell-Boltzmann distribution function, the stability criterion can be
written explicitly as
\begin{equation}
k>k_J=\left (\frac{S_dG\rho m}{k_B T}\right )^{1/2}.
\label{disrel9}
\end{equation}
More generally, using Eq. (\ref{disrel3}), it can be written as
$k>k_J=(S_d G\rho/c_s^2)^{1/2}$.

(ii) The Fourier transform of the potential of the
HMF model is given by Eq. (\ref{poig5}). In that case, Eq. (\ref{disrel2})
becomes
\begin{eqnarray}
\epsilon(0)=1+\frac{k}{2}\int\frac{f'(v)}{v}\,
dv.
\label{frigo5}
\end{eqnarray}
It can be written as
\begin{eqnarray}
\epsilon(0)=1-B,
\label{frigo6}
\end{eqnarray}
where $B$ is a sort of generalized inverse temperature defined by
\begin{eqnarray}
B=-\frac{k}{2}\int\frac{f'(v)}{v}\,
dv.
\label{frigo7}
\end{eqnarray}
The
stability criterion (\ref{disrel2b}) then reads
\begin{equation}
\epsilon(0)=1-B>0,\qquad {\rm i.e.}\qquad B<B_c=1.
\label{disrel4}
\end{equation}
For the
Maxwell-Boltzmann distribution function, we have $B=k\rho/2T$. The stability
criterion can
be written explicitly as
\begin{equation}
T>T_c=\frac{k\rho}{2}.
\label{disrel6}
\end{equation}
More generally, using Eq. (\ref{disrel3}), we have
$B=k\rho/2c_s^2$ and the stability criterion can be written as $c_s^2>k\rho/2$.

\subsection{Volterra-Fredholm equation in time with an external perturbation}
\label{sec_volterra}

Taking the Fourier transform of Eq. (\ref{i57}), we obtain
\begin{eqnarray}
\label{j7cu}
\frac{\partial \delta{\hat f}}{\partial t}+i{\bf k}\cdot{\bf v}\delta{\hat
f}=i{\bf k}\cdot \frac{\partial f}{\partial {\bf
v}}(\delta{\hat\Phi}+{\hat \Phi}_e).
\end{eqnarray}
This first order differential equation in time can be solved with the method
of
the variation of the constant, giving
\begin{eqnarray}
\label{v1}
\delta{\hat f}({\bf k},{\bf v},t)=i{\bf k}\cdot
\frac{\partial f}{\partial {\bf v}}\int_0^t dt'\, \left\lbrack \delta{\hat
\Phi}({\bf
k},t')+{\hat
\Phi}_e({\bf
k},t')\right \rbrack e^{-i{\bf k}\cdot{\bf v}(t-t')},
\end{eqnarray}
where we have assumed $\delta{\hat f}({\bf k},{\bf
v},t=0)=0$. According
to Eq.
(\ref{bof3bann}) the fluctuations of the potential are related to the
fluctuations of
the distribution function by 
\begin{eqnarray}
\delta{\hat \Phi}({\bf k},t)=(2\pi)^d{\hat u}(k)\int
d{\bf v}\,  \delta{\hat f}({\bf k},{\bf v},t).
\label{v2}
\end{eqnarray}
Substituting Eq. (\ref{v1}) into Eq. (\ref{v2}), we obtain  
\begin{eqnarray}
\delta{\hat \Phi}({\bf k},t)=(2\pi)^d{\hat u}(k)\int
d{\bf v}\,  i{\bf k}\cdot
\frac{\partial f}{\partial {\bf v}}\int_0^t dt'\, \left\lbrack \delta{\hat
\Phi}({\bf
k},t')+{\hat
\Phi}_e({\bf
k},t')\right \rbrack  e^{-i{\bf
k}\cdot{\bf v}(t-t')}.
\label{v3}
\end{eqnarray}
Integrating by parts, the foregoing equation can
be rewritten as
\begin{eqnarray}
\delta{\hat \Phi}({\bf k},t)=-(2\pi)^{2d}{\hat u}(k)k^2\int_0^t dt'\, 
(t-t') \left\lbrack \delta{\hat
\Phi}({\bf
k},t')+{\hat
\Phi}_e({\bf
k},t')\right \rbrack {\hat f}\left\lbrack {\bf k}(t-t')\right\rbrack, 
\label{v4}
\end{eqnarray}
where ${\hat f}({\bf k})$ is the Fourier transform of the distribution
function in velocity space. Eq. (\ref{v4}) is a Volterra-Fredholm
integral
equation 
equivalent to Eq. (\ref{mat1}). This integral equation was derived and studied
in Secs. 4.4 and 6.2 of
\cite{linres}. It was also discussed recently in \cite{magorrian}.

On the other hand, taking the Fourier transform in  velocity space of Eq.
(\ref{v1}) we get
\begin{eqnarray}
\label{v5}
\delta{\hat f}({\bf k},{\bf K},t)=\int \frac{d{\bf v}}{(2\pi)^d}e^{-i{\bf
K}\cdot {\bf v}} i{\bf k}\cdot
\frac{\partial f}{\partial {\bf v}}\int_0^t dt'\,\delta{\hat
\Phi}_{\rm tot}({\bf
k},t') e^{-i{\bf k}\cdot{\bf v}(t-t')},
\end{eqnarray}
where we have introduced the total fluctuation $\delta\Phi_{\rm
tot}=\delta\Phi+\Phi_e$. Integrating by parts, we obtain
\begin{eqnarray}
\label{v6}
\delta{\hat f}({\bf k},{\bf K},t)=-{\bf
k}\int_0^t dt'\,\delta{\hat
\Phi}_{\rm tot}({\bf
k},t') \left\lbrack {\bf K}+{\bf k}(t-t')\right\rbrack {\hat f}\left\lbrack {\bf
K}+{\bf k}(t-t')\right\rbrack. 
\end{eqnarray}
This equation is equivalent to Eq. (21) of \cite{magorrian} obtained by a
different method.
Now, Eq. (\ref{v2}) can be
written as
\begin{eqnarray}
\delta{\hat \Phi}({\bf k},t)=(2\pi)^{2d}{\hat u}(k)\delta{\hat f}({\bf
k},{\bf K}={\bf
0},t),
\label{v7}
\end{eqnarray}
where $\delta{\hat f}({\bf k},{\bf
K},t)$ denotes here the Fourier transform of $\delta f ({\bf r},{\bf v},t)$ in
position and velocity. Combining Eqs. (\ref{v6}) and (\ref{v7}) we find that
\begin{eqnarray}
\delta{\hat \Phi}({\bf k},t)=-(2\pi)^{2d}{\hat u}(k){\bf
k}\int_0^t dt'\,\delta{\hat
\Phi}_{\rm tot}({\bf
k},t') {\bf k}(t-t') {\hat f}\left\lbrack {\bf k}(t-t')\right\rbrack, 
\label{v8}
\end{eqnarray}
which returns Eq. (\ref{v4}).

\subsection{The resolvent of the linearized Vlasov equation}
\label{sec_resolvent}

Substituting Eq. (\ref{poto6})  into Eq.
(\ref{j3s}), we obtain
\begin{equation}
\label{iv3}
\delta\tilde{f}({\bf k},{\bf v},\omega)=\frac{{\bf k}\cdot \frac{\partial
f}{\partial {\bf v}}}{{\bf k}\cdot {\bf
v}-\omega}(2\pi)^d\frac{\hat{u}(k)}{\epsilon({\bf k},\omega)}\int
\frac{\delta\hat{f}({\bf k},{\bf v}',0)}{i({\bf k}\cdot {\bf v}'-\omega)}\,
d{\bf v}'+\frac{\delta\hat{f}({\bf k},{\bf v},0)}{i({\bf k}\cdot {\bf
v}-\omega)}.
\end{equation}
This equation relates the Fourier-Laplace transform of the
fluctuations of the distribution function to the Fourier
transform of the initial fluctuations of the distribution function. This is the
exact solution of the initial value problem for
the linearized
Vlasov equation. It can be conveniently written in terms of an operator  that
connects $\delta\tilde{f}({\bf k},{\bf v},\omega)$ to the initial
value $\delta\hat{f}({\bf k},{\bf v},0)$. One has
\begin{equation}
\label{iv4}
\delta\tilde{f}({\bf k},{\bf v},\omega)=\int d{\bf v}' R({\bf v}|{\bf v}',{\bf
k},\omega) \delta\hat{f}({\bf k},{\bf v},0),
\end{equation}
where
\begin{equation}
\label{iv5}
R({\bf v}|{\bf v}',{\bf k},\omega)=\frac{{\bf k}\cdot \frac{\partial f}{\partial
{\bf v}}}{{\bf k}\cdot {\bf v}-\omega}(2\pi)^d\frac{\hat{u}(k)}{\epsilon({\bf
k},\omega)}\frac{1}{i({\bf k}\cdot {\bf v}'-\omega)}+\frac{\delta({\bf v}-{\bf
v}')}{i({\bf k}\cdot {\bf v}-\omega)}
\end{equation}
is called the resolvent or the propagator
of the linearized Vlasov equation.
If we consider an initial condition of the form $\delta f({\bf r},{\bf
v},0)=m\delta({\bf v}-{\bf v}')\delta({\bf r}-{\bf r}')$, implying
\begin{equation}
\label{iv6}
\delta\hat{f}({\bf k},{\bf v},0)=\frac{m}{(2\pi)^d}\delta({\bf v}-{\bf
v}')e^{-i{\bf k}\cdot {\bf r}'},
\end{equation}
we find that
\begin{equation}
\label{iv7}
\delta\tilde{f}({\bf k},{\bf v},\omega)=\frac{{\bf k}\cdot \frac{\partial
f}{\partial {\bf v}}}{{\bf k}\cdot {\bf
v}-\omega}\frac{\hat{u}(k)}{\epsilon({\bf k},\omega)}m \frac{e^{-i{\bf k}\cdot
{\bf r}'}}{i({\bf k}\cdot {\bf v}'-\omega)}+\frac{m}{(2\pi)^d}\delta({\bf
v}-{\bf v}')\frac{e^{-i{\bf k}\cdot {\bf r}'}}{i({\bf k}\cdot {\bf v}-\omega)}.
\end{equation}

{\it Remark:} If we substitute Eq. (\ref{poip}) into Eq. (\ref{j3s}), we
obtain the integral equation
\begin{equation}
\label{iv8}
\delta\tilde{f}({\bf k},{\bf v},\omega)=\frac{{\bf k}\cdot \frac{\partial
f}{\partial {\bf v}}}{{\bf k}\cdot {\bf v}-\omega} (2\pi)^d\hat{u}(k)\int
\delta\tilde{f}({\bf k},{\bf v}',\omega)\, d{\bf v}'+\frac{\delta\hat{f}({\bf
k},{\bf v},0)}{i({\bf k}\cdot {\bf v}-\omega)}.
\end{equation}
This equation relates the Fourier-Laplace transform of the
fluctuations of the distribution function to the Fourier transform of
the initial fluctuations of the distribution function, so it is
equivalent to Eq. (\ref{iv3}).

\subsection{Useful identities}
\label{sec_id}

The dielectric function is defined by Eq. (\ref{lb12}), where the integral must
be performed along the Landau contour. For $\omega$
real, using the Landau prescription $\omega\rightarrow
\omega+i0^+$ and applying the Sokhotski-Plemelj formula
\begin{eqnarray}
\frac{1}{x\pm i0^+}={\rm P}\left (\frac{1}{x}\right )\mp i\pi\delta(x),
\label{plemelj}
\end{eqnarray}
where P is the principal value, we obtain the 
identity
\begin{equation}
{\rm Im}\, \epsilon({\bf k},\omega)=-\pi(2\pi)^d\hat{u}(k)\int {\bf
k}\cdot
\frac{\partial f}{\partial {\bf v}}\delta({\bf k}\cdot {\bf v}-\omega)\, d{\bf
v}.
\label{ii1}
\end{equation}
We also have
\begin{eqnarray}
\epsilon(-{\bf k},-\omega)=\epsilon({\bf k},\omega)^*.
\label{ii2}
\end{eqnarray}

\section{General results for an axisymmetric distribution of 2D
point vortices}
\label{sec_pot}

\subsection{Bare potential of interaction}
\label{sec_swift}

We consider a system of 2D point vortices interacting via a
long-range binary potential $u(|{\bf r}-{\bf r}'|)$. The
stream function is related to the vorticity by
\begin{equation}
\psi({\bf r},t)=\int u(|{\bf r}-{\bf r}'|)\omega({\bf r'},t)\, d{\bf
r}'.
\label{pot7}
\end{equation}
The potential $u(|{\bf r}-{\bf r}'|)$ is called the bare potential of
interaction. Introducing a polar system of coordinates it can be written as
\begin{equation}
u(|{\bf r}-{\bf r}'|)=u\left (\sqrt{r^2+r'^2-2rr'\cos(\theta-\theta')}\right
)=u(r,r',\phi),
\label{pot8}
\end{equation}
where $\phi=\theta-\theta'$. Because of its $2\pi$-periodicity in $\phi$, the
function $u(r,r',\phi)$ can
be decomposed
in a Fourier series of the form
\begin{equation}
u(r,r',\phi)=\sum_n e^{i n \phi}\hat{u}_n(r,r'),\qquad
\hat{u}_n(r,r')=\int_{0}^{2\pi}\frac{d\phi}{2\pi}u(r,r',\phi)\cos(n\phi).
\label{pot9}
\end{equation}
We note that ${\hat u}_{n}(r,r')$ is real, even in $n$,
and symmetric in $r$ and $r'$.  Taking the Fourier-Laplace
transform of Eq. (\ref{pot7}) and using the fact that
the integral is a product of convolution, we get
\begin{equation}
\tilde\psi(n,r,\sigma)=2\pi \int_0^{+\infty} r'dr'\,  \hat{u}_n(r,r')
\tilde\omega(n,r',\sigma).
\label{pot10}
\end{equation}
This is the
counterpart of Eq. (A5) of \cite{angleaction2} or Eq. (C4) of \cite{kinfd}
for inhomogeneous stellar
systems and of Eq. (B2) of \cite{kinfdvortex} for unidirectional flows.

In the usual model of 2D point vortices, the stream
function $\psi$ produced by the vorticity field $\omega$ is
determined by the Poisson equation  
\begin{equation}
\Delta\psi=-\omega.
\label{poiss}
\end{equation}
In that
case, the potential
of interaction is explicitly  given by
\begin{equation}
u(|{\bf r}-{\bf r}'|)=-\frac{1}{2\pi}\ln|{\bf r}-{\bf
r}'|.
\label{gtl}
\end{equation}
This corresponds to the Newtonian (or
Coulombian) interaction in two dimensions. The function $u(r,r',\phi)$ defined
by Eq. (\ref{pot8}) reads
\begin{equation}
u(r,r',\phi)=-\frac{1}{4\pi}\ln(r^2+r'^2-2rr'\cos\phi).
\label{tgl2}
\end{equation}
Taking its Fourier transform, we find that (see Appendix B of \cite{kindetail})
\begin{equation}
\hat{u}_n(r,r')=\frac{1}{4\pi |n|}\left (\frac{r_<}{r_>}\right )^{|n|},\qquad
\hat{u}_0(r,r')=-\frac{1}{2\pi}\ln r_>.
\label{pot13}
\end{equation}
Therefore, we can write the  potential
of interaction as
\begin{equation}
u(r,r',\phi)=-\frac{1}{2\pi}\ln r_>+\frac{1}{4\pi}\sum_{n\neq
0}\frac{1}{|n|}\left (\frac{r_<}{r_>}\right )^{|n|}e^{in\phi}.
\label{pot14}
\end{equation}
This corresponds to the Fourier decomposition of the logarithm $\ln|{\bf r}-{\bf
r}'|$ in
two
dimensions.

We can arrive at these results in a different manner.
Introducing a system of polar coordinates, the Poisson equation (\ref{poiss})
can be
written as
\begin{equation}
\frac{1}{r}\frac{\partial}{\partial r} r \frac{\partial \psi}{\partial
r}+\frac{1}{r^2}\frac{\partial^2\psi}{\partial\theta^2}=-\omega.
\label{pot2}
\end{equation}
Taking the Fourier-Laplace transform of this equation,  we obtain
\begin{equation}
\left\lbrack \frac{1}{r}\frac{d}{dr} r \frac{d}{d
r}-\frac{n^2}{r^{2}}\right \rbrack
\delta\tilde\psi(n,r,\sigma)=-\delta\tilde\omega(n,r,\sigma).
\label{pot3bu}
\end{equation}
The general solution of this equation is given by 
\begin{equation}
\tilde\psi(n,r,\sigma)=2\pi \int_0^{+\infty} r'dr'\,  G_{\rm bare}(n,r,r')
\tilde\omega(n,r',\sigma),
\label{pot10g}
\end{equation}
where $G_{\rm bare}(n,r,r')$ is the bare Green function determined by the
differential equation 
\begin{equation}
\left\lbrack \frac{1}{r}\frac{d}{d r} r \frac{d}{d
r}-\frac{n^2}{r^{2}}\right \rbrack G_{\rm
bare}(n,r,r')=-\frac{\delta(r-r')}{2\pi
r}.
\label{pot6}
\end{equation}
We note that the bare Green function is just a rewriting of the Fourier
transform of the bare potential of interaction:
\begin{equation}
G_{\rm
bare}(n,r,r')=\hat{u}_n(r,r').
\label{pot6blues}
\end{equation}
We have
$G_{\rm bare}(-n,r,r')=G_{\rm bare}(n,r,r')$ and $G_{\rm bare}(n,r',r)=G_{\rm
bare}(n,r,r')$. In an unbounded
domain, the differential equation (\ref{pot6}) can be solved
analytically returning the results from Eq. (\ref{pot13}).

\subsection{Dressed potential of interaction from an  external perturbation}
\label{sec_marr}

Let us determine the linear response of a 2D
incompressible axisymmetric flow to a weak external perturbation 
$\psi_e(x,y,t)$ produced by a vorticity field
$\omega_e(x,y,t)$.  Since the
perturbation is small, we can use the linearized 2D
Euler equation (\ref{ham13}). Taking its Fourier-Laplace transform and
ignoring the initial condition (or assuming that there is no initial
fluctuation) we
obtain
\begin{eqnarray}
\label{bam}
\delta{\tilde\omega}(n,r,\sigma)=-\frac{n\frac{1}{r}\frac{\partial\omega}{
\partial
r}}{n\Omega-\sigma}\delta{\tilde\psi}_{\rm
tot}(n,r,\sigma),
\end{eqnarray}
where $\delta{\psi}_{\rm tot}=\psi_e+\delta\psi$ is the total fluctuation of
the stream function acting on a point vortex. On the other hand, Eq.
(\ref{pot10}) with $\delta\omega_{\rm
tot}=\omega_e+\delta\omega$ gives
\begin{eqnarray}
\delta{\tilde\psi}_{\rm tot}(n,r,\sigma)=\int_0^{+\infty} 2\pi r'  dr'\,
G_{\rm bare}(n,r,r')\left \lbrack
{\tilde\omega}_e(n,r',\sigma)+\delta{\tilde\omega}(n,r',\sigma)\right\rbrack.
\label{j5g}
\end{eqnarray}
Combining these relations we get
\begin{equation}
\delta{\tilde\psi}_{\rm tot}(n,r,\sigma)+\int_0^{+\infty} 2\pi r'  dr'\,
G_{\rm bare}(n,r,r')\frac{n\frac{1}{r'}\frac{\partial\omega'}{
\partial
r'}}{n\Omega'-\sigma}\delta{\tilde\psi}_{\rm tot}(n,r',\sigma)
=\int_0^{+\infty} 2\pi r'  dr'\,
G_{\rm
bare}(n,r,r'){\tilde\omega}_{e}(n,r',\sigma).
\label{j5k}
\end{equation}
This is a Volterra-Fredholm equation relating the
Fourier-Laplace transform of the total stream function $\delta{\tilde\psi}_{\rm
tot}(n,r,\sigma)$
to the Fourier-Laplace transform of the external vorticity
$\tilde\omega_e(n,r,\sigma)$. The formal solution of
this integral equation reads
\begin{eqnarray}
\delta{\tilde\psi}_{\rm tot}(n,r,\sigma)=\int_0^{+\infty} 2\pi r'  dr'\,
G(n,r,r',\sigma){\tilde\omega}_e(n,r',\sigma),
\label{j5b}
\end{eqnarray}
where the Green function $G(n,r,r',\sigma)$ is defined by 
\begin{equation}
G(n,r,r',\sigma)+2\pi \int_0^{+\infty} r''dr''\, 
G_{\rm
bare}(n,r,r'')\frac{n\frac{1}{r''}\frac{\partial\omega''}{\partial
r''}}{n\Omega''-\sigma}G(n,r'',r',\sigma) =G_{\rm bare}(n,r,r').
\label{pot11ht}
\end{equation}
For any $\sigma$ the integration in Eq. (\ref{pot11ht}) must be 
performed along the Landau contour. As
a result, $G(n,r,r',\sigma)$ is a complex function which can be interpreted as
the Fourier-Laplace transform of the dressed potential of interaction:
\begin{equation}
G(n,r,r',\sigma)=\tilde{u}^d_n(r,r',\sigma).
\label{pot6blues2}
\end{equation}
It determines the
response of the system $\delta{\tilde\psi}_{\rm tot}(n,r,\sigma)$ to an external
perturbation $ {\tilde\omega}_e(n,r,\sigma)$ through Eq. (\ref{j5b}). 
If we neglect collective effects in Eq. (\ref{j5g}), we obtain
\begin{eqnarray}
\delta{\tilde\psi}_{\rm
tot}(n,r,\sigma)={\tilde\psi}_{e}(n,r,\sigma)=\int_0^{+\infty}
2\pi r' dr'\,
G_{\rm bare}(n,r,r'){\tilde\omega}_e(n,r',\sigma).
\label{j5bbare}
\end{eqnarray}
This amounts to replacing the dressed potential of interaction
$G(n,r,r',\sigma)$ by the bare potential of interaction $G_{\rm bare}(n,r,r')$
in Eq. (\ref{j5b}). Eqs.
(\ref{j5k}), (\ref{j5b}) and (\ref{pot11ht}) are the counterparts of Eqs.
(C17), (C18) and (C19) of \cite{kinfd} for inhomogeneous stellar
systems and of Eqs. (26), (27) and (28) of  \cite{kinfdvortex} for
unidirectional flows.

If the stream function is related to the vorticity by
the Poisson equation (\ref{poiss}), we have
\begin{eqnarray}
\Delta\delta\psi_{\rm tot}=-\delta\omega_{\rm tot}=-\delta\omega-\omega_e.
\label{ham18}
\end{eqnarray}
Introducing a polar system of coordinates, it can be written as
\begin{eqnarray}
\label{sb2avt}
\frac{1}{r}\frac{\partial}{\partial r}\left
(r\frac{\partial\delta\psi_{\rm tot}}{\partial
r}\right )+\frac{1}{r^2}\frac{\partial^2\delta\psi_{\rm tot}}{\partial
\theta^2}=-\delta\omega-\omega_e.
\end{eqnarray}
Taking the Fourier-Laplace transform of this equation,  we obtain
\begin{eqnarray}
\left\lbrack
\frac{1}{r}\frac{d}{dr}r\frac{d}{dr}-\frac{n^2}{r^2}\right\rbrack
\delta{\tilde\psi}_{\rm
tot}(n,r,\sigma)=-\delta{\tilde\omega}(n,r,\sigma)-{\tilde\omega}_e(n,r,\sigma).
\label{j4ju}
\end{eqnarray}
Substituting Eq.
(\ref{bam}) into Eq. (\ref{j4ju}) we find that
\begin{eqnarray}
\left\lbrack
\frac{1}{r}\frac{d}{dr}r\frac{d}{dr}-\frac{n^2}{r^2}-\frac{n\frac{1}{r}\frac{
\partial\omega } { \partial
r}}{n\Omega(r)-\sigma}\right\rbrack
\delta{\tilde\psi}_{\rm tot}(n,r,\sigma)=-{\tilde\omega}_e(n,r,\sigma).
\label{j4}
\end{eqnarray}
This differential equation, which relates
$\delta{\tilde\psi}_{\rm
tot}(n,r,\sigma)$
to
$\tilde\omega_e(n,r,\sigma)$, is equivalent to the Volterra-Fredholm equation
(\ref{j5k}).
Its formal solution is given by Eq. (\ref{j5b}) where the Green function
$G(n,r,r',\sigma)$ is defined by
\begin{eqnarray}
\left\lbrack
\frac{1}{r}\frac{d}{dr}r\frac{d}{dr}-\frac{n^2}{r^2}-
\frac{n\frac{1}{r}\frac{
\partial\omega } { \partial
r}}{n\Omega(r)-\sigma}\right\rbrack G(n,r,r',\sigma)=-\frac{\delta(r-r')}{2\pi
r}.
\label{defg}
\end{eqnarray}
If we neglect collective effects, we recover
Eq. (\ref{pot6}) for the bare potential of
interaction $G_{\rm bare}(n,r,r')$.

{\it Remark:} From Eqs. (\ref{pot10g}) and (\ref{bam}) we get
\begin{equation}
\delta{\tilde\psi}(n,r,\sigma)=-\int_0^{+\infty} 2\pi r'  dr'\,
G_{\rm bare}(n,r,r')\frac{n\frac{1}{r'}\frac{\partial\omega'}{
\partial
r'}}{n\Omega'-\sigma}\delta{\tilde\psi}(n,r',\sigma)
-\int_0^{+\infty} 2\pi r'  dr'\,
G_{\rm bare}(n,r,r')\frac{n\frac{1}{r'}\frac{\partial\omega'}{
\partial
r'}}{n\Omega'-\sigma}{\tilde\psi}_{e}(n,r',\sigma)
\label{j5c}
\end{equation}
and
\begin{eqnarray}
\label{aaa}
\delta{\tilde\omega}(n,r,\sigma)=-\frac{n\frac{1}{r}\frac{\partial\omega}{
\partial
r}}{n\Omega-\sigma}\int_0^{+\infty} 2\pi r'  dr'\,
G_{\rm bare}(n,r,r')\delta{\tilde\omega}(n,r',\sigma)
-\frac{n\frac{1}{r}\frac{
\partial\omega}{
\partial
r}}{n\Omega-\sigma}{\tilde\psi}_{e}(n,r,\sigma).
\end{eqnarray}

\subsection{Dressed potential of interaction from an initial perturbation}
\label{sec_dpicv}

In Appendix \ref{sec_marr}, we have studied the response
of an axisymmetric flow described by the 2D Euler equation to a weak external
stream function 
$\psi_e(r,\theta,t)$ by using the
linear response theory. Here, we compare these
results to those obtained when the system is isolated (i.e.,
$\psi_e(r,\theta,t)=0$) but the vorticity is slightly perturbed at $t=0$. Taking
the
Fourier-Laplace transform of the linearized 2D Euler equation (\ref{j2v}), and
assuming now that $\delta \omega(r,\theta,0)\neq 0$, we obtain
\begin{eqnarray}
\label{dq1}
\delta{\tilde\omega}(n,r,\sigma)=-\frac{n\frac{1}{r}\frac{\partial\omega}{
\partial
r}}{n\Omega-\sigma}\delta{\tilde\psi}(n,r,\sigma)+\frac{\delta{\hat\omega}(n,r
,0)}{i(
n\Omega-\sigma)},
\end{eqnarray}
where $\delta\hat{\omega}(n,r,0)$ is the Fourier transform of the
initial
perturbation $\delta \omega(r,\theta,0)$. According to Eq.
(\ref{pot10}) the fluctuations of the stream function are related to the
fluctuations of
the vorticity by
\begin{eqnarray}
\delta{\tilde\psi}(n,r,\sigma)=\int_0^{+\infty} 2\pi r'  dr'\,
G_{\rm bare}(n,r,r')\delta{\tilde\omega}(n,r',\sigma).
\label{dq2}
\end{eqnarray}
Substituting Eq. (\ref{dq1}) into Eq. (\ref{dq2}), we obtain
\begin{equation}
\delta\tilde\psi(n,r,\sigma)+2\pi \int_0^{+\infty} r'dr'\, 
G_{\rm bare}(n,r,r')\frac{n\frac{1}{r'}\frac{\partial\omega'}{\partial
r'}}{n\Omega'-\sigma}\delta\tilde\psi(n,r',\sigma) =2\pi \int_0^{+\infty}
r'dr'\,  G_{\rm bare}(n,r,r')
\frac{\delta\hat\omega(n,r',0)}{i(n\Omega'-\sigma)}.
\label{dq3}
\end{equation}
This is a Volterra-Fredholm integral equation relating the Fourier-Laplace
transform of the fluctuations of the stream function
$\delta{\tilde\psi}(n,r,\sigma)$  to the Fourier transform of the initial
fluctuations of the vorticity to $\delta{\hat\omega}(n,r,0)$. It is similar to
Eq. (\ref{j5k}). Its formal
solution is
\begin{eqnarray}
\delta{\tilde\psi}(n,r,\sigma)=\int_0^{+\infty} 2\pi r'  dr'\,
G(n,r,r',\sigma)\frac{\delta{\hat\omega}(n,r',0)}{i(
n\Omega'-\sigma)},
\label{dq4}
\end{eqnarray}
where the Green function $G(n,r,r',\sigma)$ is defined by Eq. (\ref{pot11ht}).
If we neglect collective
effects in Eq. (\ref{dq1}), the
foregoing equation reduces to
\begin{equation}
\delta\tilde\psi(n,r,\sigma) =2\pi \int_0^{+\infty} r'dr'\,  G_{\rm
bare}(n,r,r')
\frac{\delta\hat\omega(n,r,0)}{i(n\Omega-\sigma)}.
\label{dq6}
\end{equation}
This amounts to replacing the dressed potential of interaction
$G(n,r,r',\sigma)$ by the bare potential of interaction $G_{\rm bare}(n,r,r')$
in Eq. (\ref{dq4}). Eqs.
(\ref{pot11ht}), (\ref{dq2}), (\ref{dq3}) and (\ref{dq4}) are the counterparts
of Eqs. (37), (A5), (A6) and (A8) of \cite{angleaction2} or Eqs. (C19), (H4),
(H5) and (H6) of \cite{kinfd} for inhomogeneous stellar
systems and of Eqs. (28), (B2), (J4) and (J5) of \cite{kinfdvortex}  for
unidirectional flows.

For the usual interaction between point vortices we
can alternatively proceed in the following manner. The fluctuations of
stream function and vorticity are
related by the Poisson equation
\begin{eqnarray}
\label{sb1}
\Delta\delta\psi=-\delta\omega.
\end{eqnarray}
Introducing a  polar system of coordinates, it can be written as
\begin{eqnarray}
\label{sb2}
\frac{1}{r}\frac{\partial}{\partial r}\left (r\frac{\partial\delta\psi}{\partial
r}\right )+\frac{1}{r^2}\frac{\partial^2\delta\psi}{\partial
\theta^2}=-\delta\omega.
\end{eqnarray}
Taking the Fourier-Laplace transform of this equation,  we obtain
\begin{equation}
\left\lbrack \frac{1}{r}\frac{d}{dr} r \frac{d}{d
r}-\frac{n^2}{r^{2}}\right \rbrack
\delta\tilde\psi(n,r,\sigma)=-\delta\tilde\omega(n,r,\sigma).
\label{pot3}
\end{equation}
Substituting Eq. (\ref{dq1}) into Eq. (\ref{pot3}), we find
that
\begin{equation}
\left\lbrack \frac{1}{r}\frac{d}{dr} r
\frac{d}{dr}-\frac{n^2}{r^{2}}-\frac{n\frac{1}{r}\frac{\partial\omega}{
\partial
r}}{n\Omega-\sigma}\right \rbrack
\delta\tilde\psi(n,r,\sigma)=-\frac{\delta\hat\omega(n,r,0)}{i(n\Omega-\sigma)}
.
\label{pot4}
\end{equation}
This differential equation relating
$\delta{\tilde\psi}(n,r,\sigma)$
to $\delta{\hat\omega}(n,r,0)$  is equivalent to the Volterra-Fredholm integral
equation
(\ref{dq3}).
Its
formal solution  is given by Eq. (\ref{dq4}) where
the Green function $G(n,r,r',\sigma)$  is defined by Eq. (\ref{defg}).
If we neglect collective effects, we recover Eq.
(\ref{pot6}).

\subsection{Rayleigh equation}

In the absence of external  ($\omega_e=0$) or
initial ($\delta\omega(t=0)=0$) perturbation, Eqs. (\ref{j5k}) and
(\ref{dq3}) reduce to the homogeneous differential equation
\begin{equation}
\delta\tilde\psi(n,r,\sigma)+2\pi \int_0^{+\infty} r'dr'\, 
G_{\rm bare}(n,r,r')\frac{n\frac{1}{r'}\frac{\partial\omega'}{\partial
r'}}{n\Omega'-\sigma}\delta\tilde\psi(n,r',\sigma) =0.
\label{raynw}
\end{equation}
Similarly, for the usual interaction between point vortices, Eqs. (\ref{defg})
and
(\ref{pot4}) reduce to the so-called Rayleigh \cite{rayleigh} equation
\begin{eqnarray}
\left\lbrack
\frac{1}{r}\frac{d}{dr}r\frac{d}{dr}-\frac{n^2}{r^2}-\frac{n\frac{1}{r}\frac{
\partial\omega } { \partial
r}}{n\Omega(r)-\sigma}\right\rbrack
\delta{\tilde\psi}(n,r,\sigma)=0.
\label{j4ray}
\end{eqnarray}
The Rayleigh equation determines the proper
complex pulsations $\sigma_n$ of the flow as a function
of the wavenumber $n$. It plays the role of the dispersion relation in plasma
physics and stellar dynamics \cite{nicholson,bt}. It can be used
to study the linear dynamical stability of an axisymmetric flow with respect to
the 2D Euler equation \cite{drazin}. The mode $n$
is
stable when ${\rm Im}(\sigma_n)<0$ and unstable
when ${\rm Im}(\sigma_n)>0$. The condition of marginal
stability is  ${\rm Im}(\sigma_n)=0$. The Rayleigh equation can be rewritten as
\begin{eqnarray}
\left\lbrack
\frac{1}{r}\frac{d}{dr}r\frac{d}{dr}-\frac{n^2}{r^2}-\frac{\Omega''+\frac{
3\Omega'}{r}}{\Omega(r)-\sigma/n}\right\rbrack
\delta{\tilde\psi}(n,r,\sigma)=0,
\label{j4rayy}
\end{eqnarray}
where we have used Eq. (\ref{ham10}).

\subsection{Volterra-Fredholm equation in time with an external perturbation}
\label{sec_volterrab}

Taking the Fourier transform of Eq. (\ref{ham13}), we obtain
\begin{eqnarray}
\label{j7ann}
\frac{\partial \delta{\hat \omega}}{\partial t}+in\Omega\delta{\hat
\omega}=-\frac{in}{r}\frac{\partial \omega}{\partial r}\delta{\hat\psi}_{\rm
tot}.
\end{eqnarray}
This first order differential equation in time can be solved with the method of
the variation of the constant, giving
\begin{eqnarray}
\label{vv1}
\delta{\hat \omega}(n,r,t)=-\frac{in}{r}
\frac{\partial \omega}{\partial r}e^{-in\Omega t}\int_0^t dt'\, \delta{\hat
\psi}_{\rm tot}(n,r,t')e^{in\Omega t'},
\end{eqnarray}
where we have assumed $\delta{\hat \omega}(n,r,t=0)=0$. According
to Eq. (\ref{pot10}) the fluctuations of the stream function are related to the
fluctuations of
the vorticity by 
\begin{eqnarray}
\delta{\hat\psi}(n,r,t)=\int_0^{+\infty} 2\pi r'  dr'\,
G_{\rm bare}(n,r,r')\delta{\hat\omega}(n,r',t).
\label{vv2}
\end{eqnarray}
Substituting Eq. (\ref{vv1}) into Eq. (\ref{vv2}), we obtain
\begin{eqnarray}
\delta{\hat\psi}(n,r,t)=-\int_0^{+\infty} 2\pi r'  dr'\,
G_{\rm bare}(n,r,r')\frac{in}{r'}
\frac{\partial \omega'}{\partial r'}e^{-in\Omega' t}\int_0^t dt'\,
\left\lbrack \delta{\hat
\psi}(n,r',t')+{\hat
\psi}_e(n,r',t')\right\rbrack e^{in\Omega' t'}.
\label{vv3}
\end{eqnarray}
Eq. (\ref{vv3}) is a Volterra-Fredholm integral equation equivalent to Eq.
(\ref{j5k}).

\subsection{The resolvent of the linearized 2D Euler equation}
\label{sec_resolvent2}

Substituting Eq. (\ref{dq4})  into Eq.
(\ref{j3}), we obtain
\begin{eqnarray}
\label{rv1}
\delta{\tilde\omega}(n,r,\sigma)=-\frac{n\frac{1}{r}\frac{\partial\omega}{
\partial
r}}{n\Omega-\sigma}\int_0^{+\infty} 2\pi r'  dr'\,
G(n,r,r',\sigma)\frac{\delta{\hat\omega}(n,r',0)}{i(
n\Omega'-\sigma)}+\frac{\delta{\hat\omega}(n,r
,0)}{i(
n\Omega-\sigma)}.
\end{eqnarray}
This equation relates the Fourier-Laplace transform of the
fluctuations of the vorticity to the Fourier
transform of the initial fluctuations of the vorticity. This is the exact
solution of the initial value problem for
the linearized 2D Euler equation. It can be conveniently written in terms of an
operator that
connects $\delta\tilde{\omega}(n,r,\sigma)$ to the initial
value $\delta{\hat\omega}(n,r',0)$.
One has 
\begin{eqnarray}
\label{rv2}
\delta{\tilde\omega}(n,r,\sigma)=\int_0^{+\infty} 2\pi r'  dr'\,
R(n,r,r',\sigma)\delta{\hat\omega}(n,r',0),
\end{eqnarray}
where
\begin{eqnarray}
\label{rv3}
R(n,r,r',\sigma)=-\frac{n\frac{1}{r}\frac{\partial\omega}{
\partial
r}}{n\Omega-\sigma}
G(n,r,r',\sigma)\frac{1}{i(
n\Omega'-\sigma)}+\frac{\delta(r-r')}{2\pi r}\frac{1}{i(
n\Omega-\sigma)}
\end{eqnarray}
is called the resolvent or the propagator of the linearized 2D Euler equation.
If
we consider an initial condition of the form $\delta
\omega(r,\theta,0)=\gamma\delta(\theta-\theta')\delta(r-r')/r$, implying
\begin{equation}
\delta\hat{\omega}(n,r,0)=\frac{\gamma}{2\pi r}e^{-in\theta'}\delta(r-r'),
\label{rv4}
\end{equation}
we find that
\begin{equation}
\delta\tilde \omega
(n,r,\sigma)=-\frac{n\frac{1}{r}\frac{\partial\omega}{\partial
r}}{\sigma-n\Omega(r)}\gamma
e^{-in\theta'}\frac{G(n,r,r',\sigma)}{i(\sigma-n\Omega(r'))}-\gamma\frac{e^{-in
\theta'}}{2\pi r} \frac{\delta(r-r')}{i(\sigma-n\Omega(r))}.
\label{rv5}
\end{equation}

{\it Remark:} If we substitute Eq. (\ref{dq2}) into Eq. (\ref{j3}), we obtain
the
integral equation
\begin{eqnarray}
\label{rv6}
\delta{\tilde\omega}(n,r,\sigma)=-\frac{n\frac{1}{r}\frac{\partial\omega}{
\partial
r}}{n\Omega-\sigma}\int_0^{+\infty} 2\pi r'  dr'\,
G_{\rm bare}(n,r,r')\delta{\tilde\omega}(n,r',\sigma)
+\frac{\delta{\hat\omega}(n,r
,0)}{i(
n\Omega-\sigma)}.
\end{eqnarray}
This equation relates the Fourier-Laplace transform of the
fluctuations of the vorticity to the Fourier transform of
the initial fluctuations of the vorticity, so it is
equivalent to Eq. (\ref{rv1}).

\subsection{Useful identities}
\label{sec_idv}

The Green function $G(n,r,r',\sigma)$ introduced in Appendix \ref{sec_inhosrf}
is
determined by the equation 
\begin{eqnarray}
\frac{1}{r}\frac{d}{dr}\left
(r\frac{dG}{dr}\right
)-\frac{n^2}{r^2}G-\frac{n\frac{1}{r}\frac{\partial\omega}{\partial
r}}{n\Omega(r)-\sigma}G=-\frac{\delta(r-r')}{2\pi r}.
\label{hami1}
\end{eqnarray}
Multiplying Eq.
(\ref{hami1}) by
$2\pi r G(n,r,r',\sigma)^*$ and integrating over $r$ between $0$ and $+\infty$,
we get
\begin{eqnarray}
-\int_{0}^{+\infty}  \left |\frac{dG}{dr}\right |^2\,
2\pi r dr-\int_{0}^{+\infty}\frac{n^2}{r} |G|^2 2\pi\,
dr-\int_{0}^{+\infty}\frac{n\frac{\partial\omega}{\partial
r}}{n\Omega(r)-\sigma}|G|^2 2\pi\, dr=-G(n,r',r',\sigma)^*,
\label{hami2}
\end{eqnarray}
where we have integrated the first term by parts. Taking the imaginary part of
this equation, we find that
\begin{eqnarray}
{\rm Im}\,  G(n,r',r',\sigma)=- {\rm Im}\, \int_{0}^{+\infty}
\frac{n\frac{\partial\omega}{\partial
r}}{n\Omega(r)-\sigma}|G|^2(n,r,r',\sigma)2\pi\,
dr.
\label{hami3}
\end{eqnarray}
For $\sigma$ real, using the Landau prescription
$\sigma\rightarrow \sigma+i0^+$ and applying the Sokhotski-Plemelj formula
(\ref{plemelj}), we obtain the identity 
\begin{eqnarray}
{\rm Im}\,  G(n,r,r,\sigma)=-2\pi^2\int_{0}^{+\infty}
n\frac{\partial\omega'}{\partial r'} \delta(n\Omega(r')-\sigma)
|G(n,r',r,\sigma)|^2\, dr'.
\label{hami4}
\end{eqnarray}
We also have
\begin{eqnarray}
G(-n,r,r',-\sigma)=G(n,r,r',\sigma)^*.
\label{obv}
\end{eqnarray}

\subsection{The function $\chi_{\rm bare}(r,r')$}
\label{sec_nce}

In Sec. \ref{sec_monov} we have introduced the function
\begin{equation}
\chi(r,r',\Omega(r))=\sum_n |n| |G(n,r,r',n\Omega(r))|^2.
\label{poet}
\end{equation}
If we neglect collective effects, it reduces to
\begin{equation}
\chi_{\rm bare}(r,r')=\sum_n |n| G_{\rm bare}(n,r,r')^2.
\label{poetbare}
\end{equation}
Using Eqs. (\ref{pot13}) and (\ref{pot6blues}), the series in Eq.
(\ref{poetbare}) can be calculated
analytically  \cite{pre} yielding
\begin{equation}
\chi_{\rm bare}(r,r')=\frac{1}{8\pi^2}\sum_{n=1}^{+\infty}\frac{1}{n}\left
(\frac{r_<}{r_>}\right )^{2n}=-\frac{1}{8\pi^2}\ln\left\lbrack 1-\left
(\frac{r_<}{r_>}\right )^2\right\rbrack.
\label{pot15}
\end{equation}
Taking $r'=r$ in the foregoing expression, we obtain
\begin{equation}
\chi_{\rm bare}(r,r)=\frac{1}{8\pi^2}\sum_{n=1}^{+\infty}\frac{1}{n}=\frac{1}{
8\pi^2
}\ln\Lambda,
\label{pot16}
\end{equation}
where $\ln\Lambda$ is the Coulomb logarithm which must be
regularized with
appropriate cut-offs \cite{klim}. Analogous
expressions, valid in a circular domain of size $R$, are given in \cite{bbgky}.

\end{document}